\documentclass[final,11pt]{article}

\usepackage{multirow}

\usepackage[nottoc]{tocbibind}

\usepackage{geometry}
\usepackage{subcaption}
\usepackage[affil-it, auth-sc]{authblk}
\usepackage[english]{babel}
\usepackage[utf8]{inputenc}
\usepackage{cite}

\usepackage{graphicx}
\usepackage{amssymb}
\usepackage{amsthm}
\usepackage{amsmath}

\usepackage{color, soul}
\usepackage{verbatim}
\usepackage{lineno}
\usepackage{todonotes}
\presetkeys{todonotes}{inline, color=blue!30, size=\small}{}

\definecolor{darkred}{rgb}{0.9, 0.0, 0.0}
\definecolor{darkgreen}{rgb}{0.0, 0.5, 0.0}
\usepackage[colorlinks,bookmarks,linkcolor=darkred, citecolor=darkgreen]{hyperref}

\usepackage{eso-pic}

\usepackage[sort&compress,numbers]{natbib}
\usepackage{doi}

\usepackage{bm}
\usepackage{feynmf}

\usepackage{amssymb,color,paralist,amsmath,epsfig}
\usepackage{graphicx}
\usepackage{comment}
\usepackage{amsfonts}
\usepackage{relsize}
\usepackage{nicefrac,esint}
\usepackage{float}
\usepackage{cleveref}

\makeatletter
\newcommand\footnoteref[1]{\protected@xdef\@thefnmark{\ref{#1}}\@footnotemark}
\makeatother

\newcommand{\bk}{{\boldsymbol k}}

\newcommand{\beq}{\begin{equation}}
\newcommand{\eeq}{\end{equation}}

\newcommand{\ber}{\begin{eqnarray}}
\newcommand{\eer}{\end{eqnarray}}

\newcommand{\order}{{\cal O}}
\def\slash#1{#1\!\!\!/\!\,\,}
\allowdisplaybreaks

\begin{document}

\AddToShipoutPictureFG*{
    \AtPageUpperLeft{\put(-60,-60){\makebox[\paperwidth][r]{FERMILAB-PUB-19-076-T}}}  
    }

\title{Theory of elastic neutrino-electron scattering}

\author[1,2,3,*]{Oleksandr Tomalak}
\author[1,2,*]{Richard J. Hill}

\affil[1]{Department of Physics and Astronomy, University of Kentucky, Lexington, KY 40506, USA \vspace{1.2mm}}
\affil[2]{Theoretical Physics Department, Fermi National Accelerator Laboratory, Batavia, IL 60510, USA  \vspace{1.2mm}}
\affil[3]{Institut f\"ur Kernphysik and PRISMA Cluster of Excellence, Johannes Gutenberg Universit\"at, Mainz, Germany  \vspace{1.2mm}}
\affil[*]{Corresponding authors: oleksandr.tomalak@uky.edu, Richard.Hill@uky.edu}

\date{\today}

\maketitle

 Theoretical predictions for elastic neutrino-electron scattering have
 no hadronic or nuclear uncertainties at leading order making this process an
 important tool for normalizing neutrino flux.  However, the process
 is subject to large radiative corrections that differ according to
 experimental conditions.  In this paper, we collect new and existing
 results for total and differential cross sections accompanied by
 radiation of one photon, $\nu e \to \nu e (\gamma)$.  We perform
 calculations within the Fermi effective theory and provide analytic
 expressions for the electron energy spectrum and for the total
 electromagnetic energy spectrum as well as for double- and
 triple-differential cross sections with respect to electron energy, electron
 angle, photon energy, and photon angle.
 We discuss illustrative applications to accelerator-based neutrino experiments 
 and provide the most precise up-to-date values of neutrino-electron scattering
 cross sections.
 We present an analysis of theoretical error, which is dominated by the
 $\sim 0.2 - 0.4\%$ uncertainty of the hadronic correction. We also discuss
 how searches for new physics can be affected by radiative corrections.

\newpage

\tableofcontents

\newpage

\section{Introduction}
\label{sec1}

In the Standard Model of particle physics, neutrinos are massless
particles.
However, experiments with solar~\cite{Cleveland:1998nv,Hampel:1998xg,Ahmad:2002jz,Abdurashitov:2002nt,Fukuda:2001nj,Ahmed:2003kj}, atmospheric~\cite{Fukuda:1998mi,Ashie:2004mr}, reactor~\cite{Eguchi:2002dm,Araki:2004mb,Abe:2012tg,Ahn:2012nd,An:2013uza}, and
accelerator~\cite{Ahn:2002up,Michael:2006rx,Abe:2013xua} neutrinos%
\footnote{
  For the purposes of this paper,``accelerator'' neutrinos have energy large compared to the electron mass.  
}
establish that neutrinos oscillate and have nonzero mass~\cite{Pontecorvo:1957cp,Pontecorvo:1967fh},
thus providing a convincing example of physics beyond the Standard Model.
Fundamental questions about this definitive portal to new physics remain unanswered:
What is the origin of neutrino mass?
Are lepton number and CP symmetries violated?
Do sterile neutrinos exist?  
What is the absolute scale and ordering of neutrino masses?
New experiments aim to address these questions but rely on a precise
description of neutrino interactions with the ordinary matter
(electrons and nuclei) used to detect them.

Interactions with atomic nuclei compose the bulk of
neutrino scattering events at accelerator neutrino experiments.
Although interactions with atomic electrons are rarer, they
are nonetheless valuable.  The neutrino-electron scattering process plays
an important dual role: first, owing to a clean experimental
signature and a small cross-section uncertainty, the process
provides an incisive constraint on neutrino flux~\cite{Park:2015eqa,Valencia:2019mkf};
second, the bulk of next-to-leading order (NLO) radiative corrections can be evaluated
analytically and thus serve as a prototype for the more complicated
cases of neutrino-nucleon and neutrino-nucleus scattering.

Radiative corrections to elastic neutrino-electron scattering of
order $\alpha$ were calculated first in Ref.~\cite{Lee:1964jq}, where
only soft-photon bremsstrahlung was considered.  In
Ref.~\cite{Ram:1967zza}, an analytical phase-space integration
technique was developed to include hard-photon bremsstrahlung, and the
electron energy spectrum for neutrino-electron scattering
accompanied by one radiated photon was obtained. The leading-order
(LO) cross section in the low-energy limit of the Weinberg
theory~\cite{Weinberg:1967tq} was evaluated in
Ref.~\cite{tHooft:1971ucy}.
References~\cite{Sarantakos:1982bp,Bahcall:1995mm} presented the
electron energy spectrum in the limit of small electron mass accounting for
corrections of order $\alpha$ and including other electroweak NLO
radiative corrections.  The electromagnetic energy spectrum was
considered in Refs.~\cite{Bardin:1983yb,Bardin:1985fg}.
Reference~\cite{Passera:2000ug} reproduced results of
Refs.~\cite{Ram:1967zza,Sarantakos:1982bp} by numerically performing the
phase-space integration, and accounted for
the electron mass suppressed interference term; Ref.~\cite{Passera:2000ug}
also presented a numerical evaluation of the electromagnetic energy spectrum. 
The hard-photon correction to the total elastic cross section was studied in
Refs.~\cite{Green:1980uc,Bardin:1983zm}.
Different aspects of radiative corrections in
elastic neutrino-electron scattering were also discussed in
Refs.~\cite{Salomonson:1974ys,Zhizhin:1975kv,Byers:1979af,Green:1980bd,Green:1980uc,Marciano:1980pb,Aoki:1980ix,Aoki:1981kq,Hioki:1981gi,Bardin:1983yb,Bardin:1983zm,Bardin:1985fg,Mourao:1989vb,Weber:1991kf,Buccella:1992xy,Bernabeu:1994kw,Passera:2000ug,Akhmedov:2018wlf}. 
See Refs.~\cite{Marciano:2003eq,Sirlin:2012mh} for recent reviews.
 
In this work, we analytically evaluate relevant distributions and
spectra in elastic (anti)neutrino-electron
scattering starting from four-fermion effective field theory (EFT). 
We take neutrino-lepton and neutrino-quark EFT coefficients from Ref.~\cite{in_preparation}
(with $n_f=4$ active quarks at renormalization scale $\mu = 2~\mathrm{GeV}$)
and calculate real and virtual corrections in the $\overline{\mathrm{MS}}$ 
renormalization scheme within this theory.
Exploiting the technique of Ref.~\cite{Ram:1967zza}, we
evaluate the electron energy spectrum and present this
calculation in a relatively compact form.  We generalize this technique
for the evaluation of the electromagnetic energy spectrum as well as
triple- and double-differential cross sections.
We discuss a new treatment of hadronic loop diagrams;
this contribution dominates the error budget for
neutrino-electron scattering, and impacts other neutral current
neutrino processes, such as coherent neutrino-nucleus scattering~\cite{Akimov:2017ade}.  
As illustrative applications using accelerator neutrino
beams~\cite{Ayres:2007tu,Abe:2013xua,Acciarri:2015uup,Abe:2011ts}, we
consider the impact of radiative corrections on energy spectra
and compare observables employing electron energy versus total
electromagnetic energy. For possible
low-energy applications, we provide results in analytic form keeping all 
charged lepton mass terms.
The complete mass dependence could be useful in the analysis of future
reactor and solar neutrino experiments
\cite{Abe:2010hy,Wurm:2011zn,Li:2013zyd,Giunti:2014ixa,Giunti:2015gga}.
We also discuss examples where radiative corrections can impact searches for
new physics, including neutrino charge radius effects. 

The paper is organized as follows.  Section~\ref{sec:tree} considers
the kinematics of neutrino-electron scattering and computes 
the tree-level scattering process including electroweak corrections to the
low-energy four-fermion interaction.
Section~\ref{sec:virtual} computes virtual corrections to elastic
scattering. Section~\ref{sec:real} represents the bulk of the paper and  
computes QED corrections involving real
radiation.  Section~\ref{sec:results} presents
illustrative results for total cross sections and electron energy versus total electromagnetic energy spectra. 
Section~\ref{sec:summary} presents our conclusions and outlook.
In the main text of the paper, we describe the general strategy of the computations
and focus on results in the limit of small electron mass (i.e., neutrino beam energy much larger than electron mass).
Appendixes provide general expressions retaining
all electron mass terms.
Appendix~\ref{app:QCD_QED_vector_vacuum_polarization}
summarizes higher-order perturbative QCD corrections to heavy-quark loops that are
discussed in Sec.~\ref{sec:long_Range}. 
Appendix~\ref{app:plots_experiments} displays flux-averaged spectra in experimental conditions of
DUNE, MINERvA, NOvA, and T2K experiments.

\section{Neutrino-electron scattering}
\label{sec:tree}

We begin in Sec.~\ref{sec:kinematics} by reviewing
the kinematics of neutrino scattering on atomic electrons. Throughout this section
we consider general charged leptons $\ell$, but in the following sections we specialize
to the phenomenologically most relevant case of the electron, $\ell=e$. 
We introduce the relevant basis of four-fermion effective operators
in Sec.~\ref{sec:effective_interaction} and discuss their coefficients in Sec.~\ref{sec:lecs}. 

\subsection{Kinematics for neutrino-electron scattering}
\label{sec:kinematics}

\begin{figure}[t]
\begin{center}
	\includegraphics[scale=1.]{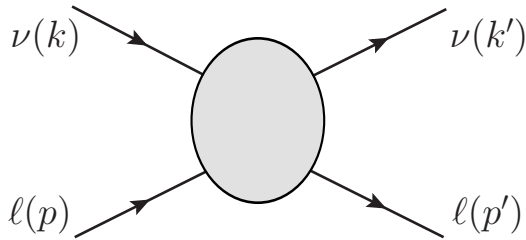}\hspace{0.cm}
\end{center}
	\caption{Neutrino-electron scattering kinematics.\label{scattering_process}}
\end{figure}

Consider the scattering of neutrinos on atomic electrons. We neglect
the atomic binding energy and momentum compared to the energy and momentum
transferred in the scattering process. 
Consequently, the initial electron is taken to be
at rest in the laboratory frame, where the kinematics is given by
$p^\mu = (m,~0)$
(initial electron with $p^2=m^2$),  $ ~ p^{\prime\mu} = (E^\prime,~\bk-\bk^\prime)$ (final charged
lepton with $p'^2 = m'^2$), $ k^\mu = (\omega,~\bk)$ (initial neutrino),
and $k^{\prime\mu} = (\omega^\prime,~\bk^\prime)$ (final neutrino); see
Fig.~\ref{scattering_process}.
The neutrino mass scale is much lower
than the electron mass and typical neutrino beam energy, and we
neglect the neutrino mass $m_\nu$ throughout. 
We will let $q^\mu = p^{\prime\mu} - p^\mu $
denote the momentum transfer to the charged lepton
and write $m_e=m$ for the electron mass. 

Elastic scattering is described by two independent kinematical variables.
It is convenient to introduce the invariant momentum transfer, 
\ber
q^2 = \left(p' - p \right)^2 \,,
\eer
and the squared energy in the center-of-mass reference frame, 
\ber
s=\left(p + k \right)^2  \,.
\eer
Note that production of heavier charged leptons in neutrino-electron scattering is
possible when the neutrino beam energy is high enough.  Using $s = m^2 + 2 m \omega \ge m'^2$
we see that 
$\omega \ge (m_\mu^2-m^2)/(2 m) \approx 10.9~\mathrm{GeV}$
to produce a muon ($m^\prime =m_\mu$), while $\omega \ge 3089~\mathrm{GeV}$ for the production of $\tau$
 ($m^\prime=m_\tau$). 

The neutrino scattering angle in the laboratory frame, $\Theta_{\nu}$, can be expressed in terms of the final neutrino energy $\omega'$ as
\ber
\cos \Theta_{\nu}  =  \frac{ \omega \omega^\prime - m (\omega - \omega^\prime) - \frac{m^2 - m'^2}{2} }{|\bf{k}|  |\bf{k}^\prime| } = 1 +\frac{m}{\omega} - \frac{m}{\omega'}- \frac{m^2 - m'^2}{2\omega \omega'}\,. \label{angle_thetanu}
\eer
The final neutrino energy varies between backward and forward scattering in the range:
\ber
 \frac{m \omega}{m+2 \omega} + \frac{m^2-m'^2}{2 \left( m + 2 \omega \right)} \le \omega' \le \omega + \frac{m^2 - m'^2}{2 m},
\eer
corresponding to the charged lepton energy range
\ber
m +  \frac{m'^2 - m^2}{2 m} \le E' \le  m + \frac{2 \omega^2}{m+2 \omega} + \frac{m'^2-m^2}{2 \left( m + 2 \omega \right)}.
\eer
The angle between recoil charged lepton direction and the neutrino beam direction, $\Theta_e$, is given by
\ber
\cos \Theta_e = \frac{ \omega E'  - m^2 - m ( \omega - E') +  \frac{m^2 - m'^2}{2} }{\omega | \bf{p}'|}, \label{angle_thetae}
\eer
and scattering is possible only in the forward cone bounded by $\Theta^\mathrm{max}_e$,
\ber
\cos \Theta^\mathrm{max}_e = \sqrt{\frac{m'^2-m^2}{m'^2} \frac{ \left(2 \omega + m \right)^2- m'^2}{4 \omega^2}}.
\eer
The scattering angle expression simplifies for the elastic process ($m'=m$) to
\ber
\cos \Theta_e = \frac{ m+\omega}{\omega} \sqrt{\frac{E'-m}{E'+m}},
\eer
when it varies between 0 and 1, i.e., the electron is scattered always into the forward hemisphere.

\subsection{Effective neutrino-charged lepton operators}
\label{sec:effective_interaction}

\begin{figure}[t]
\vspace{-0.25cm}	
\begin{center}
\begin{subfigure}{5cm}
\hspace{0.7cm}	\includegraphics[scale=1.]{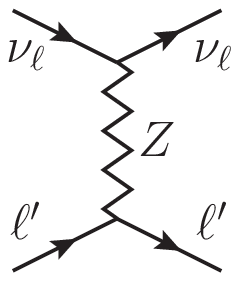}\hspace{0.cm}
\end{subfigure}
\begin{subfigure}{5cm}
\hspace{0.7cm}	\includegraphics[scale=1.]{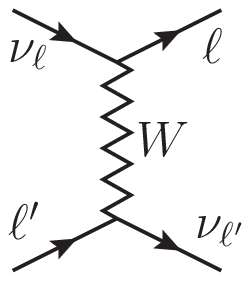}\hspace{0.cm}
\end{subfigure}
\begin{subfigure}{5cm}
\hspace{0.7cm}	 \includegraphics[scale=1.]{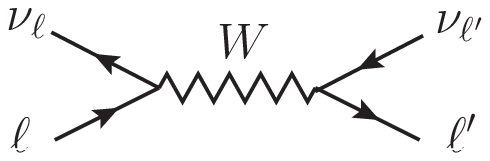}\hspace{0.cm}
\end{subfigure}
\caption{Leading-order contributions to neutrino-lepton scattering in the Standard Model. The graph with the exchange of $Z$ boson contributes to the neutrino and antineutrino scattering. $\ell$ and $\ell'$ denote charged leptons of any flavor in this figure.\label{feynman_graphs}
}
\end{center}
\end{figure}

Neutrino-electron scattering is described by the exchange of weak vector bosons $W$ and $Z$ (with masses $M_W$ and $M_Z$, respectively) in the Standard Model; cf. Fig.~\ref{feynman_graphs} for contributing Feynman diagrams.
At energies below the electroweak scale, the interactions of neutrinos and charged leptons are determined
by an equivalent effective Lagrangian~\cite{Fermi:1934hr,Feynman:1958ty,Sudarshan:1958vf}.
Neglecting corrections suppressed by $1/M_W^2$, the effective
Lagrangian consists of momentum-independent four-fermion operators.

At tree level, the matching onto this effective Lagrangian ${\cal L}_{\rm eff}$ is readily obtained,
\begin{align}
  {\cal L}_{\rm eff} &= - {g^2 \over M_W^2} (J_{W+})^\mu   (J_{W^-})_{\mu}   - {g^2 \over 2 M_Z^2} \left( J_Z\right)^\mu \left(J_Z \right)_\mu \,, 
\end{align}
where $J_{W^-}^\mu$, $J_{W^+}^\mu=J_{W^-}^{\dagger~~\mu}$, and $J_Z^\mu$
are charged and neutral currents in the Standard Model Lagrangian coupling to $W^+$, $W^-$, and $Z$, respectively,
and  $g$ is the electroweak $\mathrm{SU}(2)_\mathrm{L}$ coupling constant. 
Focusing on leptonic versus quark operators, we have 
\ber
  J_{W^-}^\mu &=& {1\over \sqrt{2}} \sum_{\ell} \bar{\ell} \gamma^\mu \mathrm{P}_\mathrm{L} \nu_{\ell} \,, \\
   J_Z^\mu &=& \frac{1}{\cos \theta_W}\sum_\ell\bigg[  \left( -\frac12 + \sin^2 \theta_W \right) \bar{\ell} \gamma^\mu \mathrm{P}_\mathrm{L} \ell
    + \sin^2 \theta_W \bar{\ell} \gamma^\mu  \mathrm{P}_\mathrm{R} \ell 
    + \frac12 \bar{\nu}_\ell \gamma^\mu \mathrm{P}_\mathrm{L} \nu_\ell \bigg] \,,
\eer
where $\mathrm{P}_\mathrm{L} = (1-\gamma_5)/{2}$
and 
$\mathrm{P}_\mathrm{R} = (1+\gamma_5)/{2}$
are projection operators onto left-handed and right-handed fermions
and $\theta_W$ denotes the weak mixing angle satisfying $M_W/M_Z = \cos\theta_W$.
After Fierz rearrangement of the charged current
contribution, the result may be written as
\begin{align}
  {\cal L}_{\rm eff} &=  
 -  \sum_{\ell,\ell^\prime}  \bar{\nu}_\ell \gamma^\mu \mathrm{P}_\mathrm{L} \nu_\ell
  \, \bar{\ell}^\prime \gamma_\mu (c_\mathrm{L}^{\nu_\ell \ell^\prime} \mathrm{P}_\mathrm{L} + c_\mathrm{R} \mathrm{P}_\mathrm{R}) \ell^\prime
  - c \sum_{\ell \ne \ell^\prime} 
  \bar{\nu}_{\ell^\prime }\gamma^\mu \mathrm{P}_\mathrm{L} \nu_{\ell}
  \, \bar{\ell} \gamma_\mu \mathrm{P}_\mathrm{L}  \ell^\prime \,, \label{effective_lagrangian_leptons}
\end{align}
with coefficients $c_\mathrm{L}^{\nu_{\ell} \ell'},~c_\mathrm{R}$, and $c$,
\begin{align}
 c_\mathrm{L}^{\nu_{\ell} \ell'} = 2 \sqrt{2} \mathrm{G}_\mathrm{F} \left( \sin^2 \theta_W - \frac12 + \delta_{\ell \ell'}\right) \,, \quad
 c_\mathrm{R} = 2 \sqrt{2} \mathrm{G}_\mathrm{F} \sin^2 \theta_W \,, \quad c = 2 \sqrt{2} \mathrm{G}_\mathrm{F}
 \,, 
\end{align}
where we have introduced the Fermi constant $\mathrm{G}_\mathrm{F} = {g^2}/(4 \sqrt{2}M_W^2)$, and  
where the Kronecker symbol $\delta_{\ell \ell'}$ satisfies $ \delta_{\ell \ell'} = 1$
for $\ell = \ell'$ and $ \delta_{\ell \ell'} = 0$ for $\ell \neq \ell'$.
Note that coefficients $c$ and $c_\mathrm{R}$ are the same for all combinations of lepton flavors,
while the coefficient $c_\mathrm{L}^{\nu_{\ell} \ell'}$ depends on whether the neutrino and charged lepton have
the same flavor.

Neglecting the neutrino magnetic moment contribution~\cite{Nahmias:1935yih,Crane:1948zz,Barrett:1950,Kulp:1952,Houtermans:1954,Cowan:1954pq,Bernstein:1963qh},
the leading-order cross section of neutrino-lepton scattering can be expressed, in all possible cases, as~\cite{tHooft:1971ucy,Hasert:1973cr,Reines:1976pv,Wolfenstein:1977ue,Wolfenstein:1979ni,Okun:1982ap,Sarantakos:1982bp,Allen:1985xx,Mikheev:1986gs,Bethe:1986ej,Parke:1986jy,Rosen:1986jy,Dorenbosch:1988is,Allen:1989dr,Ahrens:1990fp,Vidyakin:1992nf,Allen:1992qe,Derbin:1993wy,Horejsi:1993hz,Vilain:1994qy,Passera:2000ug,Auerbach:2001wg,Liu:2004ny,Daraktchieva:2005kn,Beda:2007hf,Arpesella:2008mt,Deniz:2009mu,Beda:2012zz,Formaggio:2013kya}
\ber
\frac{\mathrm{d} \sigma^{ \nu_{\ell} \ell' \to \nu_{\ell}  \ell' }_\mathrm{LO}}{\mathrm{d} \omega'}  \hspace{-0.15cm} &=&
\hspace{-0.15cm} \frac{m}{4 \pi} \left[ \left(c_{\mathrm{L}}^{\nu_{\ell} \ell'}\right)^2 \mathrm{I}_\mathrm{L}
+   c_\mathrm{R}^2 \mathrm{I}_\mathrm{R} + c_{\mathrm{L}}^{\nu_{\ell} \ell'} c_\mathrm{R}  \mathrm{I}^\mathrm{L}_\mathrm{R} \right],   \label{elastic0_xsection3}
\\
\frac{\mathrm{d} \sigma^{\bar{\nu}_{_\ell}  \ell'  \to \bar{\nu}_{_\ell}  \ell' }_\mathrm{LO}}{\mathrm{d} \omega'} \hspace{-0.15cm} &=&
\hspace{-0.15cm}  \frac{m}{4 \pi }  \left[  \left(c_{\mathrm{L}}^{\nu_{\ell}  \ell'}\right)^2 \mathrm{I}_\mathrm{R} 
+   c_\mathrm{R}^2  \mathrm{I}_\mathrm{L} + c_{\mathrm{L}}^{\nu_{\ell}  \ell'} c_\mathrm{R}  \mathrm{I}^\mathrm{L}_\mathrm{R}  \right], \label{elastic0_xsection4}
\\
\frac{\mathrm{d} \sigma^{ \nu_{\ell} \ell'  \to \nu_{\ell' }  \ell   }_\mathrm{LO}}{\mathrm{d} \omega'}  \Bigg |_{\ell \neq \ell'}\hspace{-0.15cm} &=&
\hspace{-0.15cm}   \frac{m}{4 \pi } c^2 \mathrm{I}_\mathrm{L}
, \label{elastic0_xsection5}\\
\frac{\mathrm{d} \sigma^{ \bar{\nu}_{\ell}  \ell \to \bar{\nu}_{\ell'}  \ell^\prime   }_\mathrm{LO}}{\mathrm{d} \omega'} \Bigg |_{\ell \neq \ell'}  \hspace{-0.15cm} &=&
\hspace{-0.15cm}  \frac{m}{4 \pi } c^2   \mathrm{I}_\mathrm{R}
 ,\label{elastic0_xsection6}
\eer
with kinematical factors:
\ber
\mathrm{I}_\mathrm{L}\hspace{-0.15cm} &=&
\hspace{-0.15cm} \frac{\left( k \cdot p \right) \left( k' \cdot p' \right)}{m^2 \omega^2} = 1  + \frac{m^2-m'^2}{2m \omega}  \to1, \label{elastic_I1}  \\
\mathrm{I}_\mathrm{R} \hspace{-0.15cm} &=& \hspace{-0.15cm} \frac{ \left( k \cdot p' \right) \left( k' \cdot p \right)}{m^2 \omega^2} =  \frac{\omega'^2}{\omega^2} \left(1  + \frac{m'^2-m^2}{2 m \omega'} \right) \to  \frac{\omega'^2}{\omega^2} ,  \label{elastic_I2} \\
\mathrm{I}^\mathrm{L}_\mathrm{R} \hspace{-0.15cm} &=& \hspace{-0.15cm} - \frac{ m m' \left( k \cdot k' \right)}{m^2 \omega^2} = - \frac{ m'}{ \omega} \left( 1 - \frac{\omega'}{\omega} + \frac{m^2 -m'^2}{2 m \omega} \right) \to  - \frac{m}{\omega}  \left( 1 - \frac{\omega'}{\omega} \right),  \label{elastic_I3}
\eer
where the limit of elastic process, i.e., $m'=m$, is presented in the
last step. The neutrino-energy spectra in Eqs.~(\ref{elastic0_xsection3})-(\ref{elastic0_xsection6})
are equivalent to the recoil electron energy spectra due to energy conservation: $m+ \omega = E' +
\omega'$. In particular, $\mathrm{d} \sigma / \mathrm{d} E' =
\mathrm{d} \sigma / \mathrm{d} \omega'  $. We later apply this
observation to compute differential cross sections with respect to
total electromagnetic energy in the presence of radiative corrections.
To study the angular spectrum, the differential cross section can be obtained by exploiting
\ber
\mathrm{d} E' = \frac{4 m \omega^2 \left( m + \omega \right)^2 \cos \Theta_e \mathrm{d} \cos \Theta_e}{\left[ \left( m + \omega \right)^2 - \omega^2 \cos^2 \Theta_e \right]^2}. \label{angular_differential}
\eer
We observe that the contribution from the interference term $\mathrm{I}^\mathrm{L}_\mathrm{R}$ is suppressed by the charged lepton mass.
The neutrino  and antineutrino scattering are related by the substitution $\mathrm{I}_\mathrm{L} \leftrightarrow \mathrm{I}_\mathrm{R}$ ($k\leftrightarrow k'$) or equivalently $c^{\nu_{\ell}  \ell'}_\mathrm{L} \leftrightarrow c_\mathrm{R}$.

Note that $\nu_\ell \ell \to \nu_\ell \ell $ and $ \bar{\nu}_\ell \ell \to \bar{\nu}_\ell \ell $ cross sections involving one flavor
seem to be not positive definite for energies comparable with the charged lepton mass due to the helicity-flip interference
term $c^{\nu_\ell \ell}_\mathrm{L} c_\mathrm{R} $. However, the cross section is always positive in the physical region of scattering
$ m \omega / \left(m+2 \omega\right) < \omega' < \omega$ and can vanish only in the case of forward recoil electrons with
maximum energy~$E' = m +  2 \omega^2 / \left(m+2 \omega\right)$~\cite{Segura:1993tu,Segura:1994py,Bernabeu:2003rx,Bernabeu:2004ay} in the scattering
of an electron antineutrino of energy $\bar{\omega}$:
\ber
\bar{\omega} = \left( \frac{c_{\mathrm{L}}^{\nu_{\ell} \ell}}{c_\mathrm{R}} - 1\right) \frac{m}{2}. \label{dynamical_zero}
\eer
We discuss the impact of radiative corrections on the cancellation (\ref{dynamical_zero}) in Sec.~\ref{sec:BSM}.

\subsection{Effective neutrino-lepton and neutrino-quark interactions beyond leading order}
\label{sec:lecs}

\begin{table*}[t]
\centering
\caption{Effective couplings (in units $10^{-5}~\mathrm{GeV^{-2}}$) in the Fermi theory of neutrino-fermion scattering with four quark flavors at the scale $\mu = 2~\mathrm{GeV}$. The error due to the uncertainty of Standard Model parameters is added in quadrature to a perturbative error of matching.
}
\label{results_couplings_Running}
\begin{minipage}{\linewidth}  
\footnotesize
\centering
\begin{tabular}{|l|c|c|c|c|c|c|c|c|}   
\hline          
$ c^{\nu_\ell \ell'}_\mathrm{L},~\ell = \ell'$ & $ c^{\nu_\ell \ell'}_\mathrm{L},~\ell \neq \ell'$ & $ c_\mathrm{R}$ & $ c^{u}_\mathrm{L}$ & $ c^{u}_\mathrm{R}$ & $ c^{d}_\mathrm{L}$ & $ c^{d}_\mathrm{R}$ \\
\hline
  $2.39818(33)$ &$-0.90084(32)$ &$0.76911(60)$ &$1.14065(13)$ &$-0.51173(38)$ &$-1.41478(12)$ &$0.25617(20)$  \\
\hline
\end{tabular}
\end{minipage}
\end{table*}

Higher-order electroweak and QCD contributions modify couplings in the effective Lagrangian of
Eq.~(\ref{effective_lagrangian_leptons}).
The evaluation of virtual NLO corrections to elastic neutrino-charged lepton scattering also involves
interaction with quarks and gluons; see Secs.~\ref{sec:long_Range} and~\ref{sec:hadron_physics}.
The relevant neutral current part of the effective neutrino-quark Lagrangian is
\ber
    {\cal L}^q_{\rm eff} =  - \sum_{\ell,q}  \bar{\nu}_\ell
\gamma^\mu \mathrm{P}_\mathrm{L} \nu_\ell \, \bar{q} \gamma_\mu
(c_\mathrm{L}^{q} \mathrm{P}_\mathrm{L} + c_\mathrm{R}^{q}
\mathrm{P}_\mathrm{R}) q , \label{general_EFT_Lagrangian_quarks}
\eer
with (neutrino flavor independent) 
left- and right-handed couplings
$c^{q}_\mathrm{L}$ and $c^{q}_\mathrm{R}$, respectively.
At tree level,
\begin{align} \label{eq:quark}
  c_{\rm L}^q = 2\sqrt{2} \mathrm{G}_{\rm F} \left( T^3_q - Q_q \sin^2\theta_W \right) \,,
  \quad
  c_{\rm R}^q = - 2\sqrt{2} \mathrm{G}_{\rm F} Q_q \sin^2\theta_W \,,
\end{align}
where $T^3_q$ denotes the quark isospin ($+1/2$ for $q=u,c$, $-1/2$ for $q=d,s$)
and $Q_q$ its electric charge in units of the positron charge ($+2/3$ for $q=u,c$, $-1/3$ for $q=d,s$). 
For numerical analysis, we employ low-energy effective couplings from Ref.~\cite{in_preparation}. 
For definiteness, we take inputs in four-flavor QCD ($n_f=4$)
at renormalization scale $\mu=2\,{\rm GeV}$ in the $\overline{\rm MS}$
scheme and do not distinguish between couplings to $u$ ($d$) and $c$ ($s$) quarks.\footnote{
In Ref.~\cite{in_preparation}, 
one-loop matching to the Standard Model is performed at
the electroweak scale accounting for the leading QCD corrections with one
exchanged gluon inside quark loops and neglecting masses of all
fermions except the top quark compared to the electroweak scale.
The matching is accompanied by renormalization group evolution to GeV
scales to resum large electroweak logarithms in the effective couplings.
The relation of the couplings in Table~\ref{results_couplings_Running} to 
various definitions of G$_{\rm F}$ and $\sin^2\theta_W$ is discussed 
in Ref.~\cite{in_preparation}. 
}

The effective Lagrangians of Eqs.~(\ref{effective_lagrangian_leptons}) and
(\ref{general_EFT_Lagrangian_quarks}),
and the corresponding charged current quark operators~\cite{in_preparation}, 
determine neutrino scattering rates at GeV energy scales, up to corrections
suppressed by powers of electroweak scale particle masses.
Electroweak scale physics is encoded in the values of the operator coefficients, summarized 
in Table~\ref{results_couplings_Running}.
Real photon radiation and virtual corrections
involving the photon and other light particles must still be calculated within
the effective theory.

\section{Virtual QED corrections}
\label{sec:virtual}

In this section, we present virtual corrections, considering QED vertex corrections involving virtual photons in Sec.~\ref{sec:virtual_QED}
and closed fermion loop contributions from leptons and heavy quarks in Sec.~\ref{sec:long_Range}. 
We estimate the correction coming from light-quark loops in Sec.~\ref{sec:hadron_physics}.

\subsection{QED vertex correction}
\label{sec:virtual_QED}

\begin{figure}[tb]
\begin{center}
\vspace{-0.25cm}	
\hspace{0.7cm}	\includegraphics[scale=1.]{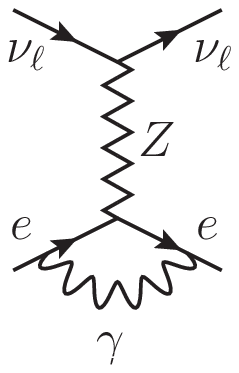}\hspace{0.cm}
\hspace{0.7cm}	\includegraphics[scale=1.]{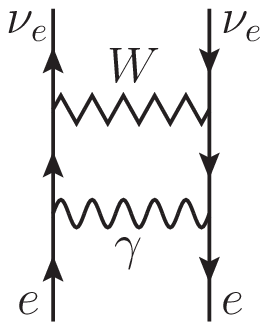}\hspace{0.cm}
\hspace{0.7cm}	\includegraphics[scale=1.]{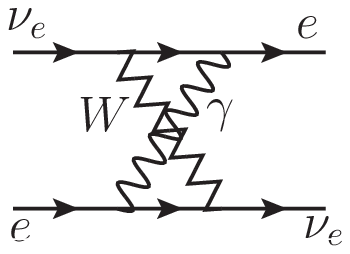}\hspace{0.cm}
\end{center}
	\caption{Virtual corrections to elastic neutrino-electron scattering in the Standard Model corresponding to the vertex correction in effective theory.\label{one_loop_QED_SM}}
\end{figure}

\begin{figure}[tb]
\begin{center}
\vspace{-0.25cm}	
\hspace{0.7cm}	\includegraphics[scale=1.]{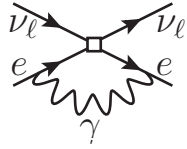}\hspace{0.cm}
\end{center}
	\caption{QED vertex correction to elastic neutrino-electron scattering in effective theory.\label{one_loop_QED}}
\end{figure}

We consider one-loop virtual corrections in elastic (anti)neutrino-electron scattering $\nu_\ell e \to \nu_\ell e$ ($ \bar{\nu}_\ell e \to \bar{\nu}_\ell e $). Within the Standard Model, the vertex correction is given by the diagrams in Fig.~\ref{one_loop_QED_SM}, while only the single
diagram in Fig.~\ref{one_loop_QED} contributes in the effective theory. The usual field renormalization factors must be applied to external legs.


First, we evaluate the one-loop vertex correction to the matrix element of left-handed (L) and
right-handed (R) charged lepton currents $\mathrm{J}^\mathrm{L,R}_\mu =  \bar{e} \left( p' \right) \gamma_\mu \mathrm{P}_\mathrm{L,R} e \left( p \right)$
from Eq.~(\ref{effective_lagrangian_leptons}). We perform the integration in $d = 4 - 2 \varepsilon$
dimensions of spacetime to regularize the ultraviolet divergence,
\ber\label{eq:dJLR}
\delta \mathrm{J}^\mathrm{L,R}_\mu = - e^2 \int \frac{i\mathrm{d}^d L }{(2 \pi)^d} \frac{\bar{e} \left( p' \right) \gamma^\lambda \left( \slash{p}' - \slash{L} + m \right)  \gamma_\mu \mathrm{P}_\mathrm{L,R}  \left( \slash{p} - \slash{L} + m \right) \gamma^\rho e \left( p \right)}{\left(L^2 - \lambda^2 \right)\left( (p - L)^2 - m^2 \right) \left((p' - L)^2 - m^2 \right)} \left( g_{\lambda \rho} - \left( 1 - \xi_\gamma \right) \frac{L_\lambda L_\rho}{L^2 - a \xi_\gamma \lambda^2}\right), 
\eer
where $\slash{k} \equiv k_\mu \gamma^\mu$ for any four-vector $k$, $\xi_\gamma$ is the photon gauge parameter, and
$a$ is an arbitrary
constant associated with the photon mass regulator.
The small photon mass $\lambda$ is introduced to regulate infrared (IR) divergences.
The corresponding field renormalization factor of external charged leptons is
\ber\label{eq:Zl}
Z_\ell = 1 - \frac{\alpha}{4\pi} \frac{\xi_\gamma}{\varepsilon}- \frac{\alpha}{4\pi} \left(  \ln \frac{\mu^2}{m^2} + 2\ln \frac{\lambda^2}{m^2} + 4 \right) + \frac{\alpha}{4\pi} \left( 1 - \xi_\gamma  \right) \left( \ln \frac{\mu^2}{\lambda^2} + 1 + \frac{a \xi_\gamma \ln  a \xi_\gamma }{1 - a \xi_\gamma }  \right). 
\eer

Neglecting Lorentz structures whose contractions with the neutrino current vanish at $m_\nu = 0$,
the resulting correction can be expressed as\footnote{Note that the vertex correction can be expressed as a modification of vector and axial currents:
\ber
\bar{e}\left( p' \right) \gamma_\mu e\left( p \right) &\to&\bar{e}\left( p' \right) \gamma_\mu e\left( p \right) + \frac{\alpha}{\pi} \bar{e}\left( p' \right) \left ( f_1  \gamma_\mu + f_2  \frac{i \sigma_{\mu \nu} q^\nu}{2 m} \right ) e \left( p \right), \\
\bar{e}\left( p' \right) \gamma_\mu \gamma_5 e\left( p \right) &\to& \bar{e}\left( p' \right) \gamma_\mu \gamma_5 e\left( p \right) + \frac{\alpha}{\pi}\left ( f_1  - f_2  \right ) \bar{e} \left( p' \right)\gamma_\mu \gamma_5 e\left( p \right).
\eer}
\ber
\left( Z_\ell - 1 \right) \mathrm{J}^\mathrm{L,R}_\mu + \delta \mathrm{J}^\mathrm{L,R}_\mu = \frac{\alpha}{\pi} \left( f_1 \mathrm{J}^\mathrm{L,R}_\mu  + f_2  \mathrm{j}^\mathrm{L,R}_\mu \right), \label{eq:f1f2}
\eer
in terms of form factors $f_1$ and $f_2$, and the additional currents $\mathrm{j}^\mathrm{L}_\mu$ and $\mathrm{j}^\mathrm{R}_\mu$:
\ber
\mathrm{j}^\mathrm{L}_\mu &=& \frac{1}{2} \bar{e} \left( p' \right) \left( \gamma_\mu \gamma_5 +  \frac{i \sigma_{\mu \nu}q^\nu}{2 m}  \right) e \left( p \right), \\
\mathrm{j}^\mathrm{R}_\mu &=& \frac{1}{2} \bar{e} \left( p' \right) \left( -\gamma_\mu \gamma_5 + \frac{i \sigma_{\mu \nu}q^\nu}{2 m} \right) e \left( p \right) \,. 
\eer
Here $\sigma_{\mu \nu} = \frac{i}{2} [ \gamma_\mu, \gamma_\nu ]$.

Using Eqs.~(\ref{eq:dJLR}) and (\ref{eq:Zl}), the UV finite and gauge-independent virtual correction is given in Eq.~(\ref{eq:f1f2}) by one-loop QED form factors~\cite{Schwinger:1949ra,Barbieri:1972as}: 
\ber
f_1 \left(\beta \right) &=& - \frac{1}{2\beta}  \left( \beta -   \frac{1}{2} \ln \frac{1+\beta}{1-\beta} \right)\ln \frac{\lambda^2}{m^2}+ \frac{1}{\beta}  \left[ \frac{3 + \rho}{8}   \ln \frac{1+\beta}{1-\beta}- \frac{1}{8}  \ln \frac{1+\beta}{1-\beta} \ln \left( 2  \frac{1 +\rho }{\rho} \right) \right] \nonumber \\
&-& \frac{1}{2\beta} \left( \mathrm{Li}_2   \frac{\beta -1+\rho}{2 \beta} - \mathrm{Li}_2  \frac{\beta+1-\rho}{2 \beta}\right) - 1,\label{DR_regularization} \\
f_2 \left(\beta \right) &=& \frac{\rho}{4 \beta} \ln \frac{1+\beta}{1-\beta},\label{F2_FF}
\eer
which are expressed in terms of the recoil electron velocity $\beta$ and the parameter $\rho$:
\ber 
\beta = \sqrt{1 - \frac{m^2}{E'^2}}, \qquad  \rho = \sqrt{1-\beta^2} = \frac{m}{E'}. \label{beta_introduced}
\eer

The vertex correction (\ref{eq:f1f2}) to the unpolarized cross section can be expressed as a sum of factorizable and nonfactorizable terms:
\ber
\mathrm{d} \sigma^{\nu_\ell e \to \nu_\ell e  }_{v} = \frac{\alpha}{\pi} \delta_v \mathrm{d} \sigma^{ \nu_\ell e \to \nu_\ell  e }_{\mathrm{LO}}
+  \mathrm{d} \sigma^{\nu_\ell e \to \nu_\ell e   }_{v,\,{\rm NF}} \label{virtual_correction} \,.
\eer
The factorizable correction is given by 
\ber
\delta_v = 2 f_1. \label{vector_form_factor_correction}
\eer
The nonfactorizable term $ \mathrm{d} \sigma^{\nu_\ell e \to \nu_\ell e   }_{v,\,{\rm NF}}$ is obtained by modifying kinematical factors $\mathrm{I}_i$ in Eqs.~(\ref{elastic0_xsection3}) and (\ref{elastic0_xsection4}) as $\mathrm{I}_i \to \mathrm{I}_i + \frac{\alpha}{\pi} f_2 \delta^{v} \mathrm{I}_i$ where
\ber
\delta^{v} \mathrm{I}_\mathrm{L} &=&\delta^{v} \mathrm{I}_\mathrm{R} =  \frac{1}{2} \mathrm{I}^\mathrm{L}_\mathrm{R} -  \frac{\omega'}{\omega}, \label{virtual_correction_nf1}  \\ 
\delta^{v} \mathrm{I}^\mathrm{L}_\mathrm{R}&=& 2  \left( \mathrm{I}_\mathrm{L} +  \mathrm{I}_\mathrm{R} - \frac{ \omega'}{\omega} \right) -  \mathrm{I}^\mathrm{L}_\mathrm{R}. \label{virtual_correction_nf2}
\eer

The resulting vertex correction to the unpolarized cross section of Eq.~(\ref{virtual_correction}) is in agreement with Refs.~\cite{Aoki:1980ix,Passera:2000ug}. In the limit of a massless electron, the Pauli form factor vanishes, $f_2 \left(\beta \right) \to 0$, and the correction becomes exactly factorizable.

\subsection{Closed fermion loops: Leptons and heavy quarks}
\label{sec:long_Range}

\begin{figure}[tb]
\begin{center}
\vspace{-0.25cm}	
\hspace{0.7cm}	\includegraphics[scale=1.]{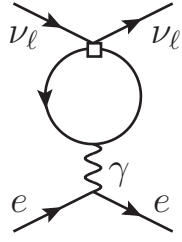}\hspace{0.cm}
\end{center}
	\caption{Long-range dynamics in elastic neutrino-electron scattering in the effective theory. Loops with all interacting fields in the theory are summed up.\label{one_loop_ET}}
\end{figure}

\begin{figure}[t]
\begin{center}
\vspace{-0.25cm}	
\hspace{0.7cm}	\includegraphics[scale=1.]{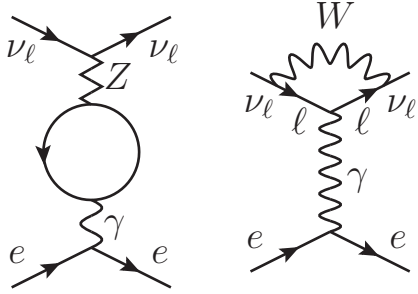}\hspace{0.cm}
\end{center}
	\caption{Standard Model diagrams giving rise to long-range dynamics in EFT: $\gamma$-$Z$ mixing and penguin-type diagram.\label{one_loop_ET_in_SM}}
 \end{figure}

In addition to the corrections involving virtual photons in Sec.
\ref{sec:virtual_QED}, we must account for the corrections with a
closed fermion loop of
Fig.~\ref{one_loop_ET}. These corrections correspond to the diagram
of penguin type and the effects of $\gamma$-$Z$ mixing in the
Standard Model; cf. Fig.~\ref{one_loop_ET_in_SM}.
They represent the EFT determination of the kinematical dependence of
electroweak corrections; cf. Refs.~\cite{Sarantakos:1982bp,Bardin:1983yb}.

In this section, we
consider the loop contribution from an arbitrary fermion with mass
$m_f$ and charge $Q_f$ (in units of the positive positron charge) and
effective left- and right-handed couplings $c^{f}_\mathrm{L}$ and
$c^{f}_\mathrm{R}$, respectively, as in
Eqs.~(\ref{effective_lagrangian_leptons}) and (\ref{general_EFT_Lagrangian_quarks}).
Note that the coupling $c^{f}_\mathrm{L}$ for charged leptons ($f = \ell$) depends on the
neutrino flavor.
This perturbative treatment applies to loops involving charged leptons or heavy quarks ($m_f \gg \Lambda_{\rm QCD}$). 
Light quarks require a nonperturbative treatment, as discussed in Sec.~\ref{sec:hadron_physics} below.
Starting from the $n_f=4$ flavor theory discussed in Sec.~\ref{sec:lecs}, we treat the charm quark as heavy
and the up, down, and strange quarks as light.

The correction can be expressed as a modification of electron left- and right-handed currents, $c_\mathrm{L,R} \mathrm{J}^\mathrm{L,R}_\mu \to c_\mathrm{L,R} \mathrm{J}^\mathrm{L,R}_\mu + c^{f}_\mathrm{L,R} \delta \mathrm{J}^\mathrm{L,R}_\mu$,
\ber
\delta \mathrm{J}^\mathrm{L,R}_\mu = Q_f  e^2 \bar{e} \left( p' \right) \gamma^\lambda e \left( p \right) \frac{-g_{\lambda \rho}}{q^2} \int \frac{i \mathrm{d}^d L }{(2 \pi)^d}  \frac{ \mathrm{Tr} \left [ \gamma^\rho \left( \slash{L} + m_f \right) \gamma_\mu \mathrm{P}_\mathrm{L,R} \left( \slash{L} - \slash{q} + m_f \right) \right ]}{ \left( L^2 -m_f^2 \right) \left( \left(L - q \right)^2 -m_f^2 \right) },
\eer
and does not depend on the photon gauge. Corrections to either left- or right-handed currents are vectorlike and may be written
\ber
\delta \mathrm{J}^\mathrm{L}_\mu =  \delta \mathrm{J}^\mathrm{R}_\mu  = Q_f \frac{\alpha}{2\pi}   \mathrm{\Pi} \left( q^2,m_f \right) \left(\mathrm{J}^\mathrm{L}_\mu+ \mathrm{J}^\mathrm{R}_\mu \right).
\eer
At renormalization scale $\mu$ in the $\overline{\mathrm{MS}}$ scheme, the form factor $\mathrm{\Pi}$ is 
\ber
\mathrm{\Pi} \left( q^2,m_f \right) = \frac{1}{3} \ln \frac{\mu^2}{m_f^2} + \frac{5}{9}  + \frac{4m_f^2}{3 q^2} + \frac{1}{3} \left( 1 + \frac{2m_f^2}{q^2} \right)  \sqrt{1-\frac{4m_f^2}{q^2}} \ln \frac{  \sqrt{1-\frac{4m_f^2}{q^2}} - 1 }{\sqrt{1-\frac{4m_f^2}{q^2}} + 1}, \label{weak_ff_MS}
\eer
and
corresponds to vacuum polarization in QED~\cite{Pauli:1936zz,Feynman:1949zx,Tsai:1960zz,Vanderhaeghen:2000ws,Heller:2019dyv}.

The resulting ``dynamical'' correction to the unpolarized cross
section, $\mathrm{d} \sigma^{\nu_\ell e \to \nu_\ell e  }_\mathrm{dyn}$,
can be expressed in the following form:
\ber
\mathrm{d} \sigma^{\nu_\ell e \to \nu_\ell e  }_\mathrm{dyn}
= \frac{\alpha}{\pi} \sum \limits_{f \ne uds}  Q_f \mathrm{\Pi} \left( q^2,~m_f \right)
\mathrm{d} \tilde{\sigma}^{ \nu_\ell e \to \nu_\ell  e }_\mathrm{dyn,\, f}
+  \mathrm{d} \sigma^{\nu_\ell e \to \nu_\ell e  }_{\mathrm{dyn},\,uds} \,. \label{virtual_correction_VP}  
\eer
The contribution from three light flavors $\mathrm{d} \sigma^{\nu_\ell e \to \nu_\ell e  }_{\mathrm{dyn},\,uds}$ is discussed below in Sec.~\ref{sec:hadron_physics}. 
The reduced cross section $\mathrm{d} \tilde{\sigma}^{\nu_\ell e \to \nu_\ell e   }_\mathrm{dyn,\,f}$
is obtained by replacing $\nu_\ell e$ couplings in Eqs.~(\ref{elastic0_xsection3}) and (\ref{elastic0_xsection4}) as
\ber
\left(c^{\nu_\ell \ell'}_\mathrm{L} \right)^2 &\to&c^{\nu_\ell  \ell'}_\mathrm{L} \left( c^{f}_\mathrm{L} + c^{f}_\mathrm{R} \right) , \label{replace1_VP}  \\ 
\left(c_\mathrm{R} \right)^2 &\to&c_\mathrm{R} \left( c^{f}_\mathrm{L} + c^{f}_\mathrm{R} \right) , \label{replace2_VP}  \\ 
c^{\nu_\ell \ell'}_\mathrm{L}  c_\mathrm{R} &\to& \frac{1 }{2} \left( c^{\nu_\ell  \ell'}_\mathrm{L} + c_\mathrm{R} \right) \left( c^{f}_\mathrm{L} + c^{f}_\mathrm{R} \right) \label{replace3_VP}.
\eer
The sum in Eq.~(\ref{virtual_correction_VP})
extends over all charged leptons ($e$, $\mu$, $\tau$) and heavy quarks ($c$) in the theory
(a factor $N_c = 3$ is obtained in the sum over colors for heavy quarks).
We also include QCD corrections due to exchanged gluons inside the quark loop; 
see Refs.~\cite{Djouadi:1987gn,Djouadi:1987di,Kniehl:1989yc,Fanchiotti:1992tu}
and Appendix \ref{app:QCD_QED_vector_vacuum_polarization} for exact expressions.

The momentum transfer in elastic neutrino-electron scattering is suppressed by the electron mass, 
\ber \label{eq:range}
0 \le -q^2 <  2 m \omega.
\eer
For neutrino beam energies smaller than $10~\mathrm{GeV}$, this implies 
$|q^2| \lesssim 0.01~\mathrm{GeV}^2 $. Consequently, the
contribution of loops with heavy quarks can be well approximated
substituting $\mathrm{\Pi} \left( q^2,~m_f \right) \to \mathrm{\Pi} \left( 0,~m_f \right)$.

\subsection{Light-quark contribution}
\label{sec:hadron_physics}

At small $q^2$, QCD perturbation theory cannot be applied to evaluate
the light-quark contribution in Fig.~\ref{one_loop_ET}.  We instead evaluate this contribution
by relating it to measured experimental quantities.

For GeV energy neutrino beams, 
momenta in the range (\ref{eq:range}) are small compared to hadronic mass scales, and
we thus evaluate the relevant hadronic tensor at $q^2=0$.  Neglecting NLO electroweak
corrections to the quark coefficients of Eqs.~(\ref{general_EFT_Lagrangian_quarks}), the light-quark contribution
in Eq.~(\ref{virtual_correction_VP}) may be written as
\ber
\mathrm{d} \sigma^{\nu_\ell e \to \nu_\ell e  }_{\mathrm{dyn},\,uds}
=  \frac{\alpha}{\pi} \left(  \hat{\Pi}_{3 \gamma}^{(3)}(0)  - 2  \sin^2 \theta_W \hat{\Pi}_{\gamma \gamma}^{(3)}(0)\right)
\mathrm{d} \tilde{\sigma}^{ \nu_\ell e \to \nu_\ell  e }_{\mathrm{dyn},\, uds} \label{eq:virtual_correction_VP_light} \,.
\eer
The reduced cross section $\mathrm{d} \tilde{\sigma}^{\nu_\ell e \to \nu_\ell e   }_{\mathrm{dyn},\,uds}$
is obtained replacing $\nu_\ell e$ couplings in Eqs.~(\ref{elastic0_xsection3}) and (\ref{elastic0_xsection4}) as
\ber
\left(c^{\nu_\ell \ell'}_\mathrm{L} \right)^2
\to  2\sqrt{2} \mathrm{G}_{\rm F}\,  c^{\nu_\ell  \ell'}_\mathrm{L}
\,, \quad \label{replace1_VPx}
c_\mathrm{R}^2
\to 2\sqrt{2} \mathrm{G}_{\rm F}\, c_\mathrm{R} 
\,, \quad \label{replace2_VPx}
c^{\nu_\ell \ell'}_\mathrm{L}  c_\mathrm{R}
\to \sqrt{2} \mathrm{G}_{\rm F}\,
\left( c^{\nu_\ell  \ell'}_\mathrm{L} + c_\mathrm{R} \right) \label{replace3_VPx}.
\eer
The quantity ${\Pi}_{\gamma \gamma}$ is defined by the vacuum correlation function, 
\begin{align}
( q^\mu q^\nu - q^2 g^{\mu\nu} ) \Pi_{\gamma\gamma}(q^2)
  = 4 i \pi^2 \int \mathrm{d} ^dx \, e^{iq\cdot x} \langle 0 | T\{ J_\gamma^\mu(x)\, J_\gamma^{\nu}(0) \} | 0\rangle \,,
\end{align}
where $J^\mu_\gamma = \sum_q Q_q \bar{q}\gamma^\mu q$ is the quark electromagnetic current.
Similarly, ${\Pi}_{3 \gamma}$ is given by 
\begin{align}
(q^\mu q^\nu - q^2 g^{\mu\nu}) \Pi_{3 \gamma}(q^2)
  = 4 i \pi^2 \int \mathrm{d} ^dx \, e^{iq\cdot x} \langle 0 | T\{ J_3^\mu(x)\, J_\gamma^{\nu}(0) \} | 0\rangle \,,
\end{align}
where $J^\mu_3 = \sum_q T^3_q \bar{q}\gamma^\mu q$ is (the third component of) the quark isospin current.
The current-current correlation functions $\hat{\Pi}_{ij}^{(3)}(0)$ 
are evaluated at $q^2=0$ for $n_f=3$ flavors, in the $\overline{\rm MS}$ scheme. 

Unlike the light-quark contribution to the photon propagator, involving only $\hat{\Pi}_{\gamma\gamma}$, 
the correction to neutral current neutrino-electron scattering involves also $\hat{\Pi}_{3\gamma}$ 
and cannot be directly related to the total hadron production cross section in $e^+ e^-$
collisions.  However, an approximate relation between $\hat{\Pi}_{\gamma \gamma}^{(3)}$
and $\hat{\Pi}_{3 \gamma}^{(3)}$ holds in the limit of $\mathrm{SU}(3)_f$
flavor symmetry for three light quarks~\cite{Jegerlehner:1985gq,Jegerlehner:2011mw}.
In general, the flavor sums read
\ber
\hat{\Pi}_{\gamma \gamma}^{(3)} &=&  \sum \limits_{i,j} Q_i Q_j \Pi^{ij}
=  \frac{4}{9} \Pi^{uu} + \frac{1}{9} \Pi^{dd} + \frac{1}{9} \Pi^{ss}
-  \frac{4}{9} \Pi^{ud} -  \frac{4}{9} \Pi^{us} +  \frac{2}{9} \Pi^{ds}, \label{pi_gamma} \\
\hat{\Pi}_{3 \gamma}^{(3)} &=& \sum \limits_{i,j} T^3_i Q_j \Pi^{ij}
= \frac{1}{2} \left( \frac{2}{3} \Pi^{uu} + \frac{1}{3} \Pi^{dd} + \frac{1}{3} \Pi^{ss}
- \Pi^{ud} -  \Pi^{us} +  \frac{2}{3} \Pi^{ds}\right) \,. \label{pi_three}
\eer
$\mathrm{SU}(3)_f$ symmetry implies $\Pi^{uu} = \Pi^{dd} = \Pi^{ss}$ and
$\Pi^{ud}=\Pi^{us}=\Pi^{ds}$, and consequently, the simple relation~\cite{Jegerlehner:1985gq}
$\hat{\Pi}_{3 \gamma}^{(3)}(0) \approx \hat{\Pi}_{\gamma \gamma}^{(3)}(0)$. 
This allows us to express the entire light-quark contribution to the unpolarized cross section
$\mathrm{d} \sigma^{\nu_\ell e \to \nu_\ell e  }_\mathrm{uds} $ in
terms of the single observable $\hat{\Pi}_{\gamma \gamma}^{(3)}(0)$. 

For numerical evaluation, we use the dispersive analysis of 
$e^+ e^-$ cross-section data and measurements of hadronic $\tau$
decays combined with a perturbative treatment of the high-energy contribution 
in Refs.~\cite{Erler:1998sy,Erler:2004in,Erler:2017knj}, 
\begin{align}\label{eq:Pigg}
  \hat{\Pi}_{\gamma \gamma}^{(3)}(0)\big|_{\mu=2\,{\rm GeV}}
  = 3.597(21) \,.
\end{align}
To estimate uncertainty due to the $\mathrm{SU}(3)_f$ symmetry approximation, we may consider
an alternative $\mathrm{SU}(2)_f$ ansatz that sets $\Pi^{uu} = \Pi^{dd},~\Pi^{ss} = 0$
and neglects disconnected, OZI-suppressed terms, $\Pi^{ud}=\Pi^{us}=\Pi^{ds} =0$.
The flavor sums (\ref{pi_gamma}) and (\ref{pi_three}) then yield
$\hat{\Pi}_{3\gamma}^{(3)} = 9 \hat{\Pi}_{\gamma \gamma}^{(3)}/10$, only a
10\% correction to the $\mathrm{SU}(3)_f$ symmetry limit.
In the final error budget, we consider a more conservative 20\%
uncertainty on this relation, 
\begin{align} \label{eq:SU3}
  \hat{\Pi}_{3 \gamma}^{(3)}(0) = \left(1 \pm 0.2 \right) \hat{\Pi}_{\gamma \gamma}^{(3)}(0) \,. 
\end{align}
Renormalization scale dependence of the light-quark contribution (\ref{eq:virtual_correction_VP_light})
is perturbatively calculable.  For $\mu\ne 2\,{\rm GeV}$, 
the additional correction corresponds with 
$ 3 \mathrm{\Pi} \left(0,~m_f = 2~\mathrm{GeV} \right)$ of Eq.~(\ref{weak_ff_MS}) for each quark
(accounting for $N_c=3$ quark colors).

The replacement $\Pi(q^2) \to \Pi(0)$ introduces an error of relative order $m \omega / m_\rho^2 \lesssim 10^{-3}$ for $\omega \lesssim {\rm GeV}$, 
where we use $m_\rho = 770\,{\rm MeV}$ as a typical hadronic scale.
This regime includes neutrinos of energy up to the TeV range produced at modern high-energy accelerators, and the 
uncertainty is contained in the error budgets (\ref{eq:Pigg}) and (\ref{eq:SU3}).
At much higher neutrino energies where $q^2$ corrections are appreciable but still 
in the nonperturbative domain, 
the same SU(3)$_f$ approximation [at momentum transfer $q^2 \neq 0$ in Eq. (\ref{eq:SU3})]
can be used to describe the light-quark contribution.%
\footnote{
  See Ref.~\cite{Blondel:2019vdq} for a discussion of $\hat{\Pi}_{\gamma\gamma}(q^2) - \hat{\Pi}_{\gamma\gamma}(0)$.}

\section{Real photon emission}
\label{sec:real}

Let us consider one-photon bremsstrahlung.  Section~\ref{sec:bremsstrahlung_general} provides basic expressions for
this process.  We then study
relevant differential observables
accounting for both soft and hard photons.
We start with the
electron energy, electron angle, and photon energy triple-differential cross
section in Sec.~\ref{sec:3xsec}. Integrating over one energy
variable, we obtain double-differential distributions in
Secs.~\ref{sec:2xsec_electron_energy_electron_angle} and \ref{sec:2xsec_electromagnetic_energy_electron_angle}.
The double-differential cross section with respect to two energy variables is described in Sec.~\ref{sec:2xsec_electromagnetic_and_electron_energy}.
We provide the distribution with respect to the photon energy and photon angle in Sec.~\ref{sec:2xsec_photon_energy_and_angle}.
Integrating it over the photon angle, we provide the photon energy spectrum in Sec.~\ref{sec:1xsec_photon_energy}.
Finally, we discuss the real soft-photon correction to elastic neutrino-electron scattering and present electron and electromagnetic energy spectra in
Secs.~\ref{sec:1xsec_electron_energy} and \ref{sec:1xsec_electromagnetic_energy}, respectively. We also provide the absolute scattering cross section in
Sec.~\ref{sec:1xsec_electron_angular}. Throughout this section, we present all expressions in the limit of the small electron
mass and provide expressions for general mass in the appendix.
For the energy spectra in Secs.~\ref{sec:1xsec_electron_energy} and~\ref{sec:1xsec_electromagnetic_energy},
we provide a general discussion of momentum regions at arbitrary mass, but present the massless limit and relegate details to the appendix.
 
\subsection{Radiation of one photon}
\label{sec:bremsstrahlung_general}

\begin{figure}[t]
\begin{center}
\vspace{-0.25cm}	
\hspace{0.7cm}	\includegraphics[scale=1.]{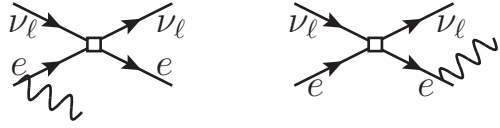}\hspace{0.cm}
\end{center}
	\caption{One-photon bremsstrahlung in elastic neutrino-electron scattering.\label{bremsstrahlung_graphs}}
\end{figure}
The one-photon bremsstrahlung amplitude $ \mathrm{T}^{1\gamma}$ (cf. Fig.~\ref{bremsstrahlung_graphs}) contains terms corresponding to
radiation from the initial electron $ \mathrm{T}_{i}^{1\gamma}$ and from the final electron $ \mathrm{T}_{f}^{1\gamma}$,
\ber
\mathrm{T}^{1\gamma} = \mathrm{T}_{i}^{1\gamma} + \mathrm{T}_{f}^{1\gamma}. 
\eer
The amplitude $ \mathrm{T}_{i}^{1\gamma}$ is obtained from the tree-level amplitude with the substitution
\ber
e \left( p \right) \to e \varepsilon^*_\rho \frac{\slash{p} - \slash{k}_\gamma + m }{ \left( p - k_\gamma \right)^2- m^2} \gamma^\rho  e \left( p \right),
\eer
where $k_\gamma$ is a photon momentum and $\varepsilon^*_\rho$ is the photon polarization vector.
The amplitude $ \mathrm{T}_{f}^{1\gamma}$ is obtained from the tree-level amplitude with the substitution
\ber
\bar{e} \left( p' \right) \to e  \varepsilon^*_\rho \bar{e} \left( p' \right)  \gamma^\rho  \frac{\slash{p}' + \slash{k}_\gamma + m' }{ \left( p' + k_\gamma \right)^2- m'^2} .
\eer

Evaluating the spin-averaged squared matrix element, $\sum \limits_{\mathrm{spin}}|\mathrm{T}^{1\gamma} |^2 $, we obtain for the 
bremsstrahlung cross sections:
\ber
\mathrm{d} \sigma^{\nu_\ell e \to \nu_\ell e \gamma }_\mathrm{LO} \hspace{-0.15cm} &=& \hspace{-0.15cm} \frac{ \alpha}{4 \pi } \frac{m \omega}{\pi^3} \left[ \left(c_{\mathrm{L}}^{\nu_{\ell} e}\right)^2 \tilde{\mathrm{I}}_\mathrm{L}
+   c_\mathrm{R}^2 \tilde{\mathrm{I}}_\mathrm{R} + c_{\mathrm{L}}^{\nu_{\ell} e} c_\mathrm{R}  \tilde{\mathrm{I}}^\mathrm{L}_\mathrm{R} \right],  \label{elastic_xsection1}   \\
\mathrm{d} \sigma^{\bar{\nu}_\ell e \to \bar{\nu}_\ell e \gamma }_\mathrm{LO}  \hspace{-0.15cm} &=& \hspace{-0.15cm} \frac{ \alpha}{4 \pi } \frac{m \omega}{\pi^3} \left[ \left(c_{\mathrm{L}}^{\nu_{\ell} e}\right)^2 \tilde{\mathrm{I}}_\mathrm{R}
+   c_\mathrm{R}^2 \tilde{\mathrm{I}}_\mathrm{L} + c_{\mathrm{L}}^{\nu_{\ell} e} c_\mathrm{R}  \tilde{\mathrm{I}}^\mathrm{L}_\mathrm{R} \right], \label{elastic_xsection2} 
\eer
where terms $\tilde{\mathrm{I}}_i$ contain the phase-space integration
\ber
 \tilde{\mathrm{I}}_i = \int \frac{R_i}{m^2 \omega^2} \delta^4 (k+p - k_\gamma- k' -p') \frac{\mathrm{d}^3 \vec{k}_\gamma}{ 2 k_\gamma} \frac{\mathrm{d}^3 \vec{k}'}{2 \omega'} \frac{\mathrm{d}^3 \vec{p}~'}{2 E'} ,
 \eer
and kinematical factors $R_i$ are expressed in terms of particle momenta as
 \ber
 R_\mathrm{L}\hspace{-0.18cm} & = & \hspace{-0.18cm} -   \mathrm{I}_\mathrm{L}\left [ \frac{p^\mu}{ \left( p \cdot k_\gamma \right)} - \frac{p'^\mu}{ \left( p' \cdot k_\gamma \right)} \right ]^2 m^2 \omega^2  + \frac{\left( k \cdot p' \right)\left( k' \cdot p' \right)}{\left( k_\gamma \cdot p' \right)} - \frac{\left( k \cdot p \right)\left( k' \cdot p \right)}{\left( k_\gamma \cdot p \right)} + \frac{\left( k \cdot p \right)\left( k' \cdot p' \right)}{\left( k_\gamma \cdot p' \right)}  - \frac{\left( k \cdot p \right)\left( k' \cdot p' \right)}{\left( k_\gamma \cdot p \right)} \nonumber \\
 & +& \frac{\left( k' \cdot p' \right) \left( k \cdot k_\gamma \right)}{\left( k_\gamma \cdot p \right)} \left( 1 + \frac{m^2}{\left( k_\gamma \cdot p \right)} - \frac{\left( p \cdot p' \right)}{\left( k_\gamma \cdot p' \right)} \right)+  \frac{\left( k \cdot p \right) \left( k' \cdot k_\gamma \right)}{\left( k_\gamma \cdot p' \right)} \left( 1 - \frac{m'^2}{\left( k_\gamma \cdot p' \right)} + \frac{\left( p \cdot p' \right)}{\left( k_\gamma \cdot p \right)} \right) , \label{bremsstrahlung_L} \\
R_\mathrm{R}\hspace{-0.18cm}   &=& \hspace{-0.18cm}-    \mathrm{I}_\mathrm{R} \left [ \frac{p^\mu}{ \left( p \cdot k_\gamma \right)} - \frac{p'^\mu}{ \left( p' \cdot k_\gamma \right)} \right]^2 m^2 \omega^2 +\frac{\left( k \cdot p' \right)\left( k' \cdot p' \right)}{\left( k_\gamma \cdot p' \right)}  - \frac{\left( k \cdot p \right)\left( k' \cdot p \right)}{\left( k_\gamma \cdot p \right)} + \frac{\left( k' \cdot p \right)\left( k \cdot p' \right)}{\left( k_\gamma \cdot p' \right)} - \frac{\left( k' \cdot p \right)\left( k \cdot p' \right)}{\left( k_\gamma \cdot p \right)}
\nonumber \\
&+& \frac{\left( k \cdot p' \right) \left( k' \cdot k_\gamma \right)}{\left( k_\gamma \cdot p \right)} \left( 1 + \frac{m^2}{\left( k_\gamma \cdot p \right)} - \frac{\left( p \cdot p' \right)}{\left( k_\gamma \cdot p' \right)} \right) + \frac{\left( k' \cdot p \right) \left( k \cdot k_\gamma \right)}{\left( k_\gamma \cdot p' \right)} \left( 1 - \frac{m'^2}{\left( k_\gamma \cdot p' \right)} + \frac{\left( p \cdot p' \right)}{\left( k_\gamma \cdot p \right)} \right), \label{bremsstrahlung_R} \\
R^\mathrm{L}_\mathrm{R} \hspace{-0.18cm} & = & \hspace{-0.18cm}  -   \mathrm{I}^\mathrm{L}_\mathrm{R} \left[ \frac{p^\mu}{ \left( p \cdot k_\gamma \right)} - \frac{p'^\mu}{ \left( p' \cdot k_\gamma \right)} \right]^2 m^2 \omega^2 - \frac{2 m m' \left( k \cdot k_\gamma \right) \left( k' \cdot k_\gamma \right)}{ \left( p \cdot k_\gamma \right) \left( p' \cdot k_\gamma \right)} \,. \label{bremsstrahlung_LR} 
\eer
Kinematical factors $\mathrm{I}_\mathrm{L},~\mathrm{I}_\mathrm{R},~\mathrm{I}^\mathrm{L}_\mathrm{R}$ are given in terms of momentum invariants in Eqs.~(\ref{elastic_I1})-(\ref{elastic_I3}), and are evaluated in the kinematics of $2 \to 3 $ scattering. Neutrino and antineutrino scattering are related by the substitution $R_\mathrm{L} \leftrightarrow R_\mathrm{R}$ (equivalently, $k\leftrightarrow k'$). The IR-divergent parts of $R_\mathrm{L}$ and $R_\mathrm{R}$ correspond to integrals $R$ and $\hat{R}$ in Ref.~\cite{Sarantakos:1982bp}, respectively.

\subsection{Triple-differential distribution}
\label{sec:3xsec}

We evaluate the bremsstrahlung cross section using the integration
technique of Ref.~\cite{Ram:1967zza} and provide expressions for the
triple-differential cross section with respect to electron angle, electron
energy, and photon energy keeping all electron mass terms in
Appendix~\ref{app:3xsec}. In the limit of small electron mass,\footnote{In the following, we denote the limit of small electron mass compared to all other relevant energy scales as $\omega \gg m$. } the
result can be approximated by the following substitutions in
Eqs.~(\ref{elastic_xsection1}) and~(\ref{elastic_xsection2})\footnote{\label{note1}Note
  that suppressed terms in the lepton mass expansion of
  $\tilde{\mathrm{I}}_\mathrm{L}$ and $\tilde{\mathrm{I}}_\mathrm{R}$
  contribute to the cross section at the same order as
  $\tilde{\mathrm{I}}^\mathrm{L}_\mathrm{R}$. For a consistent power
  counting, one has either to neglect the interference term completely
  or to expand $\tilde{\mathrm{I}}_\mathrm{L}$ and
  $\tilde{\mathrm{I}}_\mathrm{R}$ further.}:
\ber
 \tilde{\mathrm{I}}_\mathrm{L} \hspace{-0.3cm}& \underset{\omega \gg m}{\longrightarrow}  &\hspace{-0.3cm}  \left[  \frac{ \left( \omega - \omega' \right) \left( E'^2 \left( 2 - \tilde{z} \right)^2 + \omega^2 \right)}{2 |\omega - \left( \omega - \omega' \right) \left( 2 - \tilde{z} \right)|} - \frac{E' \left( E'^4 (2-\tilde{z})^2+E'^2 \omega ^2 (3 \tilde{z}-5) + E' \omega ^3 (1 - \tilde{z})+\omega ^4\right)}{2 \left(\omega - E' \right)^3}     \right. \nonumber \\
 &+&\left. \frac{E'^2 \omega' \left(2 E'^3 (1-\tilde{z}) (2-\tilde{z})+E'^2 \omega (13 + 2 \tilde{z} (2 \tilde{z}-7))  +2 E' \omega ^2 (4 \tilde{z}-7) +3 \omega ^3 \right)}{2 (\omega - E')^4}  \right. \nonumber \\
 &-&\left. \frac{E'^2 \omega'^2 \left(E'^3 (1-\tilde{z}) (2-\tilde{z})+E'^2 \omega(8 + \tilde{z} (4 \tilde{z}-11))  + E' \omega ^2\left(\tilde{z}^2+\tilde{z}-4\right) + \omega ^3\tilde{z}\right)}{2 (\omega - E')^5}  \right] {\cal{D}}, \label{IL_3D_massless} \\
\tilde{\mathrm{I}}_\mathrm{R} \hspace{-0.3cm}& \underset{\omega \gg m}{\longrightarrow}  &\hspace{-0.3cm} \frac{E'^2 \left( 1 - \tilde{z} \right)^2 + \omega'^2}{2} \left[ \frac{\omega - \omega'}{|\omega - \left( \omega - \omega' \right) \left( 2 - \tilde{z} \right)|}  - \frac{E'}{\omega - E'} \right] {\cal{D}}, \label{IR_3D_massless}  \\
 \tilde{\mathrm{I}}^\mathrm{L}_\mathrm{R} \hspace{-0.3cm}& \underset{\omega \gg m}{\longrightarrow}  &\hspace{-0.3cm} m \left[ \frac{E'^2 \left( 2 -\tilde{z} \right) \left( \tilde{z} -1 \right) + E' \left( 3 \omega' - \left( \omega + \omega' \right)\tilde{z} \right) - \omega \omega' }{|\omega - \left( \omega - \omega' \right) \left( 2 - \tilde{z} \right)|}  - \frac{E' \left( \omega - E' \left( 3 - \tilde{z} \right) \right)}{\omega - E'}\right] {\cal{D}},  \label{ILR_3D_massless}
\eer
with the phase-space factor
\ber
 {\cal{D}} = \frac{\pi^2}{\omega^3} \frac{  \mathrm{d}\tilde{z} \mathrm{d} E'  \mathrm{d} k_\gamma}{k_\gamma},
\eer
where $\omega' = \omega - k_\gamma - E'$ and the variable $\tilde{z} \leq 1$ is introduced to emphasize the forward direction of the relativistic electron:
\ber
1 - \cos \theta_e \equiv \frac{m}{\omega} \left( 1 - \tilde{z} \right).
\eer
Note the difference between the electron scattering angle in the elastic process [$\Theta_e$ of Eq.~(\ref{angle_thetae})] and in the scattering process with radiation ($\theta_e$).
At $m\to 0$, the physical region of kinematical variables is given by
\ber
0 \le E' \le \omega, \qquad 2 - \frac{\omega}{E'} \le \tilde{z} \le  1, \qquad 0 \le k_\gamma \le  \omega - E'. \label{intregion3}
\eer

In the vicinity of the elastic peak,
\ber
\tilde{z} \to \tilde{Z} =  1 - \frac{\omega'}{\omega - \omega'}, \label{elastic_peak_electron_angle}
\eer
the cross section of Eqs.~(\ref{IL_3D_massless})-(\ref{ILR_3D_massless}) diverges.
The small mass approximation in Eqs.~(\ref{IL_3D_massless})-(\ref{ILR_3D_massless}) is valid only away from this region:
\ber
 | \tilde{z} -  \tilde{Z}| \gg \frac{m}{E'} \frac{k_\gamma^2}{\left(E' + k_\gamma \right)^2} \frac{\omega'}{\omega - \omega'} .
 \eer
 For a correct description in the elastic peak region, and to obtain distributions
 (such as energy spectra) that involve integration through this region, expressions
 with an electron mass of Appendix~\ref{app:3xsec} must be used.

\subsection{Double-differential distribution in electron energy and electron angle}
\label{sec:2xsec_electron_energy_electron_angle}

Integrating the triple-differential distribution over the photon energy $k_\gamma$, we obtain the double-differential cross section with respect to the recoil electron energy and electron angle. We provide the double-differential distribution in electron energy and electron angle keeping all electron mass terms in Appendix~\ref{app:2xsec_electron_energy_electron_angle}. In the limit of small electron mass, the cross section is given by the following substitutions in Eqs.~(\ref{elastic_xsection1}) and (\ref{elastic_xsection2})\footnoteref{note1}:
\ber
  \tilde{\mathrm{I}}_i \hspace{-0.3cm}& \underset{\omega \gg m}{\longrightarrow}  &\hspace{-0.3cm} \frac{\pi^2}{\omega^2} \left(  a_i +  b_i \ln \frac{m}{2 E'} + c_i \ln \frac{E' + \omega \left( 1 - z \right)}{\omega - E'}   + d_i \ln \frac{2 \left( E' - \omega \right)^2}{ m \left( \omega z - E'  \right)}  \right)  \frac{ \mathrm{d} z \mathrm{d} E'}{\omega - E'}, \label{EM_2Dspectrum_massless}
\eer
with the coefficients $a_i,~b_i,~c_i$, and $d_i$,
\ber
 a_\mathrm{L} &=&\frac{\omega ^4 (E' (z (3-2 (7-2 z) z)+16)-\omega (8-z (8-(7-3 z) z))  )}{4 (\omega -E')^2 ( \omega z -E')} \nonumber \\
 &+& \frac{E'^2 \omega  \left(E'^2 (4-z)+E' \omega (2-(9-2 z) z)  - \omega ^2 (4-z) (5-z (z+3))\right) }{4 (\omega
   -E')^2 ( \omega z -E')}, \nonumber \\
 a_\mathrm{R} &=& \frac{- \omega^6 (1-z)^2(8-z (16-(15-4 z) z))-E' \omega ^5 \left(8-z \left(35-z \left(4 z^3-14 z^2+z+36\right)\right)\right)
   - 4 E'^2 \omega ^4}{4 (E'+ \omega (1-z))^3 (\omega z -E')}   \nonumber \\
 &+& \frac{E'^4 \omega^2 (6 + (2-5 z) z) +3 E'^3  \omega ^3 (6-(2-z) (8-z) z)-E'^2 \omega ^4 z (24-z
   (66-(46-9 z) z)) }{4 (E'+ \omega (1-z))^3 (\omega z -E')}  \nonumber \\
   &-&  \frac{E'^5 \omega (4-3 z)}{4 (E'+ \omega (1-z))^3 (\omega z -E')}, \nonumber \\
 a^\mathrm{L}_\mathrm{R} &=&m \omega \frac{E'  (\omega(2-(2-z) z)  -E' z)}{(E'+\omega (1-z)  ) (\omega z -E')}  , \nonumber \\
 b_\mathrm{L} &=& -\frac{(\omega -E') \left((E'+\omega (1-z)  )^2+\omega ^2\right)}{\omega z -E'}, \nonumber \\
 b_\mathrm{R} &=&-\frac{(\omega -E') \left((\omega -E')^2+\omega ^2(1-z)^2 \right)}{\omega z -E'} , \nonumber \\
 b^\mathrm{L}_\mathrm{R} &=& m \frac{2 (\omega -E') \left(E'^2+(\omega z - E')^2\right)}{E' (\omega z -E')}  , \nonumber \\
 c_\mathrm{L} &=& \frac{(\omega -E') \left((E'+\omega (1-z)  )^2+\omega ^2\right)}{E'+\omega (1-z)  },  \nonumber \\
 c_\mathrm{R} &=&\frac{(\omega -E') \left((\omega -E')^2+\omega ^2(1-z)^2 \right)}{E'+\omega (1-z)  }+\frac{E' \omega  (\omega
   -E') \left(- 2 \omega ^2(1-z) -E' (\omega z -E')\right)}{(E'+\omega (1-z)  )^3} , \nonumber \\
 c^\mathrm{L}_\mathrm{R} &=& m \frac{2 (\omega -E') (\omega z -2 E')}{E'+\omega (1-z)  } , \nonumber \\
 d_\mathrm{L} &=& (\omega -E') (E'+\omega (1-z)  )-\frac{E'^3 \omega }{2 (\omega -E')^2} + \frac{\omega ^3 \left(2 E'^2 (3-z)-4 E' \omega - \omega ^2 (2-z (6-(4-z) z)) \right)}{2 (E'-\omega )^2 (E'+\omega (1-z) )}, \nonumber \\
 d_\mathrm{R} &=& \frac{(E'-\omega )^2}{(E' + \omega  (1-z) )^2} d_\mathrm{L} - \frac{ \omega  (\omega z -E')^2 \left((\omega z -E')^2 + 2 \omega^2 (1-z) \right)(2-z)}{2 (E'+\omega (1-z)  )^3} , \nonumber \\
 d^\mathrm{L}_\mathrm{R} &=&m \frac{(\omega z -2 E')^2}{E'+\omega (1-z)  }.
\eer
The variable $z \leq 1$ is introduced to emphasize the forward direction of the relativistic electron,
\ber
1 - \cos \theta_e \equiv \frac{m}{E'} \left( 1 - z \right).
\eer
At $m\to0$, the physical region of kinematical variables is given by
\ber
m \le E' \le \omega, \qquad \frac{E'}{\omega} \le z \le  1. \label{intregion2Ee_Etheta}
\eer

\subsection{Double-differential distribution in electromagnetic energy and electron angle}
\label{sec:2xsec_electromagnetic_energy_electron_angle}

To obtain the distribution with respect to the electromagnetic energy and electron angle, we use the neutrino energy $\omega'$ instead of $k_\gamma$ in the triple-differential cross section, change the integration order, and integrate first over the electron energy. The final neutrino energy determines the total electromagnetic energy $E_\mathrm{EM}$: $E_\mathrm{EM} = E' + k_\gamma= m + \omega - \omega'$ and can be used to
obtain $E_\mathrm{EM}$ distributions since $\mathrm{d} E_\mathrm{EM} = - \mathrm{d} \omega'$.

In the limit of small electron mass, the neutrino energy and electron angle distribution is given by the following substitutions in Eqs.~(\ref{elastic_xsection1}) and (\ref{elastic_xsection2})\footnoteref{note1}:
\ber
 \tilde{\mathrm{I}}_i \hspace{-0.3cm}& \underset{\omega \gg m}{\longrightarrow}  &\hspace{-0.3cm} \frac{\pi^2}{\omega^3} \left(  a_i + \frac{ b_i}{|\omega - \left( 2 - \tilde{z} \right) \left( \omega - \omega' \right)|} + c_i \ln  \frac{1-\tilde{z}}{2-\tilde{z}}  + \left( d_i + \frac{e_i }{|\omega - \left( 2 - \tilde{z} \right) \left( \omega - \omega' \right)|} \right) \ln \frac{\Big|\frac{1-\tilde{z}}{2-\tilde{z}} -\frac{\omega'}{\omega}\Big|}{1-\frac{\omega'}{\omega}} \right) \nonumber \\
&& \times \mathrm{d} \tilde{z} \mathrm{d} \omega', \label{EM_2Dspectrum}
\eer
with the coefficients $a_i,~b_i,~c_i,~d_i$, and $e_i$,
\ber
 a_\mathrm{L} &=& \frac{  \left(2 \omega ^3(1 - \tilde{z}) - \omega ^2 \omega'  ( 1- 4 \tilde{z}) - 9 \omega  \omega'^2 ( 5 - 2 \tilde{z}) - \omega'^3 ( 23 - 18 \tilde{z}) \right) \omega}{4 \omega'^2}, \nonumber \\
 a_\mathrm{R} &=& \frac{(1 - \tilde{z})^2   (- \omega ( 9 - 4 \tilde{z})  + 2 \omega' (2 - \tilde{z}) ) \omega}{4 (2 - \tilde{z})^2}, \nonumber \\
 a^\mathrm{L}_\mathrm{R} &=&\frac{3- \tilde{z}}{2-\tilde{z}} m \omega , \nonumber \\
 b_\mathrm{L} &=& \frac{1}{4} \omega  (\omega -\omega') (-\omega (5 - 2 \tilde{z})  + 2 \omega'( 2 - \tilde{z}) ), \nonumber \\
 b_\mathrm{R} &=& \frac{(1-\tilde{z})^2}{(2-\tilde{z})^2}  b_\mathrm{L} , \nonumber \\
 b^\mathrm{L}_\mathrm{R} &=&\frac{ \omega( 5 - (5-2 \tilde{z}) \tilde{z})  - 2 \omega'(5 - (4 - \tilde{z}) \tilde{z}) }{2 (2 -\tilde{z})}m \omega   , \nonumber \\
 c_\mathrm{L} &=& - \frac{\omega  \left( \omega ^3 (1 -\tilde{z})  + 2  \omega ^2 \omega'\tilde{z}+\omega  \omega'^2( 31 - (37-10 \tilde{z}) \tilde{z}) +\omega'^3(18 - (26-9 \tilde{z}) \tilde{z})\right)}{2 \omega'^2} \nonumber \\
&-& \frac{\omega^2   \left( \omega - \omega' \right)^3 (1-\tilde{z})^2 }{2 \omega'^3}, \nonumber \\
 c_\mathrm{R} &=&-\frac{\omega  \left( \omega ^2 (1-\tilde{z})^2+\omega'^2\right)}{2 \omega'} , \nonumber \\
 c^\mathrm{L}_\mathrm{R} &=& (2-\tilde{z})\frac{ \omega}{\omega'} m \omega , \nonumber \\
 d_\mathrm{L} &=& \frac{(\omega -\omega') \left( \omega ^4(1-\tilde{z})^2-\omega ^3 \omega' ( 1 - (3-2 \tilde{z}) \tilde{z}) +\omega ^2 \omega'^2 (2 - (1-\tilde{z}) \tilde{z})  -    \omega  \omega'^3(3 - \tilde{z})
+ \omega'^4( 2 - \tilde{z}) \right)}{2 \omega'^3}, \nonumber \\ 
 d_\mathrm{R} &=& \frac{(\omega -\omega') \left( \left(\omega - \omega'\right) ^2 (1-\tilde{z})^2 + \omega'^2 \right)}{2 \omega'}, \nonumber \\
 d^\mathrm{L}_\mathrm{R} &=&\frac{m(\omega -\omega') ( - \omega (2 - \tilde{z})  +  \omega' ( 3 - \tilde{z}))}{\omega'} , \nonumber \\
 e_\mathrm{L} &=& -\frac{1}{2} (\omega -\omega') \left(\left(\omega - \omega'\right)^2 (2-\tilde{z})^2 + \omega^2 \right), \nonumber \\
 e_\mathrm{R} &=&-\omega' d_\mathrm{R}, \nonumber \\
 e^\mathrm{L}_\mathrm{R} &=&m \left(\omega ^2 ( 2 - (2 - \tilde{z}) \tilde{z}) -2 \omega  \omega' ( 3 - ( 3 - \tilde{z}) \tilde{z})+  \omega'^2 (5 - ( 4 - \tilde{z}) \tilde{z}) \right).
\eer
This approximation is valid only away from the elastic peak [cf. Eq.~(\ref{elastic_peak_electron_angle})] when
\ber
 | \tilde{z} -  \tilde{Z}| \gg \frac{\omega'}{\omega - \omega'} .
\eer
At $m\to 0$, the physical region is given by
\ber
0 \le \omega' \le \omega, \qquad 1 - \frac{\omega}{m} \le \tilde{z} \le 1.
\eer 

We discuss the double-differential distribution in electromagnetic energy and electron angle keeping all electron mass terms in Appendix~\ref{app:massles_2x_photon_energy_electron_angle}.

\subsection{Double-differential distribution in photon energy and electron energy}
\label{sec:2xsec_electromagnetic_and_electron_energy}

To obtain the distribution with respect to photon energy and electron energy, we can change the integration order and integrate the triple-differential cross section first over the electron scattering angle. In the limit of small electron mass, the leading terms of the photon energy and electron energy distribution are given by the following substitutions in Eqs.~(\ref{elastic_xsection1}) and (\ref{elastic_xsection2})\footnoteref{note1}:
\ber
\tilde{\mathrm{I}}_\mathrm{L} \hspace{-0.3cm}& \underset{\omega \gg m}{\longrightarrow}  &\hspace{-0.3cm} \left( \frac{-29 E'^2 + 8 E' k_\gamma \left( \frac{\omega'}{\omega} - 3 \right) + k_\gamma^2 \left(  \frac{\omega'^2}{\omega^2}  - 6 \right)}{12 E_\mathrm{EM}^2} + \frac{1}{2} \left( 1 + \frac{E'^2}{E_\mathrm{EM}^2}\right) \ln \frac{2 E' E_\mathrm{EM}}{m k_\gamma} \right)  {\cal{D}}_\gamma, \label{electron_2DspectrumEkg_IL} \\
\tilde{\mathrm{I}}_\mathrm{R} \hspace{-0.3cm}& \underset{\omega \gg m}{\longrightarrow}  &\hspace{-0.3cm} \left( \frac{-29 E'^2  + 8 E' k_\gamma  \left(  \frac{\omega}{\omega'} - 3  \right) + k_\gamma^2 \left(\frac{\omega^2}{\omega'^2} - 6    \right)}{12 E_\mathrm{EM}^2} + \frac{1}{2} \left( 1 + \frac{E'^2}{E_\mathrm{EM}^2}\right)   \ln \frac{2 E' E_\mathrm{EM}}{m k_\gamma} \right)   \frac{\omega'^2}{\omega^2}  {\cal{D}}_\gamma, \label{electron_2DspectrumEkg_IR}   \\ 
\tilde{\mathrm{I}}^\mathrm{L}_\mathrm{R} \hspace{-0.3cm}& \underset{\omega \gg m}{\longrightarrow}  &\hspace{-0.3cm}  \left( \frac{E'^2 \left( 4\frac{ E_\mathrm{EM}^2}{\omega \omega'} - 1 \right) - E' k_\gamma  \left( \frac{\omega}{\omega'} - 3  \right)  \left( \frac{\omega'}{\omega} - 3 \right) + 3 k_\gamma^2   }{2 E_\mathrm{EM}^2}   -  \left( \frac{E' E_\mathrm{EM}}{\omega \omega'} + \frac{k_\gamma^2}{E_\mathrm{EM}^2 }\right) \ln \frac{2 E' E_\mathrm{EM}}{m k_\gamma} \right) \frac{m}{E_\mathrm{EM}} \frac{\omega'}{\omega} {\cal{D}}_\gamma, \label{electron_2DspectrumEkg_ILR}  \nonumber \\
\eer
valid in the physical region, $ 0 \le E' + k_\gamma \le \omega$, with the phase-space factor ${\cal{D}}_\gamma$,
\ber
 {\cal{D}}_\gamma = \pi^2 \frac{ \mathrm{d} k_\gamma}{k_\gamma } \frac{\ \mathrm{d} E'}{ \omega}.
\eer

We discuss the double-differential distribution in photon energy and electron energy keeping all electron mass terms in Appendix~\ref{app:2xsec_electromagnetic_and_electron_energy}.

\subsection{Double-differential distribution in photon energy and photon angle}
\label{sec:2xsec_photon_energy_and_angle}

Besides the electron angle, the photon scattering angle $\theta_\gamma$ can be measured in principle. We consider the distribution with respect to the photon energy and the photon angle in the following. We present the double-differential distribution in photon energy and photon angle keeping all electron mass terms in Appendix~\ref{app:2xsec_photon_energy_and_angle}.

In the limit of small electron mass, the cross section is given by the following substitutions in Eqs.~(\ref{elastic_xsection1}) and (\ref{elastic_xsection2})\footnoteref{note1}:
\ber
  \tilde{\mathrm{I}}_i \hspace{-0.3cm}& \underset{\omega \gg m}{\longrightarrow}  &\hspace{-0.3cm} \frac{\pi^2}{\omega^3}  \left(  a_i +  b_i \ln \frac{m/2}{ \omega - k_\gamma \left( 2 - \bar{z} \right) }   \right)   \frac{ \mathrm{d} k_\gamma}{2 k_\gamma} \frac{\ \mathrm{d} \bar{z}}{\left( 2 - \bar{z} \right)^2}, \label{EM_2Dspectrum_massless_photon}
\eer
with coefficients
\ber
a_\mathrm{L} &=& -k_\gamma^3 (1-\bar{z}) (2-\bar{z})^3+k_\gamma^2 \omega (2-3 \bar{z}) (2-\bar{z})^2+4 k_\gamma \omega ^2 \bar{z}  (2-\bar{z})-2 \omega ^3 ( 1 + \bar{z}) , \nonumber \\
a_\mathrm{R} &=& \frac{ k_\gamma^2\left( - 2 k_\gamma (1-\bar{z}) (2-(2-\bar{z}) \bar{z}) (2-\bar{z}) -  \omega ( 2 - ( 2 + ( 7 - 6 \bar{z})\bar{z} )\bar{z} )  \right)}{6}  \nonumber \\
&+& \frac{\omega ^2\left(k_\gamma (2 - \bar{z}) ( 10 -  (24 - (9 + 4 \bar{z}) \bar{z} ) \bar{z} )- \omega (12 - ( 30 - (15 +2 \bar{z}) \bar{z} ) \bar{z})  \right)}{3 (2-\bar{z})^2} , \nonumber \\
a^\mathrm{L}_\mathrm{R} &=& \frac{m \left(k_\gamma^2 (3-2 \bar{z}) (1-\bar{z}) (2-\bar{z})^2-k_\gamma \omega (6-\bar{z}) \bar{z}   (2-\bar{z})+\omega ^2(6-\bar{z}) \bar{z} \right)}{2-\bar{z}}, \nonumber \\
b_\mathrm{L} &=& -\omega  \left(k_\gamma^2 (2-\bar{z})^2-2 k_\gamma \omega  (2-\bar{z})+2 \omega ^2\right), \nonumber \\
b_\mathrm{R} &=& \frac{(1-\bar{z})^2}{(2-\bar{z})^2} b_\mathrm{L}, \nonumber \\
b^\mathrm{L}_\mathrm{R} &=& \frac{2 m \left(k_\gamma^2 (1-\bar{z}) (2-\bar{z})^3-k_\gamma \omega  (2-\bar{z})+\omega ^2\right)}{2-\bar{z}},
\eer
where the variable $\bar{z} \leq 1$  is introduced to emphasize the forward direction of the photon,
\ber
1 - \cos \theta_\gamma \equiv \frac{m}{\omega} \left( 1 - \bar{z} \right).
\eer
The photon angle with respect to the~the neutrino beam direction is bounded as
\ber
\cos \theta_\gamma \geq 1 - \frac{m}{k_\gamma} \left( 1- \frac{k_\gamma}{\omega}\right),
\eer
while the physical region for the photon energy is $0 \le k_\gamma \le \omega$.

\subsection{Photon energy spectrum}
\label{sec:1xsec_photon_energy}

Integrating the double-differential distribution in the photon and electron energies over the electron energy, or the double-differential distribution in the photon energy and photon scattering angle over the angle, we obtain the photon energy spectrum. We present the photon energy spectrum keeping all electron mass terms in Appendix~\ref{app:1xsec_photon_energy}. The leading terms in the electron mass expansion are given by the following substitutions in Eqs.~(\ref{elastic_xsection1}) and (\ref{elastic_xsection2}):
\ber
 \tilde{\mathrm{I}}_\mathrm{L} \hspace{-0.3cm}& \underset{\omega \gg m}{\longrightarrow}  &\hspace{-0.3cm}  \frac{\pi^2}{\omega} g_\mathrm{L}\left( \frac{k_\gamma}{\omega} \right) \mathrm{d} k_\gamma,  \label{photon_spectrum1}\\
 \tilde{\mathrm{I}}_\mathrm{R} \hspace{-0.3cm}& \underset{\omega \gg m}{\longrightarrow}  &\hspace{-0.3cm} \frac{\pi^2}{\omega} g_\mathrm{R}\left( \frac{k_\gamma}{\omega} \right) \mathrm{d} k_\gamma,  \label{photon_spectrum2} \\
 \tilde{\mathrm{I}}^\mathrm{L}_\mathrm{R} \hspace{-0.3cm}& \underset{\omega \gg m}{\longrightarrow}  &\hspace{-0.3cm}  \frac{\pi^2}{\omega}  \frac{m}{\omega}  g^\mathrm{L}_\mathrm{R}\left( \frac{k_\gamma}{\omega} \right) \mathrm{d} k_\gamma,  \label{photon_spectrum3}
\eer
with functions $g_\mathrm{L}\left( x \right),~g_\mathrm{R}\left( x \right)$, and $g^\mathrm{L}_\mathrm{R}\left( x \right)$ derived first in the present paper,\footnoteref{note1}
\ber
g_\mathrm{L}\left( x \right) &=& \frac{\left( 1 - x \right) \left(  x^2 - 20 x - 53 \right)}{12 x} - \left(3 + \frac{1}{x}\right)\ln x - \frac{x^2 + x - 2}{2x} \ln \frac{2\omega \left( 1 - x \right)}{m}  \nonumber \\
&+& \ln \frac{2 \omega}{m} \ln x +\frac{\pi ^2}{6} - \mathrm{Li}_2 x , \label{first_kgamma} \\
g_\mathrm{R}\left( x \right) &=& -\frac{\left( 1 - x \right) \left( 37 x^2 +223 x +73 \right)}{36 x} - \left( \frac{1}{3x} + \frac{9+5x}{2}\right)\ln x + \frac{\left(1-x\right)\left( x^2 + 4 x + 1\right)}{3x} \ln \frac{2\omega \left( 1 - x \right)}{m}  \nonumber \\
&+& \left( \ln \frac{2 \omega}{m} \ln x +\frac{\pi ^2}{6} - \mathrm{Li}_2 x  \right) \left( 1 + x\right) , \label{second_kgamma} \\
g^\mathrm{L}_\mathrm{R}\left( x \right) &=&\frac{ \left( 1 - x\right) \left( 11 - 13 x\right) }{4 x } +\frac{1-2x}{2x} \ln x  -\frac{\left(1-x\right)^2}{x} \ln \frac{2\omega \left( 1 - x \right)}{m} . 
\eer

The integral of the photon energy spectrum obtained from Eqs.~(\ref{photon_spectrum1})-(\ref{photon_spectrum3}) is infrared divergent if extended to arbitrary small photon energy. The total NLO cross section is obtained by implementing an infrared regulator and including the (separately infrared divergent) virtual correction from Sec.~\ref{sec:virtual}.

\subsection{Electron energy spectrum}
\label{sec:1xsec_electron_energy}

All of our following calculations for neutrino and antineutrino scattering contain the same IR contribution arising from the soft-photon phase space, when the elastic process (without radiation) and scattering with bremsstrahlung are experimentally indistinguishable. The soft-photon contribution has to be accounted for in differential cross sections with respect to the one kinematical variable
(except for the photon energy spectrum of Sec.~\ref{sec:1xsec_photon_energy}, where one simply evaluates the spectrum above a chosen minimum photon energy).
The amplitude $\mathrm{T}^{1\gamma}_\mathrm{soft}$ for the radiation of one soft photon with energy $k_\gamma \le \varepsilon$,
 where $\varepsilon\ll m,~\omega$ denotes a cutoff regulator, can be expressed in factorizable form as
\ber
\mathrm{T}^{1\gamma}_\mathrm{soft} = \left[ \frac{\left( \varepsilon^* \cdot p' \right)}{\left( k_\gamma \cdot p' \right)} -  \frac{\left( \varepsilon^* \cdot p \right)}{\left( k_\gamma \cdot p \right)} \right] e \mathrm{T},
\eer
where $\mathrm{T}$ corresponds to the amplitude without radiation.
The corresponding contribution $  \mathrm{d} \sigma^{\nu_\ell e \to \nu_\ell e \gamma  }_\mathrm{soft} $ to the bremsstrahlung spectrum is given by
\ber
 \mathrm{d} \sigma^{\nu_\ell e \to \nu_\ell e \gamma  }_\mathrm{soft} = \frac{\alpha}{\pi} \delta_s \mathrm{d} \sigma^{ \nu_\ell e \to \nu_\ell e }_{\mathrm{LO}}, \label{soft_definition}
\eer
with the soft correction factor $\delta_s$~\cite{Lee:1964jq,Aoki:1980ix,Sarantakos:1982bp,Passera:2000ug},
\ber
\delta_s = \frac{1}{ \beta}\left(  \mathrm{Li}_2 \frac{1-\beta}{1+\beta} - \frac{\pi^2}{6} \right)-  \frac{2}{\beta} \left( \beta -   \frac{1}{2} \ln \frac{1+\beta}{1-\beta} \right)\ln  \frac{2 \varepsilon}{\lambda}  + \frac{1}{2 \beta}  \ln \frac{1+\beta}{1-\beta}\left( 1  +  \ln \frac{\rho \left(1+ \beta \right) }{4 \beta^2}   \right)+1 \,. \label{soft_result}
\eer
The velocity $\beta$ of Eq.~(\ref{beta_introduced}) (and $\rho = \sqrt{1-\beta^2}$), 
\ber
\beta = \sqrt{1 - \frac{m^2}{\bar{E}^2}}, \label{beta_both} \,
\eer
now describes either electron or electromagnetic energy spectra and $\bar{E}$ stands for the corresponding energy, i.e., $\bar{E}=E'$ or
$\bar{E}=E_{\rm EM}$.
Note the exact cancellation of the IR divergence in the sum of the vertex correction and the soft-photon emission; i.e., $\delta_s + \delta_v$ does not depend on the fictitious photon mass $\lambda$~\cite{Bloch:1937pw,Nakanishi:1958ur,Kinoshita:1962ur,Lee:1964is}. The correction of Eq.~(\ref{soft_result}) comes entirely from the first
(factorizable) terms in Eqs.~(\ref{bremsstrahlung_L})-(\ref{bremsstrahlung_LR}) and still contains an unphysical dependence on the photon energy cutoff $\varepsilon$.

For further evaluation of the electron angle distributions, we introduce the four-vector $l$~\cite{Ram:1967zza},
\ber
l = k + p - p' = \left( l_0,~\vec{f} \right),
\eer
with the laboratory frame values,
\ber
l_0 &=& m + \omega - E', \\
f^2 &=& |\vec{f}|^2 = \omega^2 + \beta^2 E'^2 - 2 \omega \beta E' \cos \theta_e. \label{parameter_f}
\eer

Besides the soft-photon correction, the first factorizable terms in Eqs.~(\ref{bremsstrahlung_L})-(\ref{bremsstrahlung_LR}) contribute from the region $ k_\gamma \ge \varepsilon$. It is convenient to split this contribution into two parts. There are no restrictions on the phase-space integration in region I: $ l^2 = l^2_0 - f^2  \ge 2 \varepsilon \left( l_0 + f\right) $. In region II: 
$ l^2  \le 2 \varepsilon \left( l_0 + f\right) $, which includes the region of scattering with elastic kinematics, the phase space of the final photon is bounded by
\ber
\cos \gamma \ge \frac{1}{f} \left( l_0 - \frac{l^2}{2 \varepsilon} \right),
\eer
where $\gamma$ is the angle between $\vec{f}$ and $\vec{k}_\gamma$.
The bremsstrahlung contribution from region I,
$\mathrm{d} \sigma^{\nu_\ell e \to \nu_\ell e \gamma  }_\mathrm{I}$, 
cancels the $\ln\varepsilon$ divergence
of the soft-photon correction.
It may be written as the sum of factorizable and nonfactorizable corrections, 
\ber
\mathrm{d} \sigma^{\nu_\ell e \to \nu_\ell e \gamma  }_\mathrm{I}
= \frac{\alpha}{\pi} \delta_\mathrm{I} \mathrm{d} \sigma^{ \nu_\ell e \to \nu_\ell e }_{\mathrm{LO}}
+ \mathrm{d} {\sigma}^{\nu_\ell e \to \nu_\ell e \gamma   }_{\mathrm{I},\,{\rm NF}}. \label{region1}
\eer
The factorizable correction $\delta_\mathrm{I}$ is obtained from the first, factorizable, terms in
Eqs.~(\ref{bremsstrahlung_L})-(\ref{bremsstrahlung_LR}), evaluating
kinematical factors $\mathrm{I}_\mathrm{L},~\mathrm{I}_\mathrm{R},~\mathrm{I}^\mathrm{L}_\mathrm{R}$
in the kinematics of the elastic $2 \to 2 $ process,
\ber
\delta_\mathrm{I} = \frac{2}{\beta} \left( \beta - \frac{1}{2} \ln \frac{1+\beta}{1-\beta} \right)  \ln \frac{2 \left( 1 + \beta \right) \varepsilon}{\beta m \left( 1 + \cos \delta_0 \right) }, 
\eer
where the angle $\delta_0$ is given by 
\ber
\cos \delta_0 = \frac{\omega^2 - \beta^2 E'^2 - l_0^2 }{2 \beta E' l_0} \,. \label{limit_of_angle_delta}
\eer
The nonfactorizable part $\mathrm{d}{\sigma}^{\nu_\ell e \to \nu_\ell e \gamma   }_{\mathrm{I},\,{\rm NF}}$
is discussed below.
The bremsstrahlung contribution from region II can be expressed in factorizable form
\ber
\mathrm{d} \sigma^{\nu_\ell e \to \nu_\ell e \gamma  }_\mathrm{II}
= \frac{\alpha}{\pi} \delta_\mathrm{II} \, \mathrm{d} \sigma^{ \nu_\ell e \to \nu_\ell e }_{\mathrm{LO}}, \label{region2}
\eer
where 
\ber
\delta_\mathrm{II} &=&   \frac{1}{\beta}\left( \left( \frac{1}{2} +  \ln \frac{\rho \left( 1 + \cos \delta_0 \right) }{4 \beta} \right) \ln \frac{1-\beta}{1+\beta} -\mathrm{Li}_2 \frac{1-\beta}{1+\beta}  - \mathrm{Li}_2 \frac{ \cos \delta_0 -1 }{\cos \delta_0 + 1} +\mathrm{Li}_2 \left(\frac{ \cos \delta_0 -1 }{ \cos \delta_0 + 1} \frac{1+\beta}{1-\beta} \right)  + \frac{\pi^2}{6} \right) \nonumber \\
&+& \ln \frac{1 - \beta \cos \delta_0}{\rho}   - 1.
\eer

Consequently, the complete electron energy spectrum is given by
\beq
\mathrm{d} \sigma^{\nu_\ell e \to \nu_\ell e \gamma }_\mathrm{LO} +  \mathrm{d} \sigma^{\nu_\ell e \to \nu_\ell e }_\mathrm{NLO}  = \left[ 1 +  \frac{\alpha}{\pi} \left(  \delta_v + \delta_s + \delta_\mathrm{I} + \delta_\mathrm{II} \right) \right] \mathrm{d} \sigma^{ \nu_\ell e \to \nu_\ell e }_{\mathrm{LO}} +     \mathrm{d} \sigma^{\nu_\ell e \to \nu_\ell e   }_v+  \mathrm{d} \sigma^{\nu_\ell e \to \nu_\ell e   }_\mathrm{dyn}
+ \mathrm{d} {\sigma}^{\nu_\ell e \to \nu_\ell e \gamma}_{\rm NF}  \label{factorizable_correction}
\eeq
and does not depend on the unphysical parameters $\varepsilon$ and $\lambda$. 
We remark that although individual corrections contain double logarithms, i.e.,
\ber \label{eq:sudakov}
\delta_v \underset{\beta \to 1}{\sim}  -\frac{1}{8} \ln^2 \left( 1- \beta \right),
\qquad
\delta_s \underset{\beta \to 1}{\sim}  -\frac{1}{4} \ln^2 \left( 1- \beta \right),
\qquad
\delta_\mathrm{II} \underset{\beta \to 1}{\sim}  \frac{3}{8} \ln^2 \left( 1- \beta \right),
\eer
the complete cross-section correction is free from such
Sudakov double logarithms~\cite{Sudakov:1954sw,Yennie:1961ad}.  In Appendix~\ref{app:coefficients_electron}, we
obtain the remaining nonfactorizable piece
$\mathrm{d} {\sigma}^{\nu_\ell e \to \nu_\ell e \gamma   }_{\rm NF}$ from the region
of hard photons ($k_\gamma \geq \varepsilon$), which contains $\mathrm{d} {\sigma}^{\nu_\ell e \to \nu_\ell e \gamma   }_{\rm I,\,NF}$ as well as the contribution beyond the first factorizable terms in Eqs.~(\ref{bremsstrahlung_L})-(\ref{bremsstrahlung_LR}), integrating the electron angle and electron energy distribution over the
variable $f$ (equivalent to the electron scattering angle $\theta_e$), 
and retaining all electron mass terms.

The resulting correction to the electron energy spectrum reproduces the result of Ref.~\cite{Sarantakos:1982bp} in the limit
$m \to 0,~E'/\omega = \mathrm{const}$.   Besides the closed fermion loop contribution
of Secs.~\ref{sec:long_Range} and \ref{sec:hadron_physics},
it is represented by the following substitutions
in Eqs.~(\ref{elastic_xsection1}) and~(\ref{elastic_xsection2}):
\ber
 \tilde{\mathrm{I}}_\mathrm{L} \hspace{-0.3cm}& \underset{\omega \gg m}{\longrightarrow}  &\hspace{-0.3cm}  \frac{\pi^2}{\omega} f_-\left( \frac{E'}{\omega} \right) \mathrm{d} E',  \\
 \tilde{\mathrm{I}}_\mathrm{R} \hspace{-0.3cm}& \underset{\omega \gg m}{\longrightarrow}  &\hspace{-0.3cm} \frac{\pi^2}{\omega}   \left( 1 - \frac{E'}{\omega} \right)^2 f_+\left( \frac{E'}{\omega}  \right) \mathrm{d} E', \\
 \tilde{\mathrm{I}}^\mathrm{L}_\mathrm{R} \hspace{-0.3cm}& \underset{\omega \gg m}{\longrightarrow}  &\hspace{-0.3cm} - \frac{\pi^2}{\omega} \frac{m}{\omega} \frac{ E' }{\omega} f^-_+\left( \frac{E'}{\omega}  \right) \mathrm{d} E',
\eer
with functions $f_-\left( x \right),~f_+\left( x \right)$~\cite{Sarantakos:1982bp},
and $f^-_+\left( x \right)$ derived first in the present paper,\footnoteref{note1}
\ber
f_-\left( x \right) &=& -\frac{2}{3} \ln \frac{2 \omega }{m}+\left(\ln \frac{1-x}{\sqrt{x}}+\frac{x}{2}+\frac{1}{4}\right) \ln \frac{2 \omega
   }{m} -\frac{1}{2} \left(  \text{Li}_2(x)-\frac{\pi ^2}{6}\right)+\frac{x^2}{24}-\frac{11 x}{12}-\frac{47}{36}\nonumber \\
&-&\frac{1}{2} \ln^2 \frac{1-x}{x} - \left(\frac{x}{2}+\frac{23}{12}\right) \ln (1-x)  + x \ln x, \label{first} \\
\left( 1 - x \right)^2 f_+\left( x \right) &=&-\frac{2}{3} (1-x)^2 \ln \frac{2 \omega }{m}+\left(\frac{x-1}{2}+(1-x)^2 \ln (1-x)\right)
   \ln \frac{2 \omega }{m} - \frac{(1-x)^2}{2} \ln
   \frac{1-x}{x^2} \ln (1-x) \nonumber \\
   &+&\left((1-x) x-\frac{1}{2}\right) \left(\text{Li}_2 \left( x \right)+\ln \frac{2 \omega x}{m} \ln x-\frac{\pi
   ^2}{6}\right)+\left(x^2+\frac{x}{2}-\frac{3}{4}\right) \ln x\nonumber \\
   &-& \frac{31-49 x}{72} (1-x) + \frac{1-x}{3} \left(5 x-\frac{7}{2}\right)\ln (1-x),  \label{second}  \\
  - x  f^-_+\left( x \right) &=& 2 + 2 \ln x + \left( x - \ln x - \frac{1}{2}\right) \ln \frac{2 \omega x}{m} + \left( \frac{3}{2} x + \frac{1}{2} - x \ln \frac{2 \omega x}{m} \right) \ln \frac{1-x}{x} + \frac{1}{2} x \ln^2 (1-x) \nonumber \\
 &+& \left( x - 1 \right) \left( \mathrm{Li}_2 \left( x \right) - \frac{\pi^2}{6} + \frac{5}{4}\right). \label{third} 
\eer

We observe that in exactly forward kinematics at electron threshold, when $E' = m$, the energy spectrum is given by the nonfactorizable contribution from the electromagnetic vertex and closed fermion loops,
\ber
\mathrm{d} \sigma^{\nu_\ell e \to \nu_\ell e \gamma }_\mathrm{LO} +  \mathrm{d} \sigma^{\nu_\ell e \to \nu_\ell e }_\mathrm{NLO} \underset{E' \to m}{\longrightarrow} \mathrm{d} \sigma^{\nu_\ell e \to \nu_\ell e }_\mathrm{NLO}  \to \mathrm{d} \sigma^{\nu_\ell e \to \nu_\ell e }_\mathrm{LO} +  \mathrm{d} \sigma^{\nu_\ell e \to \nu_\ell e   }_{v}+ \mathrm{d} \sigma^{\nu_\ell e \to \nu_\ell e   }_\mathrm{dyn}, \label{threshold_result}
\eer
with $f_2(0)=1/2$ in Eqs.~(\ref{virtual_correction}),~(\ref{virtual_correction_nf1}),~(\ref{virtual_correction_nf2}) and $\Pi \left( 0, m_f \right),~\hat{\Pi}_{\gamma \gamma}^{(3)} \left( 0 \right),~\hat{\Pi}_{3 \gamma}^{(3)} \left( 0 \right)$ of Eqs.~(\ref{virtual_correction_VP}),~(\ref{eq:virtual_correction_VP_light}). This equation provides a universal limit for electron energy and electromagnetic energy spectra.

The electron energy spectrum has the following logarithmically divergent behavior near
its endpoint $E' \leq E'_0 = m + \frac{2 \omega^2}{m+2 \omega}$:
\ber
\frac{\mathrm{d} \sigma^{\nu_\ell e \to \nu_\ell e \gamma }_\mathrm{LO} +  \mathrm{d} \sigma^{\nu_\ell e \to \nu_\ell e }_\mathrm{NLO}}{\mathrm{d} \sigma^{\nu_\ell e \to \nu_\ell e }_\mathrm{LO}}   \approx - \frac{\alpha}{\pi} \frac{2}{\beta} \left( \beta - \frac{1}{2} \ln \frac{1+\beta}{1-\beta}\right)  \ln \frac{E'_0 - E'}{m} \,, \label{logarithm_near_end_point}
\eer
as determined by infrared logarithms in Eqs.~(\ref{vector_form_factor_correction}) and~(\ref{soft_result}).

\subsection{Electromagnetic energy spectrum}
\label{sec:1xsec_electromagnetic_energy}

We evaluate the bremsstrahlung cross section with respect to the sum of electron and photon energies considering the final neutrino energy spectrum instead of the electron energy spectrum~\cite{Ram:1967zza}; see Sec.~\ref{sec:2xsec_electromagnetic_energy_electron_angle} for explanations. For the neutrino scattering angle distributions, we introduce the four-vector $\tilde{l}$,
\ber
\tilde{l} = k + p - k' = \left( \tilde{l}_0,~\vec{\tilde{f}} \right),
\eer
with the laboratory frame values,
\ber
\tilde{l}_0 &=& E_\mathrm{EM}, \\
\tilde{f}^2 &=& |\vec{\tilde{f}}|^2= \omega^2 +\omega'^2 - 2 \omega \omega'  \cos \theta_{\nu}. \label{parameter_f2}
\eer
Note the difference between the neutrino scattering angle in the elastic process [$\Theta_{\nu}$ of Eq.~(\ref{angle_thetanu})] and in the scattering with radiation ($\theta_{\nu}$).

Below the endpoint of maximal electron energy,  $E_\mathrm{EM} \le E'_0 = m + \frac{2 \omega^2}{m+2 \omega}$, we can use the same  integration technique as in Ref. \cite{Ram:1967zza}. Above the endpoint, the photon energy is bounded from below $k_\gamma \ge E_\mathrm{EM} - E'_0$, and there is no corresponding elastic process as well as no contribution from the soft region. We consider these two regions separately in the following.

\subsubsection{Below electron endpoint: $E_\mathrm{EM} \le E'_0 = m + \frac{2 \omega^2}{m+2 \omega}$}
\label{sec:below}

The contribution from the soft-photon region $k_\gamma \le \varepsilon $ is given by Eqs.~(\ref{soft_definition}) and~(\ref{soft_result}). We split the integration region with $ k_\gamma \ge \varepsilon$ for factorizable terms in Eqs.~(\ref{bremsstrahlung_L})-(\ref{bremsstrahlung_LR}) into two regions similar to Sec.~\ref{sec:1xsec_electron_energy}. In region I: $ \tilde{l}^2 - m^2 = \tilde{l}_0^2 -\tilde{f}^2 - m^2  \ge 2 \varepsilon \left( \tilde{l}_0 + \tilde{f}\right) $, there are no restrictions on the phase space.  In region II: $ \tilde{l}^2 - m^2  \le 2 \varepsilon \left( \tilde{l}_0 + \tilde{f}\right) $, the phase space of the final neutrino is restricted to
\ber
\cos \tilde{\gamma} \ge \frac{1}{\tilde{f}} \left( \tilde{l}_0 - \frac{\tilde{l}^2-m^2}{2 \varepsilon} \right),
\eer
where $\tilde{\gamma}$ is the angle between $\vec{\tilde{f}}$ and $\vec{k}_\gamma$. The correction factor from region II,
$\delta_\mathrm{II}$ [cf. Eq.~(\ref{region2})], is given by
\ber
\delta_\mathrm{II} = -  \frac{1 }{\beta} \left( \beta - \frac{1}{2} \ln \frac{1+\beta}{1-\beta}\right) \ln \frac{1+\beta}{1-\beta}.
\eer
Here $\beta$ is expressed in terms of electromagnetic energy as in Eq.~(\ref{beta_both}).
As for the electron energy spectrum, the 
bremsstrahlung contribution from region I may be written as the sum of
factorizable and nonfactorizable corrections; cf. Eq.~(\ref{region1}). 
The factorizable correction $\delta_{\rm I}$ is obtained from the first factorizable terms
in Eqs.~(\ref{bremsstrahlung_L})-(\ref{bremsstrahlung_LR}),
evaluating kinematical factors $\mathrm{I}_\mathrm{L},~\mathrm{I}_\mathrm{R},~\mathrm{I}^\mathrm{L}_\mathrm{R}$
in the kinematics of the elastic $2 \to 2 $ process,
\ber
\delta_\mathrm{I} =
\frac{2}{\beta} \left( \beta -   \frac{1}{2} \ln \frac{1+\beta}{1-\beta} \right)  \ln  \frac{\varepsilon }{m} \,. \label{region1_neutrino_IR}
\eer
In Appendix~\ref{app:coefficients_electromagnetic}
we evaluate the remaining nonfactorizable piece $\mathrm{d} {\sigma}^{\nu_\ell e \to \nu_\ell e \gamma   }_{\rm NF}$ 
of the electromagnetic energy spectrum below the electron endpoint, performing straightforward
integrations and keeping all electron mass terms. 
It accounts for the region of hard photons ($k_\gamma \geq \varepsilon$) and contains
$\mathrm{d} {\sigma}^{\nu_\ell e \to \nu_\ell e \gamma   }_{\rm I,\,NF}$ as well as the contribution beyond
the first factorizable terms in Eqs.~(\ref{bremsstrahlung_L})-(\ref{bremsstrahlung_LR}). 

The resulting correction to the electromagnetic energy spectrum reproduces the result of Refs.~\cite{Bardin:1983yb,Bardin:1985fg} in the limit
$m \to 0,~E_\mathrm{EM}/\omega = \mathrm{const}$.  
Besides the closed fermion loop contribution of Secs.~\ref{sec:long_Range} and \ref{sec:hadron_physics},
it is represented by the following substitutions in Eqs.~(\ref{elastic_xsection1}) and (\ref{elastic_xsection2}):
\ber
 \tilde{\mathrm{I}}_\mathrm{L} \hspace{-0.3cm}& \underset{\omega \gg m}{\longrightarrow}  &\hspace{-0.3cm}  \frac{\pi^2}{\omega} f_\mathrm{L}\left( \frac{E_\mathrm{EM}}{\omega} \right) \mathrm{d} E_\mathrm{EM},  \\
 \tilde{\mathrm{I}}_\mathrm{R} \hspace{-0.3cm}& \underset{\omega \gg m}{\longrightarrow}  &\hspace{-0.3cm} \frac{\pi^2}{\omega}  \left( 1 - \frac{E_\mathrm{EM}}{\omega} \right)^2 f_\mathrm{R}\left( \frac{E_\mathrm{EM}}{\omega} \right) \mathrm{d} E_\mathrm{EM}, \\
 \tilde{\mathrm{I}}^\mathrm{L}_\mathrm{R} \hspace{-0.3cm}& \underset{\omega \gg m}{\longrightarrow}  &\hspace{-0.3cm}  - \frac{\pi^2}{\omega}  \frac{m}{\omega} \frac{E_\mathrm{EM}}{\omega} f^\mathrm{L}_\mathrm{R}\left( \frac{E_\mathrm{EM}}{\omega} \right) \mathrm{d} E_\mathrm{EM},
\eer
with functions $f_\mathrm{L}\left( x \right),~f_\mathrm{R}\left( x \right)$~\cite{Bardin:1983yb,Bardin:1985fg}, and $f^\mathrm{L}_\mathrm{R}\left( x \right)$ derived first in the present work,\footnoteref{note1}
\ber
f_\mathrm{L}\left( x \right) &=&  \frac{3 x^2  - 30 x + 23 }{72} -\frac{2}{3} \ln \frac{2 \omega x}{m}  -\frac{\pi ^2}{6} , \label{firstt} \\
f_\mathrm{R}\left( x \right) &=&\frac{ - 4 x^2 - 16 x + 23 }{72 \left( 1 - x \right)^2 } -\frac{2}{3} \ln \frac{2 \omega x}{m}  -\frac{\pi ^2}{6} , \label{secondt} \\
f^\mathrm{L}_\mathrm{R}\left( x \right) &=&\frac{ x^2 + 3 x - 3 }{4 x^2 } -\frac{3}{2} \ln \frac{2 \omega x}{m}  -\frac{\pi ^2}{6}. \label{thirdt}
\eer

In exactly forward kinematics at electromagnetic energy threshold when $E_\mathrm{EM} = m$, the electromagnetic energy spectrum coincides with the electron energy spectrum; see Eq.~(\ref{threshold_result}).

Just below electron endpoint ($E_\mathrm{EM} < E'_0 = m + \frac{2 \omega^2}{m+2 \omega} \approx \omega$),
the electromagnetic energy spectrum, besides the closed fermion loop contribution,
is given by the following substitutions in the nonfactorizable correction\footnoteref{note1}:
\ber
 \mathrm{\tilde{I}}_\mathrm{L} \hspace{-0.3cm}& \underset{\omega \gg m}{\longrightarrow}  &\hspace{-0.3cm} - \frac{\pi^2 }{3} \left(  \ln \frac{4 \omega^2}{m^2} + \frac{\pi^2}{2} + \frac{1}{6} \right)  \frac{\mathrm{d} E_\mathrm{EM}}{\omega}  , \label{left_correction} \\
 \mathrm{\tilde{I}}_\mathrm{R} \hspace{-0.3cm}& \underset{\omega \gg m}{\longrightarrow}  &\hspace{-0.3cm}  \frac{\pi^2}{24}  \frac{\mathrm{d} E_\mathrm{EM}}{\omega} , \label{right_correction} \\
 \mathrm{\tilde{I}}^\mathrm{L}_\mathrm{R} \hspace{-0.3cm}& \underset{\omega \gg m}{\longrightarrow}  &\hspace{-0.3cm} \frac{\pi^2}{4}\frac{m}{\omega} \left( 3 \ln \frac{4 \omega^2}{m^2} + \frac{2\pi^2}{3} -1 \right)   \frac{\mathrm{d} E_\mathrm{EM}}{\omega}  .
\eer
Equations~(\ref{left_correction}) and (\ref{right_correction}) are in agreement with the similar limit taken
from the result of Refs.~\cite{Bardin:1983yb,Bardin:1985fg}.

\subsubsection{Above electron endpoint: $E_\mathrm{EM} > E'_0 = m + \frac{2 \omega^2}{m+2 \omega}$}
\label{sec:above}

Above the electron endpoint energy, the corresponding elastic process is kinematically forbidden.
For $\omega \gg m$, this region is relatively small but finite, 
\ber
E_\mathrm{EM} - E'_0 \le \frac{1}{1 + \frac{m}{2 \omega}} \frac{m}{2} < \frac{m}{2} \,. 
\eer
Since the photon energy is bounded from below in this region,
$k_\gamma >  E_\mathrm{EM} - E'_0 $, 
the calculation does not require IR regularization.
We present the electromagnetic energy spectrum above the electron endpoint
keeping all electron mass terms in Appendix~\ref{app:coefficients_electromagnetic2}.

The electromagnetic energy spectrum has the following logarithmically
divergent behavior just above the electron endpoint $E_\mathrm{EM} > E'_0 = m + \frac{2 \omega^2}{m+2 \omega}$:
\ber
\frac{\mathrm{d} \sigma^{\nu_\ell e \to \nu_\ell e \gamma }_\mathrm{LO} }{\mathrm{d} \sigma^{\nu_\ell e \to \nu_\ell e }_\mathrm{LO}}   \approx \frac{ \alpha}{\pi }\frac{2 }{ \beta} \left( \beta - \frac{1}{2} \ln \frac{1+\beta}{1-\beta}\right)  \ln \frac{E_\mathrm{EM} - E'_0}{m} \,.\label{EM_logarithmical_divergence}
\eer

\subsection{Absolute cross section}
\label{sec:1xsec_electron_angular}

The resulting total cross-section correction, besides closed fermion loop contributions,
in the ultrarelativistic limit is given by the following substitutions in Eqs.~(\ref{elastic_xsection1}) and (\ref{elastic_xsection2}) for $\tilde{\mathrm{I}}_\mathrm{L}$, $\tilde{\mathrm{I}}_\mathrm{R}$~\cite{Sarantakos:1982bp}, and $\tilde{\mathrm{I}}^\mathrm{L}_\mathrm{R}$ derived first in the present paper\footnoteref{note1}:
\ber
 \tilde{\mathrm{I}}_\mathrm{L} \hspace{-0.3cm}& \underset{\omega \gg m}{\longrightarrow}  &\hspace{-0.3cm}  \frac{\pi^2}{24} \left(19 - 4 \pi^2 - 16 \ln \frac{2\omega}{m} \right)   , \label{total_gamma1}  \\
 \tilde{\mathrm{I}}_\mathrm{R} \hspace{-0.3cm}& \underset{\omega \gg m}{\longrightarrow}  &\hspace{-0.3cm} \frac{\pi^2}{72} \left(19 - 4 \pi^2 - 16 \ln \frac{2\omega}{m}  \right) + \frac{\pi^2}{3}  , \label{total_gamma2} \\
 \tilde{\mathrm{I}}^\mathrm{L}_\mathrm{R} \hspace{-0.3cm}& \underset{\omega \gg m}{\longrightarrow}  &\hspace{-0.3cm} - \frac{\pi^2}{24} \frac{m}{\omega} \left(15 - 2 \pi^2 - 36  \ln \frac{2\omega}{m} \right) . \label{total_gamma3}
\eer
Factors $\tilde{\mathrm{I}}_\mathrm{L}$ and $\tilde{\mathrm{I}}_\mathrm{R}$ of Eqs.~(\ref{total_gamma1}) and~(\ref{total_gamma2}) can be obtained integrating Eqs.~(\ref{firstt}) and~(\ref{secondt}) or Eqs.~(\ref{first}) and~(\ref{second}) over the energy variable. To evaluate the factor $\tilde{\mathrm{I}}^\mathrm{L}_\mathrm{R}$, one has to regulate the logarithmic mass singularity properly or take the limit from the general expression of Appendix~\ref{app:total_crosssections}. Note the absence of double logarithms in the resulting cross-section correction in Eqs.~(\ref{firstt})-(\ref{thirdt}) and
 (\ref{total_gamma1})-(\ref{total_gamma3}),
although individual corrections contain them; cf. Eq.~(\ref{eq:sudakov}).
Note also that the total elastic cross section at leading order is given by the following substitutions in Eqs.~(\ref{elastic0_xsection3}) and (\ref{elastic0_xsection4}):
\ber
\int \mathrm{d} \omega^\prime \, \mathrm{I}_\mathrm{L}  \underset{\omega \gg m}{\longrightarrow}  \omega ,
\qquad  \int \mathrm{d} \omega^\prime \, \mathrm{I}_\mathrm{R}  \underset{\omega \gg m}{\longrightarrow}   \frac{\omega}{3} ,
\qquad  \int \mathrm{d} \omega^\prime \, \mathrm{I}^\mathrm{L}_\mathrm{R}  \underset{\omega \gg m}{\longrightarrow}   -\frac{m}{2} .
\eer
Results for the absolute cross section including the electron mass dependence are presented in Appendix~\ref{app:total_crosssections}.

\section{Illustrative results}
\label{sec:results}

Our results may be used to compute absolute and differential cross sections
for neutrino-electron scattering over a broad range of energies and experimental setups.
We focus on the application to flux normalization at accelerator-based
neutrino experiments in Secs.~\ref{sec:xsec} through \ref{sec:spectra_DUNE}
and discuss radiative corrections in the context of 
new physics searches in Sec.~\ref{sec:BSM}.

\subsection{Total cross section: Energy dependence and error analysis}
\label{sec:xsec}

\begin{figure}[htb]
          \centering
          \includegraphics[height=0.3\textwidth]{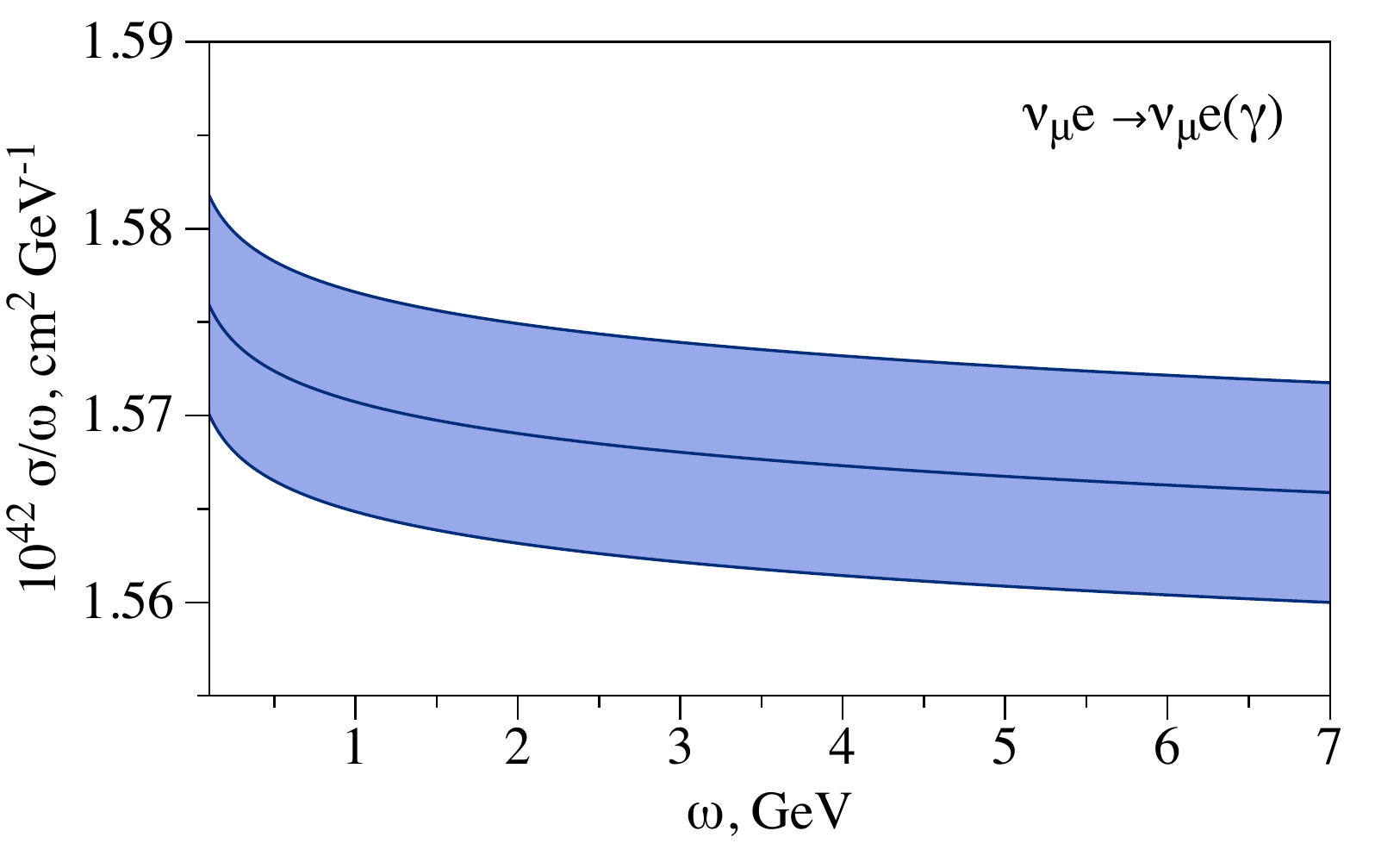}   
          \includegraphics[height=0.3\textwidth]{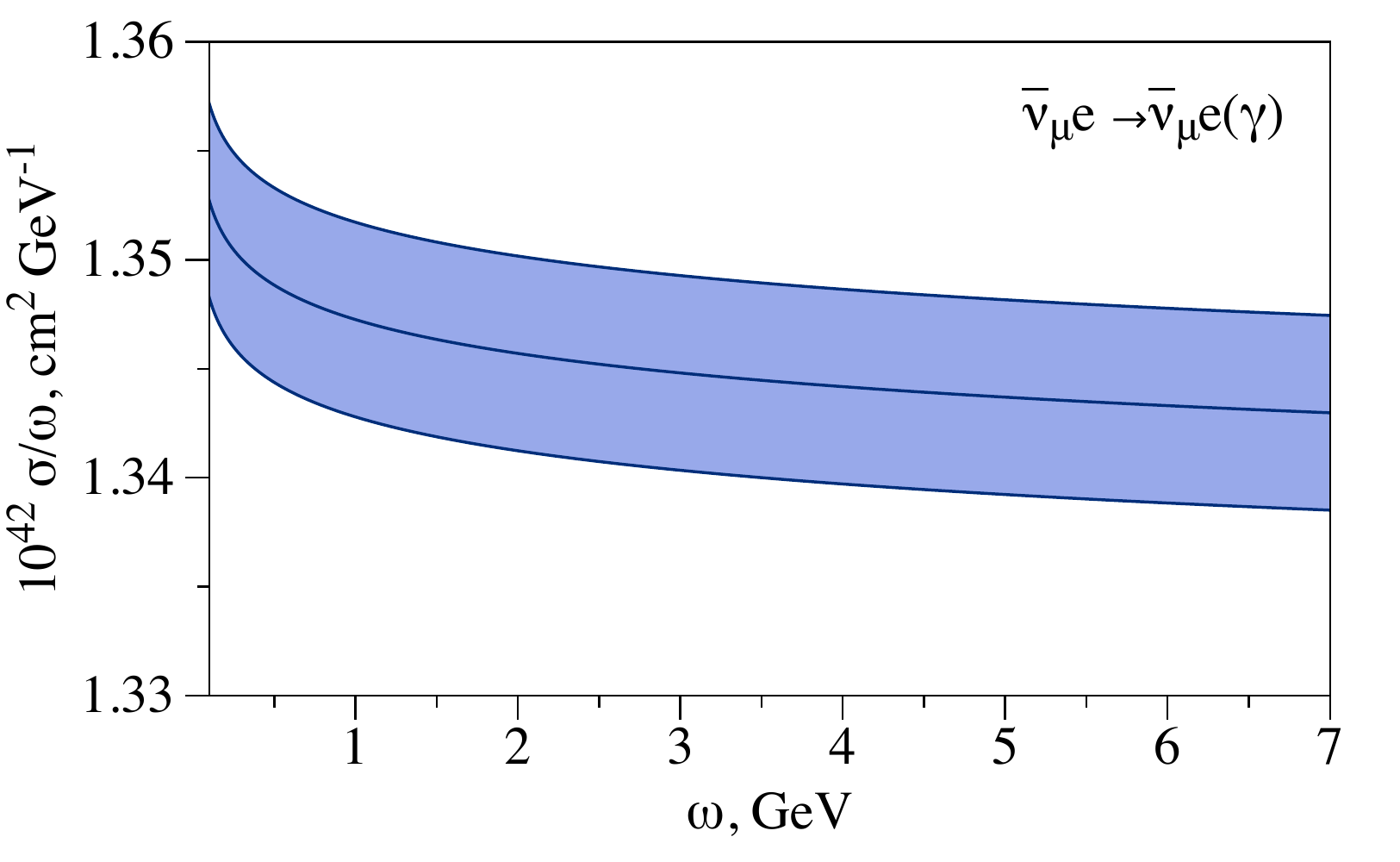}    
          \includegraphics[height=0.3\textwidth]{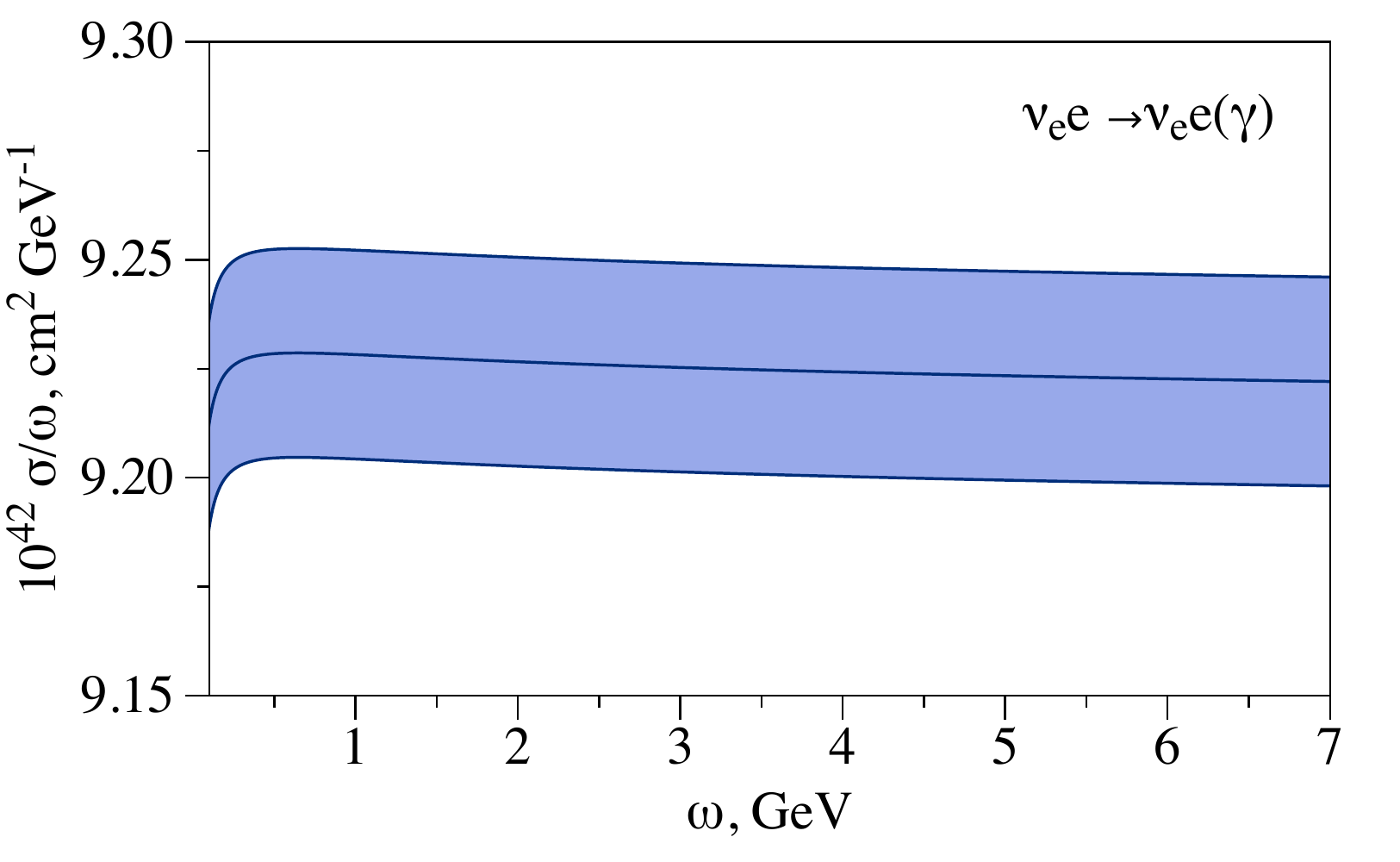}   
          \includegraphics[height=0.3\textwidth]{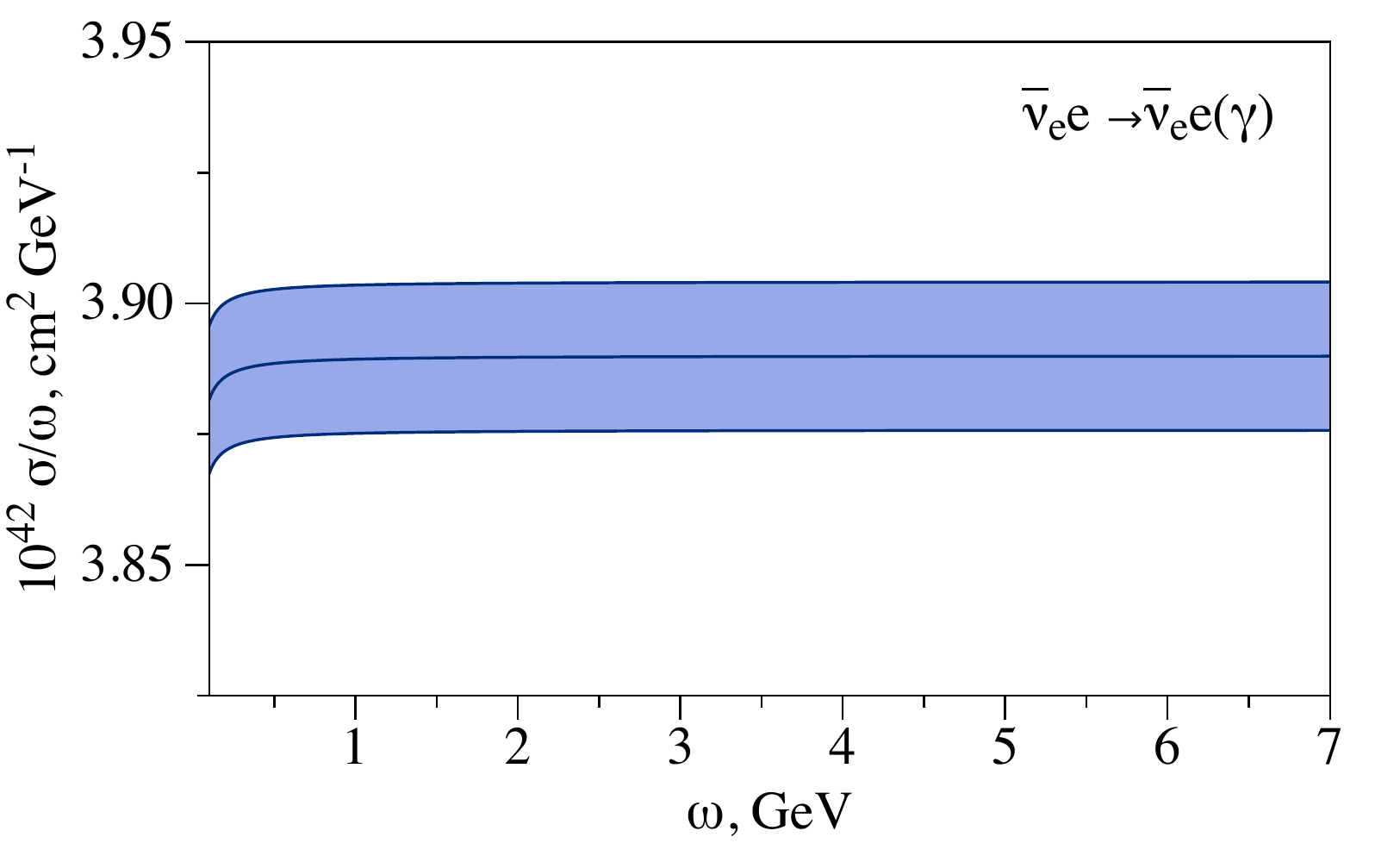}                
          \caption{Total cross section in the (anti)neutrino-electron
            scattering processes $\nu_\mu e \to \nu_\mu e(
            X_\gamma),~\nu_e e \to \nu_e e(
            X_\gamma),~\bar{\nu}_\mu e \to \bar{\nu}_\mu e(
            X_\gamma)$, and $\bar{\nu}_e e \to \bar{\nu}_e e(
            X_\gamma)$ as a function of (anti)neutrino beam energy $\omega$.
                \label{fig:nu_e_total}}
\end{figure}

The total cross sections for $\nu_\mu e,~\nu_e e,~\bar{\nu}_\mu e$, and $\bar{\nu}_e e$ scattering
are shown in Fig.~\ref{fig:nu_e_total}.
For $\omega \gg m$, cross sections grow approximately linearly with neutrino beam energy. 
As a benchmark point, we determine at $\omega = 1~\mathrm{GeV}$
\begin{align} \label{eq:sigmatot}
  \sigma^{\nu_\mu e \to \nu_\mu e (\gamma)}( \omega = 1\,{\rm GeV} )
  &= \big[ 1.5707  \times  10^{-42} \, {\rm cm}^2 \big] \times \big[ 1 \pm {0.0037}_{\rm had} \pm {0.0007}_{\rm EW} \pm {0.00007}_{\rm pert}
    \big] \,.
\end{align}
The cross section is evaluated using four-flavor QCD, with running QED and QCD couplings
$\alpha(\mu)$ and $\alpha_s(\mu)$ evaluated using two and five loop running, respectively,
with $\alpha(2\,{\rm GeV})=1/133.309$ and $\alpha_s(2\,{\rm GeV})=0.3068$.  
The uncertainties in Eq.~(\ref{eq:sigmatot}) are from the following: (i) the hadronic parameter
$\hat{\Pi}^{(3)}_{3\gamma}(0)/\hat{\Pi}^{(3)}_{\gamma\gamma}(0)$ in Eq.~(\ref{eq:SU3}) and
from $\hat{\Pi}^{(3)}_{\gamma\gamma}(0)$ in Eq.~(\ref{eq:Pigg})\footnote{The error of $\hat{\Pi}^{(3)}_{\gamma\gamma}(0)$ in Eq.~(\ref{eq:Pigg}) contributes $\pm 0.00006$.  
  };
(ii) from uncertainties in the four-fermion operator coefficients $c_\mathrm{L}^{\nu_\ell \ell^\prime}$, $c_\mathrm{R}$
in Table~\ref{results_couplings_Running};
and (iii) from higher-order perturbative corrections, estimated by varying renormalization scale
$\mu_0^2/2  < \mu^2 < 2 \mu_0^2$, where $\mu_0=2\,{\rm GeV}$.  
For simplicity, we evaluate the light-quark contribution of Eq.~(\ref{eq:virtual_correction_VP_light})
neglecting NLO electroweak corrections and renormalization group corrections to the four-fermion operator
coefficients, taking for definiteness $\mathrm{G}_\mathrm{F}=1.1663787\times 10^{-5}\,{\rm GeV}^{-2}$ and $\sin^2\theta_W = 0.23112$ in
Eqs.~(\ref{eq:virtual_correction_VP_light}) and~(\ref{replace3_VPx});
it is straightforward to include these corrections, whose  
impact is given by the few permille shift in the coefficients~\cite{in_preparation},
times the $\sim 1\%$ fractional contribution of light quarks to the cross section.  
The charm-quark contribution in Eq.~(\ref{virtual_correction_VP})
is evaluated including the $\order(\alpha_s)$ and $\order(\alpha_s^2)$ corrections from
Appendix~\ref{app:QCD_QED_vector_vacuum_polarization} and using
the $\overline{\mathrm{MS}}$ mass $\hat{m}_c(2\,{\rm GeV}) = 1.096~\mathrm{GeV}$ [corresponding to $\hat{m}_c(\hat{m_c})=1.28(2)\,{\rm GeV}$~\cite{Tanabashi:2018oca}].
The fractional uncertainty coming from the charm quark mass error is $\approx 1-2\times 10^{-5}$ and is not displayed in Eq.~(\ref{eq:sigmatot}), 
nor is the uncertainty of a similar magnitude coming from higher orders in $\mathrm{G}_\mathrm{F}$ expansion. 
The $e$-, $\mu$-, and $\tau$-lepton contributions in Eq.~(\ref{virtual_correction_VP})
are evaluated using lepton pole masses and the complete kinematic dependence of $\Pi(q^2,m_\ell)$ in
Eq.~(\ref{weak_ff_MS}).\footnote{One can safely evaluate a $\tau$-lepton contribution considering $\Pi(0,m_\tau)$ since $|q^2|  \ll m^2_\tau$.}

For $\omega \gg m$, the relative cross-section error is approximately constant, independent of neutrino energy.  
Relative uncertainties on total cross sections from  different sources are summarized in Table~\ref{error_total}.
The dominant uncertainty from the light-quark contribution in differential and absolute cross sections
can be
expressed as\footnote{It can be seen [cf. Eq.~(\ref{error_VP})] that the muon antineutrino-electron scattering
  cross section is free from hadronic uncertainty, and also effective coupling uncertainty induced by $c_R$,
  at the particular recoil antineutrino energy $\tilde{\omega}$:
\ber
\tilde{\omega} = \frac{\sqrt{ \left( c_{\mathrm{L}}^{\nu_\mu e} + c_{\mathrm{R}}\right)^2 m^2 + 8 c_{\mathrm{L}}^{\nu_\mu e}  \left( c_{\mathrm{L}}^{\nu_\mu e} + c_{\mathrm{R}}\right) m \omega - 16 c_{\mathrm{L}}^{\nu_\mu e}  c_{\mathrm{R}} \omega^2}- \left( c_{\mathrm{L}}^{\nu_\mu e} + c_{\mathrm{R}}\right)m }{4 c_{\mathrm{L}}^{\nu_\mu e}} \underset{\omega \gg m}{\longrightarrow}   \sqrt{\frac{-c_{\mathrm{R}}}{c_{\mathrm{L}}^{\nu_\mu e}}} \omega.
\eer}
\ber
\delta  \left( \frac{ \mathrm{d} \sigma^{\nu_\ell e \to \nu_\ell e  }_\mathrm{uds}}{\mathrm{d} E'} \right) \hspace{-0.05cm} &\approx& \hspace{-0.05cm}
\eta
\frac{ \mathrm{G}_\mathrm{F} m}{\sqrt{2} \pi}  \frac{\alpha}{\pi}   \hat{\Pi}_{\gamma \gamma}^{(3)} \left( 0 \right) \left | c_{\mathrm{L}}^{\nu_{\ell} e}\mathrm{I}_\mathrm{L}
+   c_\mathrm{R} \mathrm{I}_\mathrm{R} + \frac{c_{\mathrm{L}}^{\nu_{\ell} e} + c_\mathrm{R}}{2}  \mathrm{I}^\mathrm{L}_\mathrm{R} \right |,  \label{error_VP} \\
\delta   \sigma^{\nu_\ell e \to \nu_\ell e  }_\mathrm{uds} \hspace{-0.05cm} &\approx& \hspace{-0.05cm} \eta \frac{ \mathrm{G}_\mathrm{F} m \omega}{\sqrt{2} \pi}  \frac{\alpha}{\pi}  \hat{\Pi}_{\gamma \gamma}^{(3)} \left( 0 \right) \left ( \frac{2 \omega c_{\mathrm{L}}^{\nu_{\ell} e}}{m + 2 \omega} 
+   \left( 1 - \frac{m^3}{ \left( m + 2 \omega \right)^3}\right) \frac{c_\mathrm{R}}{3} - \frac{m \omega  \left(c_{\mathrm{L}}^{\nu_{\ell} e} + c_\mathrm{R}\right) }{\left( m + 2 \omega \right)^2} \right ),  \label{error_VP_absolute}
\eer
with the relative uncertainty $\eta = 
 (\hat{\Pi}_{3 \gamma}^{(3)} \left( 0 \right) / \hat{\Pi}_{\gamma \gamma}^{(3)} \left( 0 \right) - 1.0) \approx 0.2$
and the substitution $c_{\mathrm{L}}^{\nu_{\ell} e} \leftrightarrow c_\mathrm{R}$ in the case of antineutrino scattering.

\begin{table*}[t]
\centering
\caption{Relative errors of the total neutrino-electron scattering cross section. 
}
\label{error_total}
\begin{minipage}{\linewidth}  
\footnotesize
\centering 
\begin{tabular}{|l|c|c|c|c|c|c|c|}   
\hline          
& Light-quark correction &Effective couplings & Higher orders
\\
\hline
$\nu_\mu e \to \nu_\mu e(X_\gamma)$  &0.37 \% &0.068 \% & $\lesssim$ 0.008 \%
\\
$\bar{\nu}_\mu e \to \bar{\nu}_\mu e(X_\gamma)$  &0.31 \% & 0.112 \% & $\lesssim$ 0.005 \%
\\
$\nu_e e \to \nu_e e(X_\gamma)$  &0.26 \% & 0.028 \% & $\lesssim$ 0.007 \%
\\
$\bar{\nu}_e e \to \bar{\nu}_e e(X_\gamma)$  &0.36 \% & 0.044 \% & $\lesssim$ 0.007 \%
\\
\hline
\end{tabular}
\end{minipage}
\end{table*}

To illustrate the impact of radiative corrections on the total cross section, Eq.~(\ref{eq:sigmatot}) 
may be compared to the leading-order result of our calculation at scale $\mu = 2~\mathrm{GeV}$ and $\omega = 1~\mathrm{GeV}$: 
\begin{align} \label{eq:sigmatot_LO}
  \sigma^{\nu_\mu e \to \nu_\mu e}_\mathrm{LO} (\omega = 1\,{\rm GeV})
  &=  1.5971  \times  10^{-42} \, {\rm cm}^2.
\end{align}
Radiative corrections change the total cross section by $1.7\,\%$.  We turn now to a discussion of the 
energy dependence of the radiative corrections. 

\subsection{Electron and total electromagnetic energy spectra}
\label{sec:spectra}

\begin{figure}[t]
          \centering
          \includegraphics[height=0.5\textwidth]{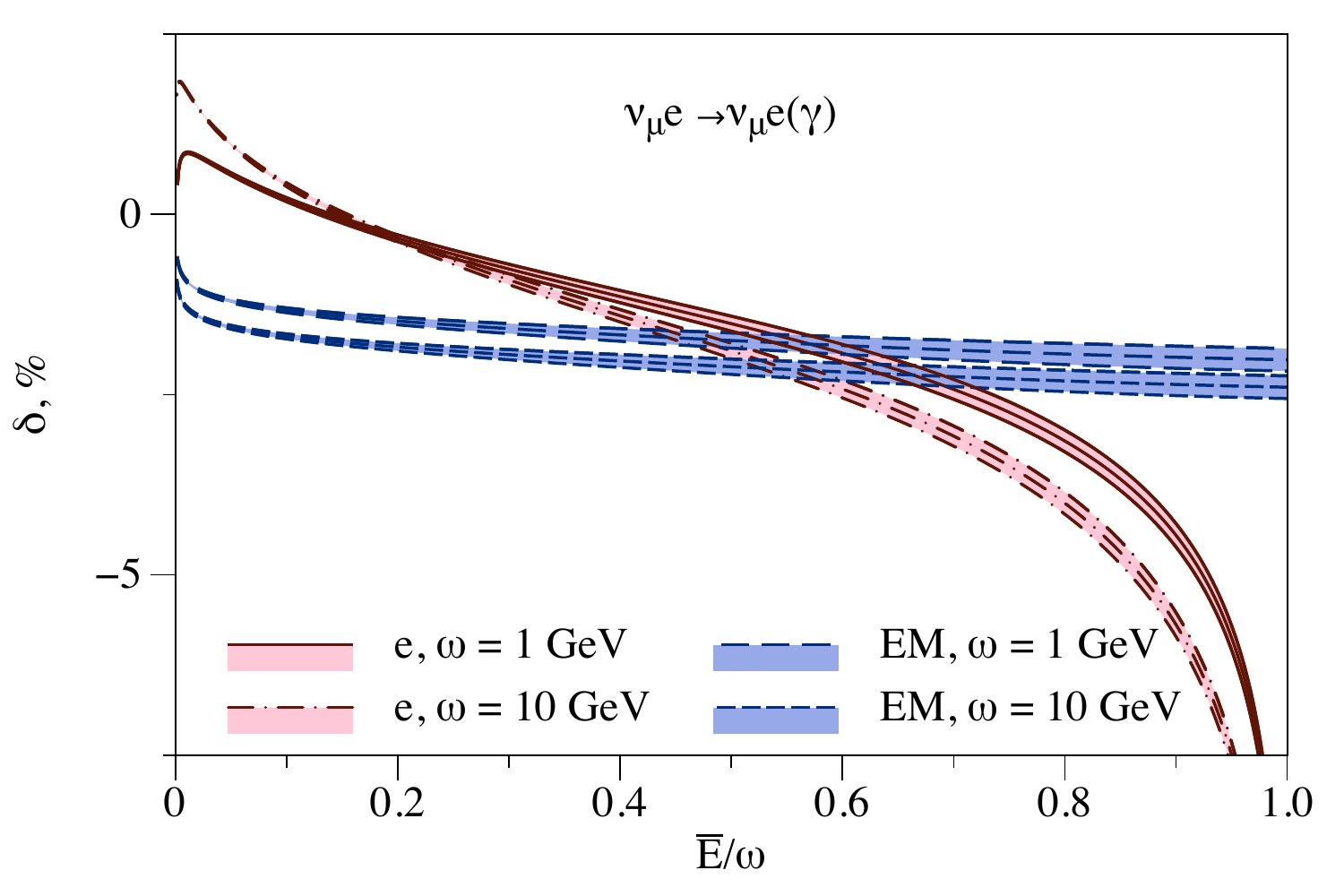}              
          \caption{Radiative corrections to the neutrino-electron
            scattering process $\nu_\mu e \to \nu_\mu e(
            X_\gamma)$ for two neutrino beam energies $\omega = 1, 10\,{\rm GeV}$. The quantity $\delta$ is defined in
            Eq.~(\ref{eq:Delta}) and strongly depends on the $\overline{\mathrm{MS}}$ scale $\mu$. Three curves for $\mu = \mu_0/\sqrt{2},~\mu = \mu_0$, and $\mu = \sqrt{2} \mu_0$ with $\mu_0=2~\mathrm{GeV}$~are presented. The solid and dash-dotted curves correspond
            with electron spectrum, i.e., $\bar{E} = E'$, dashed curves
            with electromagnetic spectrum, i.e., $\bar{E} = E' + k_\gamma$. Uncertainties are not shown on this plot with a scale-dependent quantity. Lower curves correspond to a larger value of $\mu$.
                \label{fig:nu_e_delta}}
\end{figure}
\begin{figure}[t]
          \centering
          \includegraphics[height=0.5\textwidth]{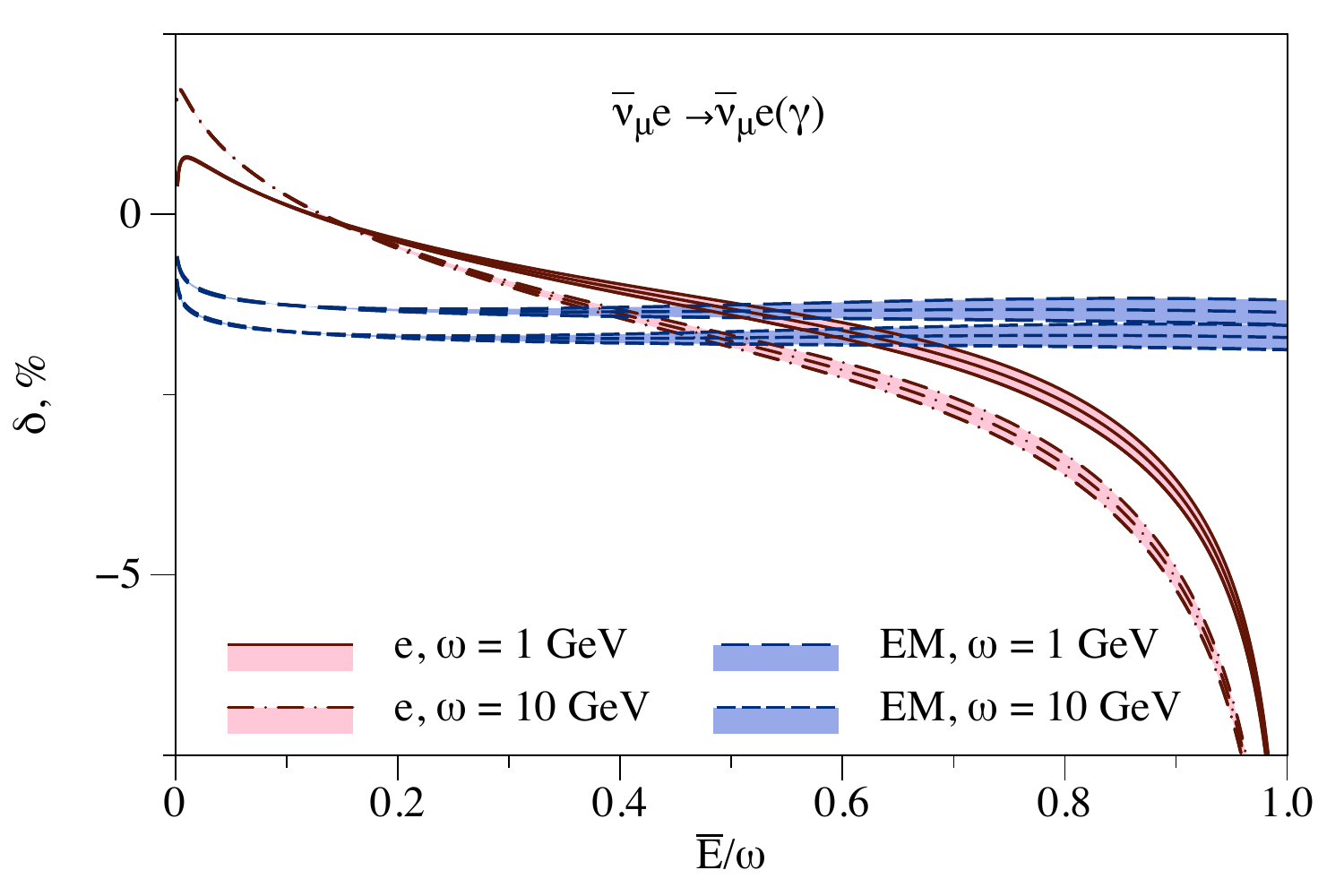}              
          \caption{Same as Fig. \ref{fig:nu_e_delta} for antineutrino-electron
            scattering process $\bar{\nu}_\mu e \to \bar{\nu}_\mu e(
            X_\gamma)$. Uncertainties are not shown on this plot with a scale-dependent quantity. Lower curves correspond to a larger value of $\mu$ for $\bar{E}/\omega \lesssim 0.07-0.1$ and to a smaller value of $\mu$ above.
                \label{fig:antinu_e_delta}}
\end{figure}

Figures~\ref{fig:nu_e_delta} and \ref{fig:antinu_e_delta} display the typical size of the radiative corrections to energy spectra with respect to the final electron energy ($\bar{E} = E'$) and with respect to the total electromagnetic energy
(i.e., the electron energy plus photon energy, $\bar{E} = E' + k_\gamma$).
We consider muon type neutrinos and antineutrinos, the primary component in the accelerator
neutrino beam. 
In these figures, we show the quantity $\delta$ representing the radiative correction
normalized to the leading-order elastic cross section:
\begin{align}\label{eq:Delta}
  \delta = \frac{ \mathrm{d} \sigma^{\nu_\ell e \to \nu_\ell e \gamma }_\mathrm{LO} +  \mathrm{d} \sigma^{\nu_\ell e \to \nu_\ell e }_\mathrm{NLO}- \mathrm{d} \sigma^{\nu_\ell e \to \nu_\ell e }_\mathrm{LO}}{\mathrm{d} \sigma^{\nu_\ell e \to \nu_\ell e }_\mathrm{LO}}.
\end{align}

The correction to the electromagnetic energy spectrum is relatively flat over a wide energy, whereas the correction to the electron energy spectrum is logarithmically divergent below the electron endpoint, cf. Eq.~(\ref{logarithm_near_end_point}). The logarithmic divergence of the electromagnetic energy spectrum above the electron endpoint [cf. Eq.~(\ref{EM_logarithmical_divergence})] is not seen in Fig.~\ref{fig:nu_e_delta} due to the small size of the region in Sec.~\ref{sec:above} compared to the scale of the figure. Both corrections start from the limit of Eq.~(\ref{threshold_result}) at $\bar{E} = m$.
Note that the correction $\delta$ depends on the renormalization scale $\mu$ since the numerator does not contain the leading-order elastic process, rather just the virtual correction to it, leaving the scale dependence of the closed fermion loops (Secs. \ref{sec:long_Range} and~\ref{sec:hadron_physics}) without cancellations.  The large renormalization scale dependence in Figs.~\ref{fig:nu_e_delta} and \ref{fig:antinu_e_delta} illustrates the
cancellations occurring between LO and NLO in arriving at the total cross section in Eq.~(\ref{eq:sigmatot}).  Other uncertainties are not shown in the figure.

\subsection{Electron angular spectrum}
\label{sec:spectra_DUNE}

\begin{figure}[t]
          \centering
          \includegraphics[height=0.5\textwidth]{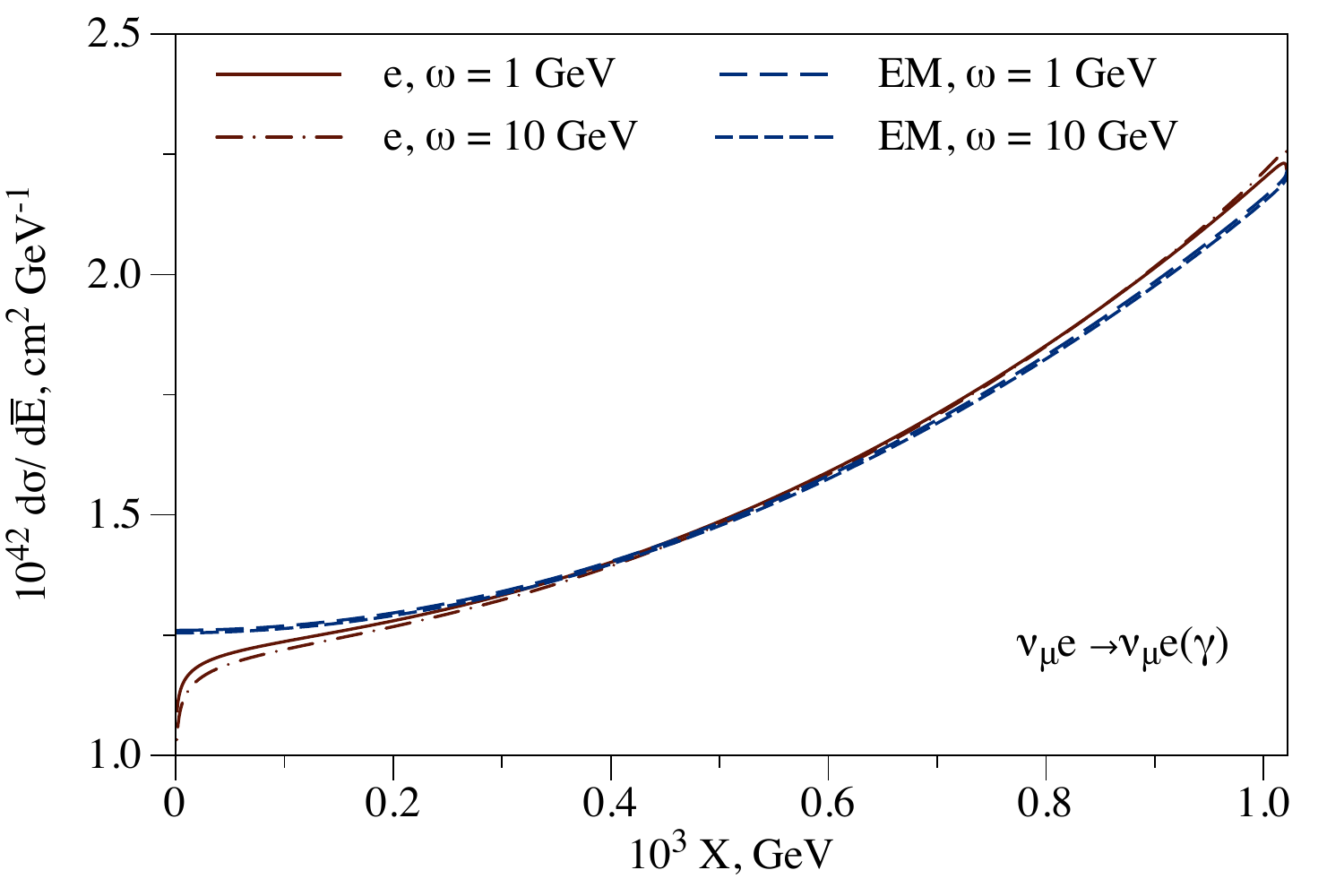}              
          \caption{Energy spectrum in the neutrino-electron
            scattering $\nu_\mu e \to \nu_\mu e(
            \gamma)$, plotted as a function of
            $X=2m(1-\bar{E}/\omega) $ for two neutrino beam energies $\omega = 1, 10\,{\rm GeV}$. The solid and dash-dotted curves correspond
            with the electron spectrum, i.e., $\bar{E} = E'$, dashed  curves
            with the electromagnetic spectrum, i.e., $\bar{E} = E' + k_\gamma$.
    \label{fig:nu_e_Etheta2}}
\end{figure}

\begin{figure}[t]
          \centering
          \includegraphics[height=0.5\textwidth]{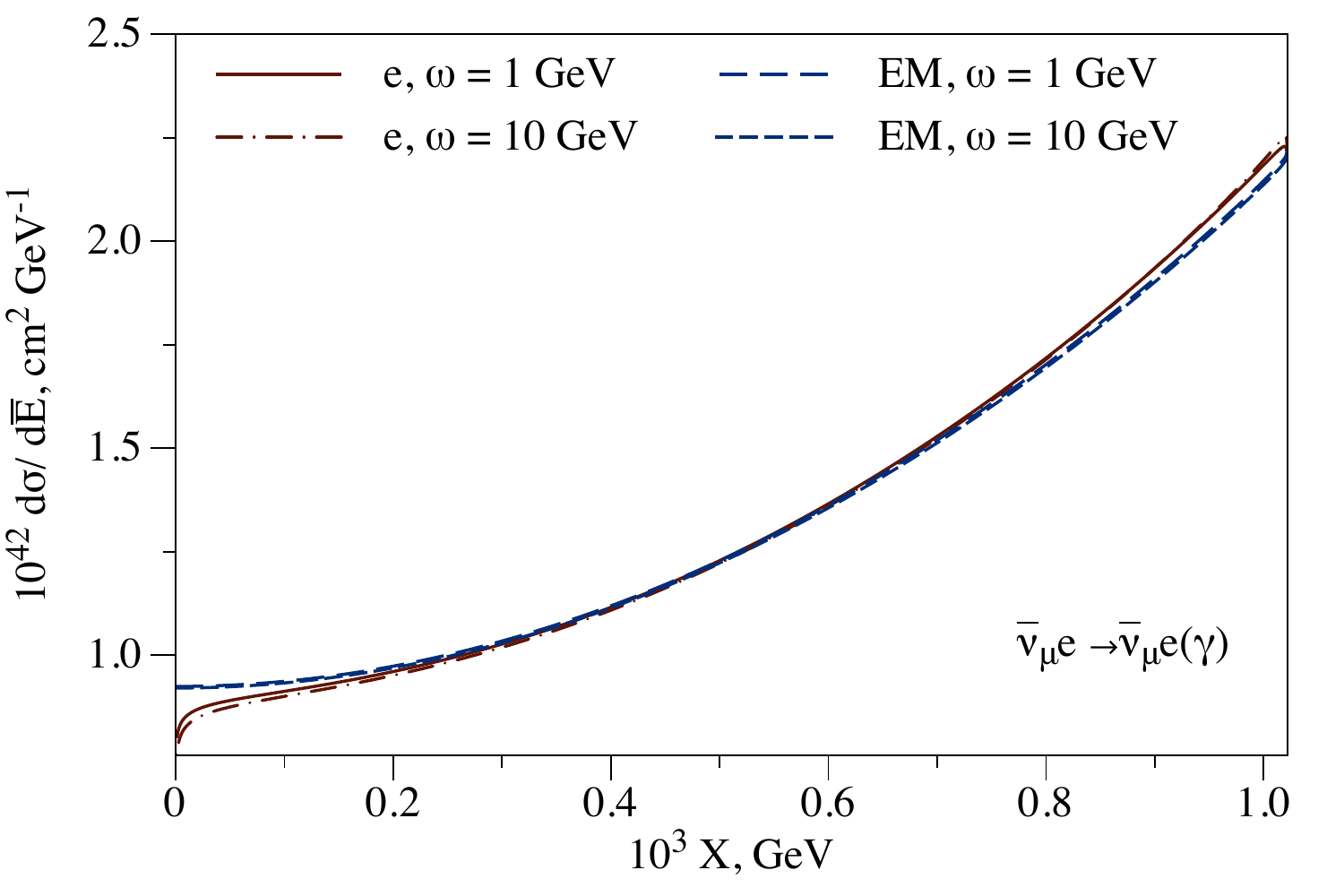}              
          \caption{Same as Fig.~\ref{fig:nu_e_Etheta2} for antineutrino-electron
            scattering process $\bar{\nu}_\mu e \to \bar{\nu}_\mu e(
            X_\gamma)$.
    \label{fig:nu_e_Etheta2_anti}}
\end{figure}

In this section, we consider the angular smearing of differential cross sections. It can be presented as a function of the variable $X$,
\begin{align}
  X = 2 m\left( 1- \frac{\bar{E}}{\omega} \right) \,,
\end{align}
which becomes $X \approx E' \theta_e^2$ for (anti)neutrinos of high energy in the case of the electron energy spectrum. We present the resulting NLO spectrum in Figs.~\ref{fig:nu_e_Etheta2} and~\ref{fig:nu_e_Etheta2_anti} for two (anti)neutrino beam energies: $\omega = 1~\mathrm{GeV}$ and $10~\mathrm{GeV}$. 
Although the electromagnetic and electron energy spectra integrate to the same total cross section,
shape effects induced by radiative corrections can potentially
impact the calibration of neutrino flux.  
For example, experimental cuts requiring a minimum observed energy will result in different numbers of
accepted events depending on which distribution (electromagnetic or electron energy) is chosen.   
In a practical analysis, neither the electron spectrum nor the electromagnetic spectrum will
perfectly represent the experimental conditions, and the more general distributions presented
elsewhere in this paper can be used. 

Results comparing $E^\prime$ and $E_{\rm EM}$ distributions 
after averaging over typical experimental flux profiles are collected in Appendix \ref{app:plots_experiments}.

\subsection{New physics considerations}
\label{sec:BSM}

In this section, we consider the impact of radiative corrections on the dynamical zero (\ref{dynamical_zero})
and isolate the dependence of the neutrino-electron scattering cross section on effective
neutrino charge radii.   Both effects are
present in the Standard Model but may also be used to search for or constrain new physics. 

Recall that an amplitude cancellation causes the tree level electron energy spectrum to vanish at the endpoint of the maximal electron
energy, $E'_0 = m + \frac{2 \omega^2}{m+2 \omega}$, 
when electron antineutrinos of a particular energy $\bar{\omega}$ scatter on electrons; cf. Eq.~(\ref{dynamical_zero}). 
This feature could have implications for novel neutrino oscillation experiments (see, e.g., Refs.~\cite{Segura:1994py,Bernabeu:2003rx}), 
and it is thus interesting to determine the impact of radiative corrections on the cancellation.
To investigate this question, it is convenient to represent the cross section near the endpoint as the factorized product of 
soft and hard functions~\cite{Hill:2016gdf}, $d\sigma \sim S(\mu) H(\mu)$.  The soft 
function accounts for infrared divergences and real photon emission.  Using the
explicit forms for virtual corrections from Sec.~\ref{sec:virtual}, the hard function through first order 
in $\alpha$ takes the form 
\begin{align}
H \propto \bigg( {\omega\over \bar{\omega}} - 1 \bigg) 
\bigg(  {\omega\over \bar{\omega}} - 1 + \order(\alpha) \bigg)  \,.
\end{align}
For $\omega/\bar{\omega} - 1 = \order(\alpha)$, the cross section including radiative corrections
is suppressed by $\order(\alpha^2)$.
The electromagnetic energy spectrum is equal to the electron energy spectrum at tree level
and vanishes at the same kinematic point. 
However, radiative corrections now receive a contribution from ``hard'' real photon emission,
and the electromagnetic spectrum in the vicinity of $\omega = \bar{\omega}$ and $E_{\rm EM}=E_0^\prime$
is nonvanishing at first order in $\alpha$. 
For general $\omega \ne \bar{\omega}$, the electromagnetic energy spectrum vanishes at
the endpoint $E_{\rm EM} = m+ \omega$ and is discontinuous at $E_{\rm EM} = E_0^\prime$; 
at $\omega = \bar{\omega}$ the discontinuity is replaced by a kink. 

Neutrino charge radii~\cite{Bernstein:1963jp,Bardeen:1972vi,Lee:1973fw,Dolgov:1981hv}
may be systematically defined and computed with low-energy effective
field theory~\cite{in_preparation}, where new physics contributions are represented as%
\footnote{
  In terms of weak scale matching coefficients, this corresponds to a contribution to 
  the neutrino-photon coupling in Ref.~\cite{in_preparation}, 
  $\delta \tilde{c}^{\nu_\ell \gamma} = (e^2/6)  \delta r_{\nu_\ell}^2$.
  The ``charge radius" as a low-energy observable quantity is unambiguously 
  defined in terms of four-Fermi coefficients in Ref.~\cite{in_preparation}.
  For a diagrammatic formulation of neutrino charge radii in the Standard Model
  see Ref.~\cite{Bernabeu:2004jr} and references therein. 
  }
\begin{align}
\delta c_{\mathrm{L}}^{\nu_{\ell} e} = \delta c_R^{\nu_\ell e} =  {e^2\over 6} \delta r_{\nu_\ell}^2 \,.
\end{align}
The impact on neutrino-electron scattering is given by 
\ber
\delta \left( \frac{ \mathrm{d} \sigma^{\nu_\ell e \to \nu_\ell e  }}{\mathrm{d} \bar{E}} \right) \hspace{-0.05cm}
&=& \hspace{-0.05cm}
  \frac{m\alpha}{3}\, \delta r^2_{\nu_\ell}  \left | c_{\mathrm{L}}^{\nu_{\ell} e}\mathrm{I}_\mathrm{L}
+   c_\mathrm{R} \mathrm{I}_\mathrm{R} + \frac{c_{\mathrm{L}}^{\nu_{\ell} e} + c_\mathrm{R}}{2}  \mathrm{I}^\mathrm{L}_\mathrm{R} \right |,  \label{error_radius} \\
\delta \sigma^{\nu_\ell e \to \nu_\ell e  } \hspace{-0.05cm} &=& \hspace{-0.05cm} \frac{m\omega\alpha}{3}\,
\delta r^2_{\nu_\ell} \left ( \frac{2 \omega c_{\mathrm{L}}^{\nu_{\ell} e}}{m + 2 \omega} 
+   \left( 1 - \frac{m^3}{ \left( m + 2 \omega \right)^3}\right) \frac{c_\mathrm{R}}{3} - \frac{m \omega  \left(c_{\mathrm{L}}^{\nu_{\ell} e} + c_\mathrm{R}\right) }{\left( m + 2 \omega \right)^2} \right ),  \label{error_radius_absolute}
\eer
with the substitution $c_{\mathrm{L}}^{\nu_{\ell} e} \leftrightarrow c_\mathrm{R}$ in the case of antineutrino scattering.

\section{Conclusions and outlook}
\label{sec:summary}

In this work, we have presented analytical results for elastic (anti)neutrino-electron
scattering starting from four-fermion effective field theory. 
Total cross sections, the electron and electromagnetic energy spectra,
as well as double- and triple-differential cross sections were presented in a relatively compact form.
Our results can be applied to improve constraints of neutrino flux measurements via elastic
neutrino-electron scattering.
All expressions were obtained for finite electron mass and can also be 
used in low-energy applications such as
oscillation measurements with solar and reactor (anti)neutrinos.

Next-to-leading order corrections with bremsstrahlung of one photon are typically 
of order few percent and depend on the experimental setup. For instance,
as discussed in Sec.~\ref{sec:spectra_DUNE}, electron and electromagnetic energy spectra differ significantly.
Although these two spectra integrate to the same total cross section, kinematical cuts can alter inferred
flux constraints if radiative corrections are not matched correctly to experimental conditions.  
Future precise measurements of the electron angular spectrum in neutrino-electron 
scattering can provide energy-dependent neutrino flux constraints.  Our results provide a complete
description of the kinematic dependence of radiative corrections needed to control uncertainties
in neutrino energy reconstruction.
We have discussed the impact of radiative corrections on cross sections
and energy distributions in searches for physics beyond the Standard Model
in Sec.~\ref{sec:BSM}.

We provided a complete error budget for neutrino-electron scattering observables.  
The light-quark contribution to the radiative correction is the 
dominant source of uncertainty.  We have expressed this contribution in terms of
well-defined Standard Model observables, independent of ``constituent quark'' models
used in previous treatments, and determined the relevant hadronic parameter,
denoted $\hat{\Pi}^{(3)}_{3\gamma}(0)$, using SU(3)$_{f}$ symmetry to relate it
to the experimentally constrained parameter $\hat{\Pi}^{(3)}_{\gamma\gamma}(0)$.  
To further pin down the uncertainty of this light-quark contribution, one can
evaluate a closed fermion loop contribution within the dispersion
relation approach decomposing $e^+ e^-$ cross-section data and
measurements of hadronic $\tau$ decays into flavor
components~\cite{Wetzel:1981vt,Papadopoulos:1984rt,Papadopoulos:1984zc,Jegerlehner:1985gq,Jegerlehner:2011mw,Erler:2017knj}
or perform a calculation in lattice QCD~\cite{Burger:2015lqa,Ce:2018ziv}.

We note that due to the restrictive kinematics of neutrino-electron scattering ($|q^2|< 2m \omega$ for the elastic process)
the light-quark contribution enters
as a single constant, representing the $q^2\to 0$ limit of the relevant hadronic tensor.  
This single constant will also impact (and may be constrained by) other low $q^2$ processes such as
coherent neutrino-nucleus scattering.

Besides its phenomenological relevance, the neutrino-electron
scattering process provides an analytically calculable prototype for
the more complicated case of neutrino-nucleus scattering~\cite{HMT}.
In general, radiative corrections can be decomposed (``factorized'')
into soft and hard functions using effective field
theory~\cite{Hill:2016gdf}.\footnote{An application of this formalism to the discussion of the dynamical zero 
in $\bar{\nu}_e e$ scattering was described in Sec.~(\ref{sec:BSM}).} 
The soft functions depend on experimental
configuration but are independent of hadronic physics and  describe
universal large logarithms that are present in general kinematics.
The hard functions are independent of experimental configuration and
describe hadronic physics. In neutrino-electron scattering the
analogous hard functions are perturbatively calculable, whereas in
neutrino-nucleus scattering they must be parametrized and
experimentally constrained.

\section*{Acknowledgments}
We thank K.~McFarland for useful discussions.
O. T. thanks Matthias Heller for useful discussions regarding radiative corrections in QED. O. T. acknowledges the Fermilab theory group for warm hospitality and support. The work of O. T. is supported by the Visiting Scholars Award Program of the Universities Research Association. The work is supported by the U.S. Department of Energy, Office of Science, Office of High Energy Physics, under Award No. DE-SC0019095
and by the Deutsche Forschungsgemeinschaft DFG through the Collaborative Research Center
[The Low-Energy Frontier of the Standard Model (SFB 1044)]. Fermilab is operated by Fermi Research Alliance, LLC under Contract No. DE-AC02-07CH11359 with the United States Department of Energy.  FeynCalc~\cite{Mertig:1990an,Shtabovenko:2016sxi}, LoopTools~\cite{Hahn:1998yk}, JaxoDraw~\cite{Binosi:2003yf}, Mathematica~\cite{Mathematica} and DataGraph were extremely useful in this work.

\appendix

\section{QCD correction to QED vacuum polarization}
\label{app:QCD_QED_vector_vacuum_polarization}

For quark loop contributions in Sec. \ref{sec:long_Range}, we include the leading QCD correction due to one exchanged gluon inside the quark loop. This correction modifies the form factor $\mathrm{\Pi}$ in Eq.~(\ref{weak_ff_MS}) as~$\mathrm{\Pi} \to \mathrm{\Pi} + \mathrm{\Pi}^\mathrm{QCD}$ with $\mathrm{\Pi}^\mathrm{QCD}$ from Refs.~\cite{Djouadi:1987gn,Djouadi:1987di,Kniehl:1989yc,Fanchiotti:1992tu}\footnote{Note that the color factor applies as $N_c \left( \mathrm{\Pi} + \mathrm{\Pi}^\mathrm{QCD}\right)$.}:
\ber
\mathrm{\Pi}^\mathrm{QCD} = \frac{\alpha_s}{3\pi} \left( \ln \frac{\mu^2}{m_f^2} - 4 \zeta \left(3\right) + \frac{55}{12} + \frac{4 m^2_f}{q^2} V_1 \left( \frac{q^2}{4m_f^2} \right) \right),
\eer
where $\alpha_s$ is a strong coupling constant, $\zeta \left( s\right)$ denotes the Riemann zeta functions, and the function $V(r)$ is given by (for spacelike momentum transfer, $r < 0$)
\ber
V\left(r\right) &=&\sqrt{1-\frac{1}{r}} \left( \frac{8}{3} \left( r + \frac{1}{2} \right) \left(  \mathrm{Li}_2 \left( r_-^2\right) - \mathrm{Li}_2 \left( r_-^4\right) +\ln \frac{- 64 \left( 1 - r \right)^2 r }{r_+^3} \ln r_+\right)  - 2 \left( r + \frac{3}{2}\right) \ln r_+ \right)  \nonumber \\
&+& 4 \left( r - \frac{1}{4 r} \right) \left( 2 \mathrm{Li}_3 \left( r_-^2\right) - \mathrm{Li}_3 \left( r_-^4\right) + \frac{8}{3} \left( \mathrm{Li}_2 \left( r_-^2\right) - \mathrm{Li}_2 \left( r_-^4\right)  \right)  \ln r_+\right) +  \frac{13}{6} + \frac{\zeta \left( 3\right)}{r}\nonumber \\
&+& \frac{16}{3}  \left( r - \frac{1}{4 r} \right)  \ln \frac{8 (1-r) \sqrt {-r }}{r_+^3} \ln^2 r_+  - 8 \left( r - \frac{1}{6} - \frac{7}{48 r} \right) \ln^2 r_+,
\eer
with notations $r_\pm = \sqrt{1-r} \pm r$.
As discussed at the end of Sec.~\ref{sec:long_Range},
the relevant limit for neutrino-electron scattering is $-q^2 \to 0$, corresponding with
\ber
\mathrm{\Pi}^\mathrm{QCD} \Bigg |_{q^2 \to - 0}= \frac{\alpha_s}{3\pi} \left(  \ln \frac{\mu^2}{m_f^2} + \frac{15}{4} \right).
\eer

For practical evaluation of a $c$-quark contribution, we take the well-convergent expression in terms of $\overline{\mathrm{MS}}$ quark mass $\hat{m}_c$ from Refs.~\cite{Chetyrkin:1994js,Chetyrkin:1996cf,Chetyrkin:1997un,Erler:1998sy},
\ber
\mathrm{\Pi} &=& \frac{1}{3} \ln \frac{\mu^2}{\hat{m}_c^2} +  \frac{\alpha_s}{3\pi} \left(  -\ln \frac{\mu^2}{\hat{m}_c^2} + \frac{13}{12} \right) \nonumber \\
&+& \frac{\alpha_s^2}{3\pi^2} \left( \frac{655}{144} \zeta \left(3 \right) - \frac{3847}{864} - \frac{5}{6} \ln \frac{\mu^2}{\hat{m}_c^2} - \frac{11}{8}  \ln^2 \frac{\mu^2}{\hat{m}_c^2} + n_f \left( \frac{361}{1296} - \frac{1}{18} \ln \frac{\mu^2}{\hat{m}_c^2}+ \frac{ 1}{12} \ln^2 \frac{\mu^2}{\hat{m}_c^2}\right) \right), \label{eq:MSbar_vaccum_polarization}
\eer
where $n_f = 4$ denotes the number of active quarks. The correction of order $\alpha_s^2$ in Eq.~(\ref{eq:MSbar_vaccum_polarization}) does not change our results within significant digits.

\section{Triple-differential distribution}
\label{app:3xsec}

We evaluate the bremsstrahlung cross section following Ref.~\cite{Ram:1967zza}. For the electron angle distributions, we introduce the four-vector $l$,
\ber
l = k + p - p' = \left( l_0,~\vec{f} \right),
\eer
with the laboratory frame values,
\ber
l_0 &=& m + \omega - E', \\
f^2 &=& |\vec{f}|^2 = \omega^2 + \beta^2 E'^2 - 2 \omega \beta E' \cos \theta_e. \label{parameter_f}
\eer
Note the difference between the electron scattering angle in the elastic process [$\Theta_e$ of Eq.~(\ref{angle_thetae})] and in the scattering with radiation ($\theta_e$).

The triple-differential cross section with respect to the electron angle, electron energy, and photon energy is given by the following substitutions in Eqs.~(\ref{elastic_xsection1}) and (\ref{elastic_xsection2}):
\ber
 \tilde{\mathrm{I}}_\mathrm{L} &\to& \left(  \frac{ l^2 f^2 \left( \left( m + \rho k_\gamma \right) l^2 - 2 \rho m^2 k_\gamma \right)}{4 \sqrt{d}}  -  \frac{ m \omega f^2 \left( \left( m + \rho k_\gamma \right) \left( \rho \left(  l^2 - 2 m l_0 \right) - 2 m^2 \right) + 2 \rho m^3 \right)}{\rho \sqrt{d}} \right. \nonumber \\
 &+& \left.  \frac{ \left(\rho m \left( k_\gamma^2 - m^2 \right) - \rho \left( m + k_\gamma \right) \left( l^2 - 2 m l_0 + m^2\right) + \left( 2m +3 k_\gamma \right) m^2 \right) \sigma}{8 \rho^2 m k_\gamma f^2}   -\frac{\left( m + k_\gamma \right) \sigma^2}{32 \rho^2 m k_\gamma f^4} \right. \nonumber \\
 &-& \left. \frac{ \left( 1 - \rho\right) m \left( m + k_\gamma \right) l^2 \left( l^2 - 4 k_\gamma l_0 + 4 k_\gamma^2 \right)   \left( \rho \left( m+\omega\right) l^2 + \left( 1 - \rho \right) m \left( m+\omega\right)^2 - \left( 1 + \rho \right) m \omega^2 \right)}{16 \rho^2 k_\gamma m f^4}\right. \nonumber \\
 &-&\left. \frac{\left( l^2 - 2 m \omega \right)^2}{8 k_\gamma} + \frac{k_\gamma \left( l^2 - m \left( m+2 \omega \right)\right)}{2}  -\frac{\rho l^4 - 2 m \left( 2 \rho \omega + \left( 2 + \rho \right) m \right) l^2 + 8 m^2 \omega \left( \rho l_0 + 2 m \right)}{8 \rho m}\right. \nonumber \\
 &-&\left.   \frac{\rho^2  k_{\gamma } m^4 \omega^2 f^4 \sigma}{d^{3/2}} -  \frac{ l^6 \left( m + k_\gamma \right) \left( l^2 - 4 k_\gamma l_0 + 4 k_\gamma^2 \right) }{64 k_\gamma m f^4} \right) {\cal{D}}_m, \label{IL_3D} \\
\tilde{\mathrm{I}}_\mathrm{R} &\to&  \left( \frac{f^2 \left( \left(\rho k_{\gamma }+m\right) \left( l^2 - 2 m l_0 \right)^2 + 4 m^2 \left( \rho k_\gamma  \left( \left( l_0 - k_\gamma \right)^2 - m \left( 2 l_0 - k_\gamma \right) \right)  + m \left( l_0 - k_\gamma \right)^2 \right) \right)}{4 \sqrt{d}} \right. \nonumber \\
   &+&\left. \frac{ \rho m^2 f^2 l^2  k_\gamma }{2 \sqrt{d}}- \frac{ \left( l^2 - 2 m l_0 \right)^2}{8 k_\gamma}  - \frac{l^4}{8 m}  -\frac{\rho^2  k_{\gamma } m^4 f^4 \left(l_0 - k_{\gamma }\right)^2  \sigma}{d^{3/2}} + \frac{\left( 2 l_0 - m \right) l^2}{4} \right. \nonumber \\
   &+&\left. \frac{1}{2} k_\gamma m \left( 2 l_0 - k_\gamma - m \right)  - m l_0 \left( l_0 - m \right) \right) {\cal{D}}_m, \label{IR_3D}  \\
 \tilde{\mathrm{I}}^\mathrm{L}_\mathrm{R} &\to& \left( \frac{m \left( m \left(\left( 1 - \rho \right) \left( 4 m^2 l_0^2 - 2 m \left( l_0 + m \right) l^2 \right) -\rho l^4 \right) + 4 k_\gamma  \rho \left( - f^2 - m ^2 + \frac{m l_0}{2} \right) l^2   \right)}{8 \rho k_{\gamma } f^2}  \right.  \nonumber \\
 &-&\left. \frac{ f^2 m \left(\rho^2 \left(l^2+m^2-2m k_\gamma \right) \left(l^2-m^2\right) + 4 \rho k_\gamma m^3  +\left( 2 - \rho \right)^2  m^4 \right)}{2 \rho \sqrt{d}}    + \frac{\rho  (1 -\rho) k_{\gamma } m^6 f^4 \sigma}{d^{3/2}} \right.  \nonumber \\
 &+ &\left. \frac{\rho^2  k^2_{\gamma } m^5 f^4 \sigma}{d^{3/2}} + \frac{m^3  \left( l^2_0 + f^2 + \left( \rho - 1 \right) m l_0 \right)  }{2 \rho f^2}  \right) {\cal{D}}_m,  \label{ILR_3D}
\eer
with the kinematical notations
\ber
\sigma &=& \rho \left( \omega^2 - f^2 - E'^2 +m^2 \right) \left(l^2-2 k_{\gamma } l_0 \right) + 4 k_{\gamma } m f^2  , \\
d &=& \beta^2 m^2 l^2 \omega^2 \left(  l^2 + 4 k_{\gamma }^2 - 4 k_{\gamma } l_0\right)  \sin^2 \theta_{e} + \frac{\sigma^2}{4},
\eer
where the phase-space factor ${\cal{D}}_m$ is given by
\ber
 {\cal{D}}_m = \frac{\pi^2}{m^2 \omega^2} \mathrm{d} f \frac{ \mathrm{d} k_\gamma}{k_\gamma } \frac{\ \mathrm{d} E'}{ \omega} . \label{measure_3D}
\eer
The physical region of variables corresponding to the radiation of hard photons
with energy $k_\gamma \geq \varepsilon$ ($\varepsilon \ll m,~\omega$),
is the following (see Sec.~\ref{sec:1xsec_electron_energy} for a description of hard- and soft-photon regions):
\ber
m + \frac{2\varepsilon^2}{m-2\varepsilon} &\le& E' \le m + \frac{2 \left( \omega - \varepsilon \right)^2}{m + 2 \left( \omega - \varepsilon \right)}, \label{intregion1} \\
| \omega - |\vec{p}~'| | &\le& f \le  l_0 - 2 \varepsilon, \label{intregion2} \\
 \frac{ l_0 - f }{2} &\le& k_\gamma \le \frac{ l_0 + f }{2}. \label{intregion3}
\eer

We keep the exact dependence on the unphysical parameter $\varepsilon$ which is important in the evaluation of the electron energy spectrum in Sec.~\ref{sec:1xsec_electron_energy}. Our integration region in Eqs.~(\ref{intregion1})-(\ref{intregion3}) corresponds to region I in Sec.~\ref{sec:1xsec_electron_energy}.

\section{Double-differential distribution in electron energy and electron angle}
\label{app:2xsec_electron_energy_electron_angle}

Integrating Eqs.~(\ref{IL_3D})-(\ref{ILR_3D}) over the photon energy $k_\gamma$, we obtain the double-differential cross section with respect to the recoil electron energy and angle. The result is given by the following substitutions in Eqs.~(\ref{elastic_xsection1}) and (\ref{elastic_xsection2}):
\ber 
\tilde{\mathrm{I}}_i &\to& \frac{\pi^2 m}{\omega^3} \left( a_i  + b_i \ln  \frac{1+\beta}{1-\beta} + c_i \ln \frac{l_0+f}{l_0-f} + d_i  \ln \frac{l_0 - \beta f \cos \delta - \sqrt{g}}{l_0 - \beta f \cos \delta + \sqrt{g}} \right) \mathrm{d} f \mathrm{d} E'  \nonumber \\
&-&\frac{\pi^2}{\omega}\frac{2}{\beta} \left( \beta -   \frac{1}{2} \ln \frac{1+\beta}{1-\beta} \right)  \mathrm{I}_i \frac{ \mathrm{d} f^2}{l^2} \mathrm{d} E', \label{electron_2Dspectrum}
\eer
with $ g = \left( f \cos \delta - \beta l_0\right)^2 + \rho^2  f^2 \sin^2 \delta $ and the angle $\delta$ between vectors $\vec{l}$ and $\vec{p}~'$,
\ber
\cos \delta = \frac{\omega^2 - \beta^2 E'^2 - f^2 }{2 \beta E' f}. \label{angle_delta}
\eer
Kinematical factors $\mathrm{I}_\mathrm{L},~\mathrm{I}_\mathrm{R},~\mathrm{I}^\mathrm{L}_\mathrm{R}$ in Eq.~(\ref{electron_2Dspectrum}) correspond to the $2 \to 2 $ process.

The coefficients in integrals $\tilde{\mathrm{I}}_i$ are given by
\ber
a_\mathrm{L} &=&\frac{f \left( 2\omega - m \right)}{2 m^2} - \frac{\beta  \cos \delta \left(\frac{3}{4}  l^2  - \frac{f^2}{2 \rho} - E' l_0 - 2 m \omega \right)}{ \rho m^2} -\frac{
   \left(1 + \beta^2 \cos^2 \delta\right) f (l_0+2 m)}{4 \rho^2 m^2}  \nonumber \\
   &-& \frac{\beta^2 l^2 \left(1-3 \cos^2 \delta\right) \left( l_0-4 m \right)}{8 \rho^2 m^2 f} - \frac{(f-\beta l_0 \cos \delta )
   \left(l^2-m \left(\frac{5}{2} l_0 -m +  3 E'\right)\right)}{2 \rho  m^3} , \nonumber \\
b_\mathrm{L} &=& \frac{f l^2}{4 \beta  m^3}-\frac{ \omega f}{\beta  m^2}, \nonumber \\
c_\mathrm{L} &=& \frac{ \beta^2 l^2 \left(1-3 \cos^2 \delta \right) \left(l^2-4 l_0 m\right)}{16 \rho^2 f^2 m^2}-\frac{l^2 \left(l^2-2 s\right)}{8 m^4} - \frac{\omega \left( m + \omega \right)}{m^2} \nonumber \\ 
&-& \frac{\beta  \cos \delta \left( l^2 \left( l^2 - 4 l_0 m - m E'  + m^2 \right)+ 4 m^2 \omega l_0 \right)}{4 \rho f 
   m^3},\nonumber \\
d_\mathrm{L} &=& -\frac{\rho f  \left(\left(l^2-s\right)^2 + s \left(s-2 m^2\right)\right)}{8 \sqrt{g} m^4 }-\frac{ f \rho \omega}{2 \sqrt{g} m }, \nonumber \\
a_\mathrm{R} &=&\frac{3 \beta^2 \rho^2 f^3 l^2 \sin^2 \delta 
   \left(\frac{l_0-\beta  f \text{cos$\delta $} }{4 m \rho}+1\right)}{2 g^2 m}-\frac{f \left(l^2-\frac{11 l_0 m}{2}+m^2\right)}{2 m^3} +  \frac{\beta^2 f \left(2 f^2+3 l_0^2\right)}{2 g m}\nonumber \\
  &+&\frac{f^2 \rho \left(\beta  \text{cos$\delta $} \left(f^2+2 l_0 \left(\omega- 6 E' \right)\right)+f \left(2 m-3 l_0\right)\right)}{4 g m^2} , \nonumber \\
b_\mathrm{R} &=&\frac{ l^2 f}{4  \beta  m^3}-\frac{l_0 f}{\beta  m^2} , \nonumber \\
c_\mathrm{R} &=& -\frac{ l^2 \left(l^2-4 m \left(l_0-\frac{m}{2}\right)\right)}{8 m^4} -\frac{l_0 \left( l_0 - m \right)}{m^2}, \nonumber \\
d_\mathrm{R} &=& \frac{\beta  f l^2 \left(\left(2 l_0 \rho-m \left(-2 \rho^2+\rho+1\right)\right) \left(\beta  l_0- f \text{cos$\delta $}
   \right)-\frac{1}{2} \beta  \rho l^2 \right)}{4 g^{3/2} m^2} - \frac{\rho^2 f l_0\left(f \left(f-\beta l_0 \text{cos$\delta $} \right)+\frac{1}{2} \beta^2 l^2 \right)}{g^{3/2} m}   \nonumber \\
&-&
\frac{\rho f \left( l^2 \left(l^2-4 l_0 m+2 m^2\right)+8 l_0 m^2 \left(\omega - 2 E' \right)\right)}{8 \sqrt{g}
   m^4} +  \frac{3 \beta^2  \rho^3 f^3 l^2
     \sin^2 \delta  \left(l^2+ 4 E' \left(l_0-\beta  f   \cos \delta \right)\right)}{16
   g^{5/2} m^2}, \nonumber \\
a^\mathrm{L}_\mathrm{R} &=& \frac{ \beta f (\beta  l_0- f \cos \delta) }{g}-\frac{\beta \cos \delta}{\rho}, \nonumber \\
b^\mathrm{L}_\mathrm{R} &=& - \frac{\rho f l^2}{2 \beta m^3 }, \nonumber \\
c^\mathrm{L}_\mathrm{R} &=& \frac{\beta l_0 \cos \delta}{2 \rho f } -  \frac{l^2 - m \left( m + E'\right)}{2 m^2}  , \nonumber \\
d^\mathrm{L}_\mathrm{R} &=& \frac{\rho f \left( \left(m + 2 E' \right) m - l^2 \right) }{2 m^2 \sqrt{g}}-\frac{\rho^2f^2  (f-\beta l_0 \cos \delta) }{2 g^{3/2}}.  \nonumber
\eer

\section{Double-differential distribution in electromagnetic energy and electron angle}
\label{app:massles_2x_photon_energy_electron_angle}

To obtain neutrino energy (equivalently, electromagnetic energy) and electron angle distribution, Eqs.~(\ref{IL_3D})-(\ref{ILR_3D}) can be integrated over the electron energy, exploiting the energy conservation: $k_\gamma = m + \omega - \omega' - E'$.
The integration measure of Eq.~(\ref{measure_3D}) is replaced as
\ber
 {\cal{D}}_m = \frac{\pi^2}{m^2 \omega^2} \mathrm{d} f \frac{ \mathrm{d} k_\gamma}{k_\gamma } \frac{\mathrm{d} E'}{ \omega} \to \frac{\pi^2}{m^2 \omega^2} \mathrm{d}  \cos \theta_e \frac{ \mathrm{d} \omega'}{k_\gamma } \frac{ \beta E'  \mathrm{d} E'}{ f} .
\eer
The physical integration region is contained in
 \ber
0 &\le& \omega' \le  \omega ,  \label{massles_2x_photon_energy_electron_angle__large_omegap}    \\
0 &\le& \cos \theta_e \le 1 ,   \label{massles_2x_photon_energy_electron_angle__large_cos} \\
m &\le& E' \le m \frac{\left( m + \omega \right)^2 + \omega^2 \cos^2 \theta_e}{\left( m + \omega \right)^2 - \omega^2 \cos^2 \theta_e},\label{massles_2x_photon_energy_electron_angle__large_Ep}
\eer
which is actually larger than the physical region. The extraneous regions I and II are above the electron endpoint ($E_\mathrm{EM} \ge E'_0 = m + \frac{2 \omega^2}{m+2 \omega}$) and below it ($E_\mathrm{EM} \le E'_0 = m + \frac{2 \omega^2}{m+2 \omega}$):
 \ber
 \mathrm{region~I} :\qquad \qquad \qquad \qquad \qquad \qquad \qquad \qquad \qquad \qquad 0 &\le& \omega' \le  \frac{m \omega }{ m + 2 \omega},     \label{massles_2x_photon_energy_electron_angle_subtraction_first} \\
 \frac{2 \sqrt{ \omega'  \left( m - \omega' \right) \left( \omega -  \omega'  \right)  \left( m + \omega-  \omega'  \right)} }{m \omega}  &\le& \cos \theta_e \le 1  ,   \\
E'_- \left( \omega' \right) &\le& E' \le E'_+ \left( \omega' \right),   \\
 \nonumber \\ 
 \mathrm{region~II} :\qquad \qquad \qquad \qquad \qquad \qquad \qquad \qquad ~~~ \frac{m \omega}{ m + 2 \omega}&\le& \omega' \le \omega,   \\
 \frac{m+\omega}{\omega} \sqrt{\frac{\omega-\omega'}{2 m + \omega - \omega'}} &\le&\cos \theta_e \le1 ,  \\
 E'_- \left( \omega' \right) &\le& E' \le m \frac{\left( m + \omega \right)^2 + \omega^2 \cos^2 \theta_e}{\left( m + \omega \right)^2 - \omega^2 \cos^2 \theta_e}.  \label{massles_2x_photon_energy_electron_angle_subtraction_last}
 \eer
Here $ E'_\pm \left( \omega' \right) $ stand for two solutions ($E_+ \ge  E_-$) of 
\ber
\cos \theta_e = \frac{E' (m+\omega -2 \omega' )-m^2+m (2 \omega' -\omega )+2 \omega'  (\omega -\omega' )}{\omega  \sqrt{E'^2-m^2}}.
\eer

The presentation here in terms of a larger region~(\ref{massles_2x_photon_energy_electron_angle__large_omegap})-(\ref{massles_2x_photon_energy_electron_angle__large_Ep}) and subtractions~(\ref{massles_2x_photon_energy_electron_angle_subtraction_first})-(\ref{massles_2x_photon_energy_electron_angle_subtraction_last}) is designed as a simple description of the actual physical region. In practice, one may perform the integration over this larger region and use subtractions above the electron endpoint $E_\mathrm{EM} \ge E'_0 = m + \frac{2 \omega^2}{m+2 \omega}$; or one may break up the integration region~(\ref{massles_2x_photon_energy_electron_angle__large_omegap})-(\ref{massles_2x_photon_energy_electron_angle__large_Ep}) and integrate once only over the physical region.

\section{Double-differential distribution in photon energy and electron energy}
\label{app:2xsec_electromagnetic_and_electron_energy}

To obtain the distribution with respect to the photon energy and electron energy, Eqs.~(\ref{IL_3D})-(\ref{ILR_3D}) can be integrated first over the variable $f$ after the change of the integration order. The kinematical region of electron energy is bounded as
\beq 
m \le E' \le m + \frac{2 \omega^2}{m + 2  \omega }.
\eeq
The physical region of $f$ for different values of $k_\gamma$ is then given by
\ber
0 &\le& k_\gamma \le \frac{l_0 - | \omega - |\vec{p}~'||}{2},~\quad~~~~ l_0 - 2 k_\gamma \le f \le  l_0 ; \\
\frac{l_0 - | \omega - |\vec{p}~'||}{2} &\le& k_\gamma \le \frac{l_0 + | \omega - |\vec{p}~'|}{2},~\quad~~~ | \omega - |\vec{p}~'| | \le f \le  l_0 ; \\
\frac{l_0 + | \omega - |\vec{p}~'||}{2} &\le& k_\gamma  \le l_0, ~\quad \quad \quad ~~~~~~~~~~~~- l_0 + 2 k_\gamma  \le f \le  l_0 .
\eer

\section{Double-differential distribution in photon energy and photon angle}
\label{app:2xsec_photon_energy_and_angle}

We evaluate the bremsstrahlung cross section with respect to the photon energy and photon angle considering the final photon energy spectrum instead of the electron spectrum~\cite{Ram:1967zza}, see Sec.~\ref{sec:2xsec_electromagnetic_energy_electron_angle} for explanations. For the photon scattering angle (with respect to the neutrino beam direction) distributions, we introduce the four-vector $\bar{l}$,
\ber
\bar{l} = k + p - k_\gamma = \left( \bar{l}_0,~\vec{\bar{f}} \right),
\eer
with the laboratory frame values,
\ber
\bar{l}_0 &=& m + \omega - k_\gamma, \\
\bar{f}^2 &=& |\vec{\bar{f}}|^2= \omega^2 + k_\gamma^2 - 2 \omega k_\gamma  \cos \theta_\gamma, \label{parameter_f2}
\eer
where $\theta_\gamma$ denotes the photon scattering angle.

The photon energy spectrum accounting for electron mass terms is given by the following substitutions in Eqs.~(\ref{elastic_xsection1}) and (\ref{elastic_xsection2}):
\ber
 \tilde{\mathrm{I}}_i \to \frac{\pi^2}{2 m \omega^2} \left( a_i \left( \bar{l}^2 - m^2 \right) +   b_i \ln \frac{m^2}{\bar{l}^2} \right) \frac{\bar{f} \mathrm{d} \bar{f}}{\left(\bar{l}^2-s\right)^2} \mathrm{d} k_\gamma ,
\label{photon_spectrum_non_universal_part_2D}
\eer
with $s = m^2 + 2 m \omega$ and coefficients $a_i$ and $b_i$ in Eq.~(\ref{photon_spectrum_non_universal_part_2D}):
\ber
a_\mathrm{L}&=&\frac{\left(\bar{l}^2-m^2\right)^2 \left(2 \bar{l}^2 (k_\gamma \bar{l}_0+m (2 \bar{l}_0-m)) + m \left(-2 \bar{l}_0^2 (2
   \omega + m )+\bar{l}_0 m (6 \omega + m)-3 m^2 \omega \right)\right)}{4 k_\gamma^2 \bar{l}^2 m \omega } \nonumber \\
  &-&\frac{ 4 m \omega ^3 \left(m (2 k_\gamma-\omega+m )+\bar{l}^2\right)- \omega  (\omega -k_\gamma)
   \left(3 \bar{l}^4-6 \bar{l}^2 \bar{l}_0 m -m^3 (2 \bar{l}_0 - 5 m)\right)}{k_\gamma^2 \bar{l}^2 } \nonumber \\
   &-&\frac{2   \omega ^2 \left(\bar{l}^4- \bar{l}^2 m (5 \bar{l}_0 - 3 m)+m^2 (2 \bar{l}_0-3 m) (\bar{l}_0-2 m)\right)}{k_\gamma^2 \bar{l}^2} -\frac{\left(\bar{l}^2-m^2\right)^2 \bar{l}^2 (k_\gamma+m)}{4 k_\gamma^2 m^2 \omega }, \nonumber \\
\frac{a_\mathrm{R}/\left(\bar{l}^2-m^2\right)}{\bar{f}^2-(\bar{l}_0-m)^2 }  &=&-\frac{m \left(\bar{l}^2-s\right) \left(-4 k_\gamma^3 (11 \omega +17 m)-4 k_\gamma^2 \left(35 \omega ^2 +103 m^2 \right)-3
   k_\gamma m^2 (12 \omega +29 m )\right)}{12 k_\gamma^2 \bar{l}^6 \omega } \nonumber \\
   &-&\frac{2 m^3 s \left(k_\gamma^2 \left(344 \omega ^2+1116 m \omega + 537 m^2\right)+m \left(312  \omega
   ^3+501 m \omega ^2+342 m^2 \omega +72 m^3 \right)\right)}{3 \bar{l}^6 \omega  \left(\bar{l}^2-m^2\right) \left(\bar{f}^2-(\bar{l}_0-m)^2\right)} \nonumber \\
     &-&\frac{4 m^2 s^2 \left(6 k_\gamma^2 m \left(27 \omega ^2+93 m \omega +46 m^2\right)+k_\gamma s \left(51
   \omega ^2+154 m \omega +108 m^2\right)-3 s m \omega ^2\right)}{3 \bar{l}^6 \omega  \left(\bar{l}^2-m^2\right) \left(\bar{l}^2-s\right) \left(\bar{f}^2-(\bar{l}_0-m)^2\right)} \nonumber \\
       &-& \frac{m^3 \left(\bar{l}^2-s\right) \left(4 k_\gamma^2 \left(64 \omega ^2+197 m \omega +96 m^2\right)+k_\gamma m \left(2950 \omega ^2+3376 m
   \omega +1191 m^2\right)\right)}{3 \bar{l}^6 \omega  \left(\bar{l}^2-m^2\right) \left(\bar{f}^2-(\bar{l}_0-m)^2\right)} \nonumber \\
         &-&\frac{m^2 \left(\bar{l}^2-s\right) \left(-4 k_\gamma^2 \omega +k_\gamma \left(268 \omega ^2+794 m \omega +384 m^2\right)+2 m^2 \left(785 \omega+327
   m \right)\right)}{6 \bar{l}^6 \omega  \left(\bar{f}^2-(\bar{l}_0-m)^2\right)} \nonumber \\
           &-&\frac{2 m^2 s \left( k_\gamma^2 m \left(590 \omega ^3+2106 m \omega ^2+2144 m^2 \omega +617 m^3\right)+92 k_\gamma m \omega ^4-2
   s^2 \omega ^2\right)}{3 k_\gamma \bar{l}^6 \omega  \left(\bar{l}^2-m^2\right) \left(\bar{f}^2-(\bar{l}_0-m)^2\right)} \nonumber \\
             &-&\frac{m^2 \left(\bar{l}^2-s\right) \left(2 \omega ^2 k_\gamma  \left(162 \omega+579 m\right)-m \left(16
   \omega ^3-18 m \omega ^2-105 m^2 \omega - 64 m^3\right)\right)}{6 k_\gamma \bar{l}^6 \omega  \left(\bar{f}^2-(\bar{l}_0-m)^2\right)} \nonumber \\
                 &-&\frac{4 m^3 \left(\bar{l}^2-s\right) \left(193 k_\gamma^2 \omega ^3-m \left(4 \omega ^2 + 2 m \omega-3 m^2\right) \left(\omega
   ^2+m \omega +m^2\right)\right)}{3 k_\gamma \bar{l}^6 \omega  \left(\bar{l}^2-m^2\right) \left(\bar{f}^2-(\bar{l}_0-m)^2\right)}+\frac{11 \left(\bar{l}^2-m^2\right)}{6 \bar{l}^4 }  \nonumber \\
                   &-&\frac{m^3 \left(\bar{l}^2-s\right) \left( 184 \omega ^4+740 m \omega ^3+1344 m^2 \omega ^2+1167 m^3 \omega + 405 m^4\right)}{3 \bar{l}^6 \omega 
   \left(\bar{l}^2-m^2\right) \left(\bar{f}^2-(\bar{l}_0-m)^2\right)} + \frac{106 m^2 \left(\bar{l}^2-s\right)}{3 \bar{l}^6}\nonumber \\
                     &+&\frac{\left(\bar{l}^2-m^2\right) \left(k_\gamma^2+ 11 k_\gamma  m + \omega ^2 + 6 m^2\right)}{3 k_\gamma \bar{l}^4 \omega }+ \frac{(k_\gamma+m) \left(\bar{l}^2-m^2\right) \left(\bar{f}^2-(\bar{l}_0-m)^2\right)^2}{12 k_\gamma^2 m^2 \omega  \left(\bar{l}^2-s\right)^2}\nonumber \\
                  &-& \frac{8 k_\gamma m^3 s^3 \left(48 k_\gamma s +\omega ^2 (27 k_\gamma-m) \right)}{3 \bar{l}^6
   \omega  \left(\bar{l}^2-m^2\right) \left(\bar{l}^2-s\right)^2 \left(\bar{f}^2-(\bar{l}_0-m)^2\right)}  -\frac{1}{6 k_\gamma^2}  -\frac{m^5 \left(\bar{l}^2-s\right)}{12 k_\gamma^2 \bar{l}^6 \omega } , \nonumber \\
a^\mathrm{L}_\mathrm{R} &=&\frac{\bar{l}^2 m \left(8 k_\gamma^2+14 k_\gamma \bar{l}_0-9 k_\gamma m-2 \bar{l}_0 m\right)}{2
   k_\gamma^2 \omega }-\frac{m^4 s^2 (2 k_\gamma-m)^2  (2 \bar{l}_0-m)}{4 k_\gamma^2 \bar{l}^4 \omega 
   \left(\bar{l}^2-s\right)}\nonumber \\
   &+& \frac{m^3  \left(8 k_\gamma^3 (2 \omega +5 m)-8 k_\gamma^2 \left(6 (\omega + m)^2+m (\omega +2 m)\right)+4
   k_\gamma \left(8 (\omega + m)^3-3 s \omega \right)\right)}{4 k_\gamma^2 \bar{l}^2 \omega } \nonumber \\
   &+& \frac{m^2  \left(-4 k_\gamma^3+2 k_\gamma^2 (8 \omega -7 m)-k_\gamma \left(28
   \omega ^2+34 m \omega +15 m^2\right)+m \left(3 (\omega + m)^2-\omega ^2\right)\right)}{2 k_\gamma^2 \omega }\nonumber \\
   &-& \frac{m^3  \left(8 k_\gamma^2 \left(16 k_\gamma \omega + 13 k_\gamma m +m^2\right)-s (2
   k_\gamma - m)^2+2 k_\gamma m (4 k_\gamma - m)^2\right)}{4 k_\gamma^2 \omega  \left(\bar{l}^2-s\right)} \nonumber \\
   &-& \frac{\bar{l}^4 (4 k_\gamma - m) }{4 k_\gamma^2 \omega } - \frac{m^3 s  \left(2 s + 3 m^2\right)}{4 k_\gamma^2 \bar{l}^2 \omega } , \nonumber \\
b_\mathrm{L} &=& -\left( \bar{l}^2 - m^2  \right)^2 - 4 m^2 \omega \left( \omega+2 m \right), \nonumber \\
\frac{b_\mathrm{R}}{ \bar{f}^2-(\bar{l}_0-m)^2} &=& \frac{16 k_\gamma m^3 (\omega + m) \left((\omega +2 m)^2+4 m \omega \right)}{\omega 
   \left(\bar{l}^2-s\right)^2}+\frac{\bar{l}^2  \left(k_\gamma (\omega + 2 m )+m^2\right) }{k_\gamma \omega }\nonumber \\
   &+& \frac{m  \left(2 k_\gamma^2 (\omega + m)+(2 \omega + 3 m ) \left(k_\gamma (\omega + 4 m
   )+m^2\right)\right) }{k_\gamma \omega } \nonumber \\
   &-&\frac{8 m^2\left(\bar{l}_0 \left((\omega+2 m )^2+m (\omega -m)\right)-2 (\omega + m)^2 (\omega+4 m
   )\right)}{\omega  \left(\bar{l}^2-s\right)} , \nonumber \\
b^\mathrm{L}_\mathrm{R} &=&-\frac{\bar{l}^4 \left(m^2 \left(-12 k_\gamma \omega+20 \omega  (\omega + m) +7 m^2\right)-2 \bar{l}^2 m (\bar{l}_0+\omega
   )+\left(\bar{l}^2-s\right)^2\right)}{k_\gamma \omega  \left(\bar{l}^2-s\right)} \nonumber \\
   &-&\frac{2 \bar{l}^2 m^2 \left(2 k_\gamma^2 m (2 \omega + 3 m )+k_\gamma m \left(7 m^2+10 m \omega +12 \omega ^2\right)-4 s \left((\omega+m
   )^2+\omega ^2\right)\right)}{k_\gamma \omega  \left(\bar{l}^2-s\right)} \nonumber \\
   &-&\frac{m^2 \left(4 k_\gamma^2 m^3 (8 \omega + 5 m )-8 k_\gamma m s \left((\omega + m)^2+m^2\right)+s^2 \left((2 \omega + m)^2 + 2 m^2\right)\right)}{k_\gamma \omega  \left(\bar{l}^2-s\right)} .\nonumber
\eer

\section{Photon energy spectrum}
\label{app:1xsec_photon_energy}

The photon energy spectrum accounting for electron mass terms is given by the following substitutions in Eqs.~(\ref{elastic_xsection1}) and (\ref{elastic_xsection2}):
\ber
 \tilde{\mathrm{I}}_i &\to& \frac{\pi^2}{\omega^3} \Bigg[ a_i +   b_i \ln \frac{k_\gamma}{\omega} + c_i \ln \frac{2 \bar{l}_0 - m}{m} - d_i \ln \frac{2 k_\gamma}{2\omega + m} \ln \frac{2 \bar{l}_0 - m}{m} \Bigg. \nonumber \\
 &+& \Bigg. d_i   \sum \limits_{\sigma_1,~\sigma_2 =\pm} \Re \Bigg( \mathrm{Li}_2 \frac{\bar{l}_0+ \sigma_1 \sqrt{\bar{l}_0^2 - m^2}}{\bar{l}_0+ \sigma_2 \sqrt{\left( \bar{l}_0-m\right)^2 - 2 m k_\gamma}} -  \mathrm{Li}_2 \frac{\bar{l}_0 + \sigma_1\left(\bar{l}_0-m\right)}{\bar{l}_0+ \sigma_2 \sqrt{\left(\bar{l}_0-m\right)^2 - 2 m k_\gamma}}  \Bigg) \Bigg] \mathrm{d} k_\gamma,
\label{photon_spectrum_non_universal_part}
\eer
with coefficients $a_i,~b_i,~c_i$, and $d_i$ in Eq.~(\ref{photon_spectrum_non_universal_part}):
\ber
a_\mathrm{L} &=& \frac{(\omega - k_\gamma ) \left(2 k_\gamma^3-k_\gamma^2 m+6 k_\gamma m^2-2 \omega ^2 (53 k_\gamma+2
   m)-\omega  (8 k_\gamma-m) (5 k_\gamma+3 m)-3 m^3\right)}{24 k_\gamma^2},\nonumber \\
\frac{a_\mathrm{R}}{\omega - k_\gamma} &=&   \frac{m^5}{24 k_\gamma^2 (2 \omega + m)^2} +\frac{m^3 \left(-36
   k_\gamma^2-10 k_\gamma m+m^2\right)}{96 k_\gamma^3 (2 \omega + m)} -\frac{m \left(k_\gamma^3 m+\left(k_\gamma-\frac{m}{4}\right) \left(4
   k_\gamma^3-2 k_\gamma^2 m-m^3\right)\right)}{24 k_\gamma^3 (2 \omega-2 k_\gamma+m )} \nonumber \\
&+&\frac{m^2 \omega }{12 (2 \omega-2 k_\gamma+m )^2} -  \frac{\omega ^2 (73 k_\gamma+2 m)}{36 k_\gamma^2}-\frac{m (656
   \omega + 897 m )}{144 k_\gamma}-\frac{37 k_\gamma}{36}-\frac{892 \omega + 1184 m}{144},  \nonumber \\
a^\mathrm{L}_\mathrm{R} &=& \frac{m^2 \omega  (\omega -k_\gamma) (2 \omega+3 m )}{8 k_\gamma^2 (2 \omega + m)}-\frac{m (\omega -k_\gamma) \left(26
   k_\gamma^2 - k_\gamma (22 \omega -13 m )+3 m^2\right)}{8 k_\gamma^2},\nonumber \\
b_\mathrm{L}    &=&-\frac{\omega ^2 (3 k_\gamma (2 \omega + m)+2 \omega  (\omega + m))}{k_\gamma (2 \omega + m)}, \nonumber \\
b_\mathrm{R} &=& \frac{4 m \omega ^4 (\omega + m)}{3
   k_\gamma (2 \omega + m)^3} -\frac{6 \omega ^4}{(2 \omega + m)^2}+ \frac{\omega ^2 \left(3 k_\gamma^2+14 \omega ^2\right)}{3 k_\gamma (2 \omega + m)}-\frac{14 \omega ^3}{3 k_\gamma} \nonumber  \\ 
   &-&\frac{\omega  \left(8 k_\gamma m+3
   k_\gamma (k_\gamma+\omega )-2 \left(\omega ^2-m^2\right)\right)}{k_\gamma}, \nonumber \\
b^\mathrm{L}_\mathrm{R} &=& -\frac{m \omega  \left(k_\gamma (2 \omega + m) (2 \omega+3 m )-2 \omega ^3+3 m^2 \omega +m^3\right)}{k_\gamma (2 \omega + m)^2},\nonumber \\
c_\mathrm{L}    &=&\frac{m^2 \left(4 \omega ^2 - 3 m^2\right)}{16 k_\gamma (2 \omega + m)} - \frac{8 k_\gamma^3 \omega + 2 k_\gamma^2 \left(4 \omega ^2 -m^2\right) -16 k_\gamma \omega ^2 (\omega + m)-m^4}{16
   k_\gamma^2} , \nonumber \\
c_\mathrm{R} &=&-\frac{m^4 \left(36 \omega ^2+30 m \omega +7 m^2\right)}{24
   k_\gamma (2 \omega + m)^3} +\frac{3 m^4}{8 (2 \omega + m)^2}-\frac{3 m^3}{2 (2 \omega + m)}-k_\gamma \left(\frac{\omega ^2}{2 \omega + m}+\frac{\omega}{2} +\frac{15 m}{4}\right) \nonumber \\
   &+&\frac{\omega ^3-k_\gamma^3}{3 k_\gamma} + \frac{m \left(72 \omega ^2+204 m \omega +123 m^2\right)}{48 k_\gamma} +\frac{m^4}{48 k_\gamma^2}-\frac{13 m^2}{8}+\omega  (\omega + 3 m ),\nonumber \\
c^\mathrm{L}_\mathrm{R} &=& \frac{m \omega  \left(8 k_\gamma (\omega + m) (2 \omega + m)-\omega  \left(8 \omega ^2 +12 m \omega+ 3 m^2\right)\right)}{2 k_\gamma
   (2 \omega + m)^2}-\frac{m (2 k_\gamma-m) \left(8 k_\gamma (k_\gamma+m)+3 m^2\right)}{16 k_\gamma^2}, \nonumber \\
d_\mathrm{L}    &=& - \omega^2 , \nonumber \\
d_\mathrm{R} &=& -\frac{k_\gamma^2 (2 \omega+ 3 m )+2 k_\gamma (\omega + 2 m )^2+2 m^2 (\omega + m )}{2 k_\gamma},\nonumber \\
d^\mathrm{L}_\mathrm{R} &=& -\frac{m^2 (3 k_\gamma+m)}{2 k_\gamma}.\nonumber
\eer

\section{Electron energy spectrum}
\label{app:coefficients_electron}

The nonfactorizable contribution to the
electron energy spectrum $\mathrm{d} {\sigma}^{\nu_\ell e \to \nu_\ell e \gamma   }_{\,{\rm NF}}$ from
Eq.~(\ref{factorizable_correction}), 
is given by the following substitutions in Eqs.~(\ref{elastic_xsection1}) and (\ref{elastic_xsection2}):
\ber
 \tilde{\mathrm{I}}_i &\to& \frac{\pi^2}{\omega^3} \left( z_i +   y_i \ln \frac{\frac{2\omega}{m}}{-1+\frac{\rho}{1+\beta}\left(1 + \frac{2\omega}{m} \right)} + x_i \ln \frac{\frac{2 l_0}{m}}{1 + \frac{2\omega}{m}  - \frac{1+\beta}{\rho}} + r_i   \ln \frac{1 - \frac{1+\beta}{\rho}}{\frac{1+\beta}{1-\beta} - \frac{1+\beta}{\rho}\left(1 + \frac{2\omega}{m} \right)}\right) \mathrm{d} E' \nonumber \\
&+& \frac{\pi^2}{\omega^3} \left(  q_i \ln  \frac{1+\beta}{1-\beta} + v_i \left(\mathrm{Li}_2 \frac{1+\beta}{\rho} - \mathrm{Li}_2 \left(1 + \frac{2\omega}{m} \right) +  \mathrm{Li}_2 \frac{\left(1 + \frac{2\omega}{m} \right)\rho}{1+\beta}- \frac{\pi^2}{6} \right)\right) \mathrm{d} E'.
\label{electron_spectrum_non_universal_part}
\eer
Exact expressions for coefficients $z_i,~y_i,~x_i,~r_i,~q_i$, and $v_i$ in Eq.~(\ref{electron_spectrum_non_universal_part}) are given by
\ber
v_\mathrm{L} &=& \frac{1}{2} \left(\frac{m^2}{2}+2 m \omega +\omega ^2\right), \qquad v_\mathrm{R} = \frac{1}{2} \left(l_0^2+\frac{\beta ^2+\rho}{\rho^2} m^2\right), \qquad v^\mathrm{L}_\mathrm{R} = \frac{1}{2} m \left(2 l_0 - m \right), \nonumber \\
x_\mathrm{L} &=& - \frac{2}{15} \frac{\omega^5}{m^3} + \frac{1}{3} \frac{\omega^3}{m} + \left( \frac{1+3\beta^2}{3\rho^3} - \frac{4 \beta^4-11\beta^2+7}{3\rho^4} \right) \omega^2 + \left(  \frac{2}{\rho^3} - \frac{\beta^4-\beta^2+2}{\rho^4} \right) m \omega \nonumber \\
&+& \left(  \frac{-7\beta^4+14\beta^2-22}{15\rho^4} + \frac{15 \beta^4 -25 \beta^2+22}{15\rho^5} \right) m^2   , \nonumber \\
x_\mathrm{R} &=&-\frac{l_0^2 \left(35 l_0 m^2-10 l_0^2 m+2 l_0^3-30 m^3\right)}{15 m^3}, \nonumber \\
x^\mathrm{L}_\mathrm{R} &=&\frac{3 l_0 m^2-3 l_0^2 m-2 l_0^3+3 m^2 \omega }{3 m}, \nonumber \\
y_\mathrm{L} &=& \frac{1}{2} \omega  (\omega -m), \nonumber \\
y_\mathrm{R} &=& \frac{- \omega^4 - 2 \left( 5 - \frac{1}{\rho} \right) m \omega^3+\frac{12 \beta^2 + 11 \rho - 16}{\rho^2} m^2 \omega^2+ \frac{6 \beta ^2 +9 \rho - 10}{\rho^2} m^3 \omega+ \frac{\beta ^2+2 \rho-2}{\rho^2}m^4}{\left(m+2\omega \right)^2},   \nonumber \\
y^\mathrm{L}_\mathrm{R} &=&m E'  \left(1-\frac{ \left(m + 2 \omega \right)^2- m \omega }{E' \left(m + 2 \omega\right)}\right),\nonumber \\
r_\mathrm{L} &=& \left( -\frac{2+\beta}{3} \frac{\rho}{ \left(1+\beta \right)^2} +\frac{1}{6} \frac{4 + \beta }{1+\beta} \right)  \omega^2 + \left( \frac{\beta - \rho^2}{\rho \left( 1 + \beta \right)} +\frac{1}{2} \left( 1+ \frac{1}{\left( 1 + \beta \right)^2} \right)\right)  m \omega \nonumber \\
&+& \left( -\frac{(17 \beta^2 +36\beta+22) \rho}{30 (1 + \beta )^3} + \frac{14 \beta^2 +43 \beta+44}{60 (1 + \beta )^2}\right)  m^2, \nonumber \\
r_\mathrm{R} &=&\left(  -\frac{2+\beta}{3} \frac{\rho}{ \left(1+\beta \right)^2} +\frac{1}{6} \frac{4 + \beta }{1+\beta}\right)   \omega'^2 + \left( \frac{\beta^2  -5 \beta + 1}{3 \rho \left( 1 + \beta \right)} +\frac{1}{6} \frac{7 \beta^2+8 \beta -2 }{\left( 1 + \beta \right)^2}\right)  m \omega' \nonumber \\
&+& \left( \frac{ -23 \beta^3 +14\beta^2+41\beta-2}{30\rho (1 + \beta )^2}+\frac{-28 \beta \rho^2 +43 \beta^2+2}{30 \rho^2 (1 + \beta )}\right) m^2, \nonumber \\
r^\mathrm{L}_\mathrm{R} &=&  \frac{1}{3} \left(7 + \frac{5 \beta}{2}+\frac{2\beta^2-4 \beta -7}{\rho}\right) \frac{m^2}{1+\beta} + \left( 1 + \frac{1-2\rho}{1+\beta}\right)m \omega, \nonumber \\
q_\mathrm{L} &=&\left(  \frac{1}{2\beta} \frac{\rho}{1+\beta} - \frac{1+\beta}{2\beta}\right)  \omega^2 +  \frac{\beta}{2 \rho} m \omega +  \frac{1-\rho}{2 \beta}m^2, \nonumber \\
q_\mathrm{R} &=& \left(  \frac{1}{2\beta} \frac{\rho}{1+\beta} - \frac{1+\beta}{2\beta}\right) \omega'^2 +  \left(2 - \frac{1}{1 + \beta} - \frac{2-\beta}{2 \rho} \right)m \omega' \nonumber \\
&+&  \left(\frac{4 \beta ^3+\beta ^2-4 \beta +2}{4 \beta \rho^2} + \frac{-\beta ^3+2 \beta ^2+\beta -1}{2 \beta \rho (1 + \beta )} \right)m^2, \nonumber \\
q^\mathrm{L}_\mathrm{R} &=& \frac{\left( 1 - \beta \right) \omega^2 - 2 \rho m \omega +  \left(1+\frac{\beta}{2}\right) m^2}{\beta} \frac{l_0-\omega}{m}+ 
   \beta m E',  \nonumber \\  
z_\mathrm{L} &=&\frac{z^{\omega^4}  \omega^4 + z_\mathrm{L}^{\omega^3} m \omega^3 + z_\mathrm{L}^{\omega^2} m^2 \omega^2 +z_\mathrm{L}^{\omega}  m^3 \omega + z_\mathrm{L}^{0} m^4 }{m^2},  \nonumber \\
z_\mathrm{R} &=& \frac{ 2 z^{\omega^4}  \omega^5 + z_\mathrm{R}^{\omega^4} m \omega^4 + z_\mathrm{R}^{\omega^3} m^2 \omega^3 +z_\mathrm{R}^{\omega^2}  m^3 \omega^2 +  z_\mathrm{R}^{\omega} m^4 \omega +  z_\mathrm{R}^{0} m^5}{m^2 \left(m+2\omega\right)}, \nonumber \\
   z^\mathrm{L}_\mathrm{R} &=&\frac{2 l_0+9 m}{6} \left( l_0  - \frac{\rho \omega}{ 1+\beta}  \right), \nonumber \\
z^{\omega^4} &=& \frac{1}{15} - \frac{1}{15} \frac{\rho}{1+\beta}, \qquad \qquad \qquad ~  z_\mathrm{L}^{0}  = \frac{25 \beta^2 -49}{60 \rho^3} \left( 1 - \frac{1}{\rho}\right) - \frac{8 \beta^2}{15 \rho^2}, \nonumber \\ 
z_\mathrm{L}^{\omega^3}  &=& \frac{3 - \beta  }{30\rho} -\frac{3 +2\beta}{30 \left( 1+ \beta \right)} , \qquad \qquad
z_\mathrm{L}^{\omega^2}  = \frac{7 \beta^2 + 8 \beta - 23}{30 \left( 1+ \beta \right) \rho } - \frac{15 \beta^2 + 6 \beta - 23}{30 \rho^2}, \nonumber \\ 
z_\mathrm{L}^{\omega}  &=& \frac{-20 \beta^3+51 \beta^2 + 38 \beta -105}{60 \rho^3} - \frac{55 \beta^3+54 \beta^2 -82 \beta -105}{60 \rho^2 \left( 1 + \beta \right)}, \nonumber \\
z_\mathrm{R}^{\omega^4} &=& - \frac{8}{15 \rho}+ \frac{1}{15} \frac{8 - \beta }{1+ \beta} , \qquad \qquad
z_\mathrm{R}^{\omega^3}  = \frac{113 \beta^2 - 2 \beta - 133 }{30 \left( 1 + \beta \right)\rho} -\frac{143 \beta^2 - 34 \beta - 133}{30 \rho^2} ,  \nonumber \\
z_\mathrm{R}^{\omega^2}  &=& - \frac{339 \beta ^3-805 \beta ^2-353 \beta +851}{60 \rho^3} +\frac{-760 \beta ^3-825 \beta ^2+778 \beta +851}{60 \rho^2 \left( 1 + \beta \right)}, \nonumber \\ 
z_\mathrm{R}^{\omega}  &=& \frac{\beta  ((433-45 \beta ) \beta +44)-439}{30 \rho^3} + \frac{\beta  (\beta  (27 \beta  (11 \beta +1)-730)-29)+439}{30 \rho^4},\nonumber \\ 
z_\mathrm{R}^{0}  &=& \frac{270 \beta^2 -269}{60 \rho^3} + \frac{309 \beta^4 - 839 \beta^2 +538}{120 \rho^4},\nonumber
\eer
where $l_0 = m + \omega - E'$ and $\omega' = l_0$. Our result agrees numerically with Refs.~\cite{Aoki:1981kq,Passera:2000ug}. Integrated over the electron energy, it agrees with the total cross section of Appendix~\ref{app:total_crosssections}.

\section{Electromagnetic energy spectrum below electron endpoint}
\label{app:coefficients_electromagnetic}

For the remaining nonfactorizable contribution to the
electromagnetic energy spectrum $\mathrm{d} {\sigma}^{\nu_\ell e \to \nu_\ell e \gamma}_{\rm NF}$,
it is convenient to express the result as 
\begin{align}
  \mathrm{d} {\sigma}^{\nu_\ell e \to \nu_\ell e \gamma}_{\rm NF}
  = \frac{\alpha}{\pi} \delta_{\gamma} \mathrm{d} \sigma^{ \nu_\ell e \to \nu_\ell e }_{\mathrm{LO}}
+ \left( \mathrm{d} {\sigma}^{\nu_\ell e \to \nu_\ell e \gamma   }_{\,{\rm NF}} \right)^\prime,
\end{align}
where in the first term the cross section of the elastic process is expressed as a function of the final state neutrino energy,
and 
\ber
\delta_\gamma &=& \frac{1}{2\beta}  \ln \frac{1-\beta}{1+\beta}\left( 1  +  \ln \frac{\rho^{17/2} }{4 \beta^4 \left( 1-\beta \right)^{9/2}}   \right)  - 1 - 2 \ln \frac{1-\rho}{\rho} \nonumber \\
&-& \frac{1}{\beta}\left( \mathrm{Li}_2  \frac{-\rho^3}{\left(1+\beta \right)^3}  + \frac{1}{2} \mathrm{Li}_2  \frac{1-\beta}{1+\beta}  -  \mathrm{Li}_2 \frac{\rho}{1+\beta}  + \frac{\pi^2}{6}  \right).
\eer
As for the electron energy spectrum, individual corrections contain double logarithms,
\beq
\delta_v \underset{\beta \to 1}{\sim} -\frac{1}{8} \ln^2 \left( 1- \beta \right), \qquad
\delta_s \underset{\beta \to 1}{\sim} -\frac{1}{4} \ln^2 \left( 1- \beta \right), \qquad
\delta_\mathrm{II} \underset{\beta \to 1}{\sim} \frac{1}{2} \ln^2 \left( 1- \beta \right), \qquad
\delta_\gamma \underset{\beta \to 1}{\sim} -\frac{1}{8} \ln^2 \left( 1- \beta \right),
\eeq
but the complete electromagnetic energy spectrum is free from Sudakov double logarithms~\cite{Sudakov:1954sw}.
The residual nonfactorizable piece of the bremsstrahlung contribution,
$\left( \mathrm{d} {\sigma}^{\nu_\ell e \to \nu_\ell e \gamma   }_{\,{\rm NF}} \right)^\prime$
is given by the following substitutions in Eqs.~(\ref{elastic_xsection1}) and (\ref{elastic_xsection2}):
\ber
 \mathrm{\tilde{I}}_i \to \frac{ \pi^2  }{\omega^3} \left( a_i  + b_i \ln  \frac{1+\beta}{1-\beta} + c_i \ln \frac{2 - \rho }{ 1-\beta}  \right) \mathrm{d} \omega',\label{neutrino_spectrum_non_universal_part}
\eer
where coefficients $a_i,~b_i$, and $c_i$ can be expressed in terms of the initial and final neutrino energies, $\omega$ and $\omega'$, respectively, in the following form:
\ber
f_\mathrm{L} \left( \omega \right)& = & f^{\omega^2}  \omega^2 + f^{\omega} m \omega +  f^{0} m^2, \nonumber \\
f_\mathrm{R}  \left( \omega \right)=f_\mathrm{L} \left( - \omega' \right) & = & f^{\omega^2}  \omega'^2 - f^{\omega} m \omega' +  f^{0} m^2, \nonumber
\eer
with dimensionless coefficients,
\ber
c^{\omega^2} &=&  \frac{3 \beta^2+1}{3 \rho^{3}}-\frac{7 \beta^2+8}{3 \rho^2}, \nonumber \\
c^{ \omega }  &=& \frac{2 \left(\beta^2+4\right)}{\rho^3} + \frac{17 \beta^4+22 \beta^2-55}{8 \rho^4}, \nonumber \\
c^{0} &=& \frac{112 - 15 \beta^4-85 \beta^2}{15 \rho^5} + \frac{31 \beta^4+118 \beta^2-449}{60 \rho^4} , \nonumber \\
b^{\omega^2}  &=& \frac{(\beta-3) (2 \beta-1) \rho}{6 (1 - \beta)^2 \beta}+\frac{\beta+14}{6 \left(1- \beta \right)} , \nonumber \\
b^{ \omega } &=& \frac{((\beta-4) \beta-2) \rho}{2 (1 - \beta)^2 \beta} + \frac{55-\beta (17 \beta+30)}{16 \left( 1-\beta\right)^2}+\frac{1}{\beta}, \nonumber \\
b^{0}  &=&  \frac{\rho (\beta (\beta+1) (23 - 2 \beta)-45)}{30 (1-\beta)^{3} \beta} + \frac{-31 \beta^3-88 \beta^2+89 \beta+180}{120 (1 - \beta)^2 \beta}, \nonumber \\
a^{\omega^2}  &=& \frac{\rho\left(11 \beta^2+21\right)}{3 \left(\beta^4+2 \beta^2-3\right)} + \frac{2 \left(3 \beta^4+8 \beta^2-15\right)}{3 \left(\beta^4+2 \beta^2-3\right)} , \nonumber \\
a^{\omega }  &=& \frac{23 \beta^4+34 \beta^2-73}{4 \rho^3 \left(\beta^2+3\right)} + \frac{-2 \beta^4+13 \beta^2+73}{-4 \beta^4-8 \beta^2+12} , \nonumber \\
a^{0} &=& \frac{85 \beta^2-163}{30 \rho^3} + \frac{15 \beta^4-166 \beta^2+163}{30 \rho^4} .\nonumber
\eer
The interference part of the energy spectrum is determined by
\ber
a^\mathrm{L}_\mathrm{R} &=& \left( - \frac{\rho}{2-\rho} \frac{2 \omega \omega'}{m^2} - \frac{1}{3 \rho} + 4 \right) \omega^2 {\mathrm{I}}^\mathrm{L}_\mathrm{R},\nonumber  \\
b^\mathrm{L}_\mathrm{R} &=&   \left( \frac{1+\beta}{2 \beta}\frac{2 \omega \omega'}{m^2}  - \frac{1}{3 \beta^2} + \frac{1}{3 (1-\beta )} + \frac{7}{6} - \frac{1+\beta}{\rho} \left(\frac{1}{3 \beta^2} + \frac{7}{6 \beta }+ \frac{1}{6} \right) \right) \omega^2 {\mathrm{I}}^\mathrm{L}_\mathrm{R}, \nonumber \\
c^\mathrm{L}_\mathrm{R} &=&  \left( - \frac{2 \omega \omega'}{m^2}  + \frac{\beta^4-5 \beta^2+2}{3 \beta^2 \rho^2} + \frac{2 \left(1 + 4 \beta^2\right)}{3 \beta^2 \rho} \right) \omega^2 {\mathrm{I}}^\mathrm{L}_\mathrm{R},\nonumber
\eer
where ${\mathrm{I}}^\mathrm{L}_\mathrm{R}$ is given by Eq.~(\ref{elastic_I3}).

Our result agrees with the numerical evaluation in Ref.~\cite{Passera:2000ug}.

\section{Electromagnetic energy spectrum above electron endpoint}
\label{app:coefficients_electromagnetic2}

The electromagnetic energy spectrum above the electron endpoint can be conveniently expressed as a sum of the factorizable and nonfactorizable corrections, 
\begin{align}
  \mathrm{d} {\sigma}^{\nu_\ell e \to \nu_\ell e \gamma   }
  = \frac{\alpha}{\pi} \delta_{\gamma} \mathrm{d} \sigma^{ \nu_\ell e \to \nu_\ell e }_{\mathrm{LO}}
+ \left( \mathrm{d} {\sigma}^{\nu_\ell e \to \nu_\ell e \gamma   } \right)^\prime \,. 
\end{align}
The factorizable part is given by
\ber
\delta_\gamma &=&  \frac{1}{\beta}\left( - \frac{\pi^2}{3} + \frac{7}{8} \ln^2 \frac{1+\beta}{1-\beta} + 2 \ln \left( 1 + \frac{2 \omega}{m} \right)  \ln \frac{1+\beta}{1-\beta}  - \frac{3}{2} \ln  \frac{1+\beta}{1-\beta} \ln \frac{2-\rho}{1-\beta} +  2 \mathrm{Li}_2 \frac{\rho}{1+\beta}   \right. \nonumber \\
&+& \left.  \ln \frac{2 - \rho  \left( 1 + \frac{2 \omega}{m} \right) }{\rho  \left( 1 + \frac{2 \omega}{m} \right)}\ln \left( \frac{1+\beta}{1-\beta} \frac{1+ \beta - \rho  \left( 1 + \frac{2 \omega}{m} \right)}{ - 1 + \beta + \rho  \left( 1 + \frac{2 \omega}{m} \right)} \right)  - \mathrm{Li}_2 \frac{\rho \left(1 + \frac{2 \omega }{m}\right)}{1 + \beta }  - \mathrm{Li}_2\frac{2- \rho  \left( 1 + \frac{2 \omega}{m} \right)}{1+ \beta }   \right. \nonumber \\
&+& \left. \mathrm{Li}_2 \frac{2-\rho}{1+\beta }     + \Re \left( \mathrm{Li}_2 \frac{\rho \left(1 + \frac{2 \omega }{m}\right)}{1 - \beta }  + \mathrm{Li}_2\frac{2- \rho  \left( 1 + \frac{2 \omega}{m} \right)}{1- \beta } - \mathrm{Li}_2 \frac{2-\rho}{1-\beta }  \right) \right) \nonumber \\
&+& 2  \ln\left(  \frac{ 2 - \rho \left( 1 + \frac{2\omega}{m}\right) }{1 - \rho}  \frac{2 \omega \omega' + m \left( \omega' - \omega \right)}{-m^2} \right),
\eer
where the elastic cross section $\mathrm{d} \sigma^{ \nu_\ell e \to \nu_\ell e }_{\mathrm{LO}}$ is expressed in terms of $\omega'$. 
The nonfactorizable part is given by the following substitutions in Eqs.~(\ref{elastic_xsection1}) and (\ref{elastic_xsection2}):
\ber
 \mathrm{\tilde{I}}_i \to \frac{\pi^2  }{\omega^3} \left( a_i  + b_i \ln \frac{2 - \rho \left( 1 + \frac{2\omega}{m}\right)}{  \rho} + c_i \ln \left( 1 + \frac{2\omega}{m}\right) + d_i \ln \frac{2 - \rho \left( 1 + \frac{2\omega}{m}\right)}{  2-\rho} \right) \mathrm{d} \omega', \label{neutrino_spectrum_non_universal_part2}
\eer
with coefficients $a_i,~b_i,~c_i$, and $d_i$,
\ber
a_\mathrm{L} &=&\frac{\omega  \left(\frac{30 m^4 (2 \omega + m)}{m + 2 \omega - 2 \omega'}-\frac{15 m^5}{m-2 \omega '}-15 m^4+4 \left(109 m^2+78 m \omega +2 \omega
   ^2\right) \omega '^2-2 (m-2 \omega ) \left(11 m^2+4 \omega ^2\right) \omega ' \right)}{120 m^3} \nonumber \\
   &+& \frac{\omega  \omega '^3 \left(8 \omega -7 m-4
   \omega '\right)}{5 m^3}, \nonumber \\
b_\mathrm{L} &=& \frac{2 \omega ^5}{15 m^3} - \frac{\omega ^3}{3 m} - 2 \omega ^2, \nonumber \\
c_\mathrm{L} &=& -b_\mathrm{L} + \frac{m^2}{60}-\frac{9 m \omega }{8}-\frac{11 \omega ^2}{3}, \nonumber \\
d_\mathrm{L} &=&\frac{m^2 \left(6 \omega '^2 + 6 \omega  \omega '- 5 \omega ^2\right)+m \left(\omega -\omega '\right)
   \left(\omega ^2+7 \omega  \omega '+13 \omega '^2\right)+3 \omega '^2 \left(\omega -\omega '\right)^2}{3 m^2} \nonumber \\
   &+& \frac{2 m^5-135 m^4 \omega -16
   \left(\omega -\omega '\right)^3 \left(\omega ^2+3 \omega  \omega '+6 \omega '^2\right)}{120 m^3}, \nonumber \\
a_\mathrm{R} &=& a_\mathrm{L} \left( \omega \leftrightarrow - \omega' \right),  \nonumber \\
b_\mathrm{R} &=& b_\mathrm{L} \left( \omega \leftrightarrow  \omega' \right) + b_\mathrm{L} \left( \omega \leftrightarrow - \omega' \right)+ c_\mathrm{L} \left( \omega \leftrightarrow - \omega' \right) - d_\mathrm{L} \left( \omega \leftrightarrow -  \omega' \right) + 2 \omega'^2, \nonumber \\
c_\mathrm{R} &=& d_\mathrm{R} + \frac{\omega'^3}{3m} - \frac{2\omega'^5}{15m^3}, \nonumber \\
d_\mathrm{R} &=& d_\mathrm{L} \left( \omega \leftrightarrow - \omega' \right), \nonumber \\
a^\mathrm{L}_\mathrm{R} &=&\frac{4 \omega  \omega ' \left(\left(\omega -\omega ' - 3 m \right)^2 - 13 m^2\right)}{3 m \left(2
   E_{\text{EM}} - m \right)} ,\nonumber \\
b^\mathrm{L}_\mathrm{R} &=& 2 \left( \frac{\omega ^3}{3 m} -m \omega '-\omega ^2 + \omega  \omega ' \right),\nonumber \\
c^\mathrm{L}_\mathrm{R} &=&\frac{2 m^2}{3}-\frac{2 \omega ^3}{3 m}+m \left(3 \omega -\omega '\right)+2 \omega  \left(\omega -\omega '\right),\nonumber \\
d^\mathrm{L}_\mathrm{R} &=&c^\mathrm{L}_\mathrm{R}  + \frac{2}{3} \omega' \left( \frac{\omega'^2}{m} - 3 E_\mathrm{EM} \right), \nonumber
\eer
where $E_\mathrm{EM} = m + \omega - \omega'$, and as explained in Sec.~\ref{sec:effective_interaction} $\mathrm{d} \sigma / \mathrm{d} E' = \mathrm{d} \sigma / \mathrm{d} \omega'  $. Our result agrees with a numerical evaluation of Ref.~\cite{Passera:2000ug}. The total cross section from both regions of Secs. \ref{sec:below} and \ref{sec:above} is in agreement with Ref.~\cite{Bardin:1983zm}. Correcting obvious typos, the function $\mathrm{\tilde{I}}_\mathrm{L}^\mathrm{R}$ and only the function $\mathrm{\tilde{I}}_\mathrm{L}$ of Eq.~(\ref{neutrino_spectrum_non_universal_part2})  with the interchange $ \mathrm{\tilde{I}}_\mathrm{L} \leftrightarrow \mathrm{\tilde{I}}_\mathrm{R}$ are in agreement with Ref.~\cite{Bardin:1983zm}.
For all other kinematical factors of Secs. \ref{sec:below} and \ref{sec:above},
we find nontrivial discrepancies with Ref.~\cite{Bardin:1983zm}.

\section{Absolute cross section}
\label{app:total_crosssections}

The total cross-section correction including both real and virtual contributions, besides the closed fermion loop correction of Secs.~\ref{sec:long_Range} and~\ref{sec:hadron_physics}, is given by the following substitutions in Eqs.~(\ref{elastic_xsection1}) and~(\ref{elastic_xsection2})~\cite{Bardin:1983zm}:
\ber
\frac{ \tilde{\mathrm{I}}_\mathrm{L}}{\pi^2} &\to& \left( 1 + R\right) \mathrm{L}_2 + \frac{r^2 \left( 1 - r \right)}{2} \ln^2 R + 4 \left(  1 - R \right) \ln r - \left( r^2-\frac{r}{2}+\frac{3 R}{2}+\frac{10}{3}\right) \ln R-\frac{r}{2}+\frac{19\left( 1 - R \right)}{24}  ,\nonumber \\ \\
\frac{ \tilde{\mathrm{I}}_\mathrm{R}}{\pi^2} &\to& -4 r^2 (2 r+1) \mathrm{L}_3 + \left( 8 r^2+\frac{R^3}{3}+2 R+\frac{1}{3} \right) \mathrm{L}_2 -\frac{7 r^3}{6} \ln^2 R  + \left( 8 r-\frac{8 R^3}{9}+\frac{R^2}{3}-\frac{16 R}{3}+\frac{17}{9} \right) \ln r \nonumber \\
&-& \left( \frac{31 r^2}{3}-\frac{7 r}{3}-\frac{R^3}{18}+\frac{35 R}{6}+\frac{5}{3}\right) \ln R -\frac{7 r}{6}-\frac{11 R^3}{8}+\frac{13 R^2}{12}+\frac{73 R}{36}+\frac{43}{72},  \\
 \frac{\tilde{\mathrm{I}}^\mathrm{L}_\mathrm{R}}{\pi^2}&\to&- 4 r^3 \mathrm{L}_3 - \left( -4 r^2+2 r+R^2-R\right) \mathrm{L}_2 -r^2 \left( 2 + 5 r \right) \ln^2 R + \left( 4 r+3 R^2-7 R \right) \ln r\nonumber \\
 &+& 7 \left( -2 r^2+ r-R \right) \ln R -5 r+\frac{13 }{4} R^2 + \frac{15 }{4} R,
\eer
with additional definitions,
\ber
\mathrm{L}_2 &=& \frac{ \mathrm{Li}_2 \left( 1 - 1/ R^2\right) - \mathrm{Li}_2 \left( 1 - R^2\right) }{2}+\Re \left( \mathrm{Li}_2 \left( 1 + \frac{1}{R}\right) - \mathrm{Li}_2 \left( 1 + R\right)\right)+ \mathrm{Li}_2 \left(- \frac{1}{r}\right) + 2 \ln R \ln r, \nonumber \\ \\
\mathrm{L}_3 &=& \frac{\mathrm{Li}_3\left( 1 - 1/R^2 \right)  + \mathrm{Li}_3 \left( 1 - R^2 \right) }{2}  + 2 \left(  \mathrm{Li}_2 \left( -R \right) + \frac{1}{2} \mathrm{Li}_2  \left( R^2 \right)  \right)\ln R   - \mathrm{Li}_2 \left(- \frac{1}{r}\right) \ln R  - \ln^2 R \ln r  \nonumber \\
&-&\frac{\mathrm{Li}_3 \left( R^2 \right)}{4}   - \mathrm{Li}_3 \left( -R \right) - \mathrm{Li}_3 \left( -\frac{1}{R} \right) + \ln \left[ \left( 1 - R^2 \right) \left( 1 + R \right) \right] \ln^2 R- \pi^2 \ln \frac{1+R}{2 \sqrt{R}} - \zeta \left( 3 \right) \nonumber \\
&+&  \Re \left( 2 \left( \mathrm{Li}_3\left( 1 + \frac{1}{R} \right) +   \mathrm{Li}_3\left( 1 + R \right) \right) - \frac{1}{4} \mathrm{Li}_3 \left(\frac{1}{R^2} \right)  - 4 \mathrm{Li}_3 \left( 2 \right)  \right),  \\
R &=& \frac{m}{m + 2 \omega}, \qquad r = \frac{m}{2 \omega}.
\eer
Note that the total elastic cross section at leading order is given
by the following substitutions in Eqs.~(\ref{elastic0_xsection3}) and (\ref{elastic0_xsection4}), 
\ber
\int \mathrm{d} \omega^\prime \, \mathrm{I}_\mathrm{L} \to \omega (1 - R) , \qquad  \int \mathrm{d} \omega^\prime \, \mathrm{I}_\mathrm{R} \to \frac{\omega(1 - R^3)}{3},  \qquad \int \mathrm{d} \omega^\prime \mathrm{I}^\mathrm{L}_\mathrm{R} \to -\frac{ \omega R^2}{r}.
\eer

The ``dynamical'' correction of Secs.~\ref{sec:long_Range} and~\ref{sec:hadron_physics} to the total unpolarized cross section, 
$\sigma^{\nu_\ell e \to \nu_\ell e  }_\mathrm{dyn}$, can be expressed in the following form:
\ber
\sigma^{\nu_\ell e \to \nu_\ell e  }_\mathrm{dyn}
= \frac{\alpha}{\pi} \sum \limits_{\ell'}
\tilde{\sigma}^{ \nu_\ell e \to \nu_\ell  e }_\mathrm{dyn,\, \ell'}
+   \frac{\alpha}{\pi} \left(  \hat{\Pi}_{3 \gamma}^{(3)}(0)  - 2  \sin^2 \theta_W \hat{\Pi}_{\gamma \gamma}^{(3)}(0) +\frac{ Q_u \left( c^{u}_\mathrm{L} + c^{u}_\mathrm{R} \right) }{2 \sqrt{2} \mathrm{G}_{\rm F}} N_c \Pi \left( 0,~m_c\right) \right)
 \tilde{\sigma}^{ \nu_\ell e \to \nu_\ell  e }_{\mathrm{dyn},\, q}\,. \label{virtual_correction_VP_total}  
\eer
The reduced cross section due to the lepton $\ell'$ loop contribution $\tilde{\sigma}^{ \nu_\ell e \to \nu_\ell  e }_\mathrm{dyn,\, \ell'}$ is obtained by replacements of Eqs.~(\ref{replace1_VP})-(\ref{replace3_VP}) and the following substitutions in Eqs.~(\ref{elastic0_xsection3}) and (\ref{elastic0_xsection4}), 
\ber
Q_{\ell'} \int \mathrm{d} \omega^\prime \, \mathrm{\Pi} \left( q^2,~m_{\ell'} \right) \mathrm{I}_\mathrm{L} &\to& - \frac{\omega (1 - R)}{3} \ln \frac{\mu^2}{m^2_{\ell'}} +  \frac{\omega R}{3 r} \left( R^3_l \ln \frac{R_l+1}{R_l-1} - 2 R^2_l -\frac{2}{3} \right),  \\
Q_{\ell'} \int \mathrm{d} \omega^\prime \, \mathrm{\Pi} \left( q^2,~m_{\ell'} \right) \mathrm{I}_\mathrm{R} &\to& -\frac{\omega(1 - R^3)}{9} \ln \frac{\mu^2}{m^2_{\ell'}} +\frac{m R^3 R^3_l}{24 r^4} \left( \frac{3m^2+15 m \omega + 25 \omega^2}{3 \omega^2}-R^2_l \right)  \ln \frac{R_l +1}{R_l -1}    \nonumber  \\
&+& \frac{\omega R^3}{8 r^2} \left( 1 + \frac{5 + R^2_l}{6 r} \right)  \left( \left( R^2_l - 1 \right)^2 \ln^2 \frac{R_l +1}{R_l -1} - 4 R_l \ln \frac{R_l +1}{R_l -1} \right)+ \frac{m R^3 R^4_l}{24 r^4} \nonumber \\
&-& \frac{\omega R^3}{18 r^3} \left( \frac{6 m^2 + 39 m \omega + 53 \omega^2}{2 \omega^2} R^2_l +  \frac{m^2 - 10 m \omega - 18 \omega^2}{\omega^2}\right) , \\
Q_{\ell'} \int \mathrm{d} \omega^\prime \, \mathrm{\Pi} \left( q^2,~m_{\ell'} \right) \mathrm{I}^\mathrm{L}_\mathrm{R} &\to& \frac{ \omega R^2}{3 r} \ln \frac{\mu^2}{m^2_{\ell'}} +  \frac{ \omega R^2}{24 r}  \left( 3  \left( R^2_l - 1 \right)^2 \ln^2 \frac{R_l+1}{R_l-1} + 4 \left( R^2_l - 3 \right) R_l \ln \frac{R_l+1}{R_l-1} \right) \nonumber \\
&+&  \frac{ \omega R^2}{24 r} \left( \frac{112}{3} - 20 R^2_l\right),
\eer
with the vanishing in the limit $R_l \to \infty $ terms, beyond the first $\mu$-dependent contributions, where 
\ber
R_l = \sqrt{1 + 4 \frac{m^2_{\ell'}}{m^2} \frac{r^2}{R}}.
\eer
The reduced cross section arising from the quark loop contributions $\tilde{\sigma}^{\nu_\ell e \to \nu_\ell e   }_{\mathrm{dyn},\,q}$
is obtained replacing $\nu_\ell e$ couplings in Eqs.~(\ref{elastic0_xsection3}) and (\ref{elastic0_xsection4}) as
\ber
\int \mathrm{d} \omega^\prime \, \left(c^{\nu_\ell \ell'}_\mathrm{L} \right)^2 \mathrm{I}_\mathrm{L} &\to& 2 \sqrt{2} \mathrm{G}_{\rm F} c^{\nu_\ell  \ell'}_\mathrm{L} \omega (1 - R), \qquad  \int \mathrm{d} \omega^\prime \, c_\mathrm{R}^2 \mathrm{I}_\mathrm{R} \to 2\sqrt{2} \mathrm{G}_{\rm F} c_\mathrm{R} \frac{\omega(1 - R^3)}{3}, \nonumber \\
  \qquad \int \mathrm{d} \omega^\prime c^{\nu_\ell \ell'}_\mathrm{L}  c_\mathrm{R} \mathrm{I}^\mathrm{L}_\mathrm{R} &\to& -\sqrt{2} \mathrm{G}_{\rm F} \left( c^{\nu_\ell  \ell'}_\mathrm{L} + c_\mathrm{R} \right) \frac{ \omega R^2}{r}.
\eer

\section{Averaged over flux neutrino cross sections}
\label{app:plots_experiments}

\begin{figure}[H]
          \centering
          \includegraphics[height=0.5\textwidth]{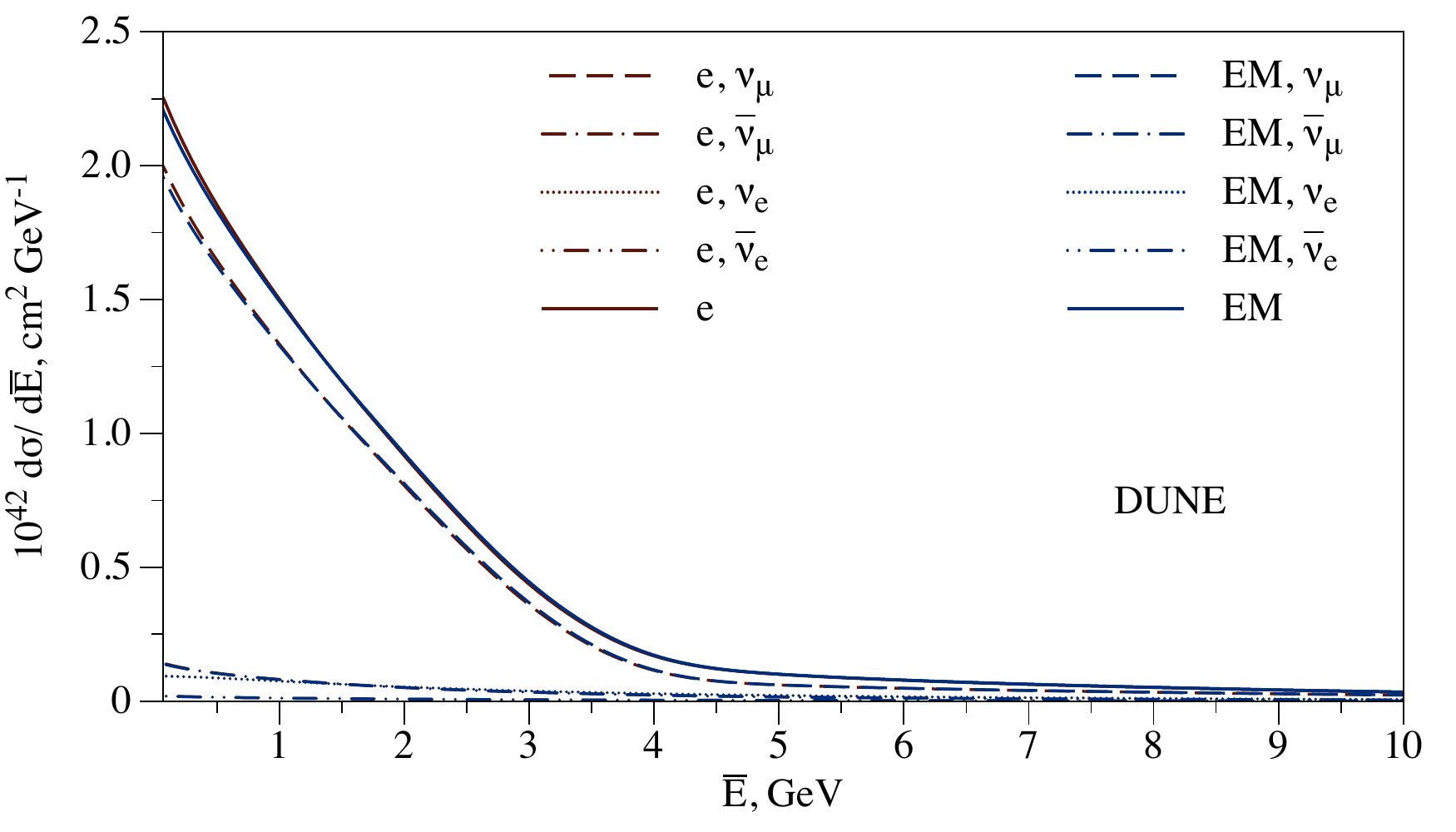}              
          \caption{Electron (e) and electromagnetic (EM) energy spectra in elastic neutrino-electron scattering for the neutrino beam mode of DUNE experiment. The electron energy spectrum is above at low energy. Electron and muon (anti)neutrino contributions are shown.
    \label{fig:experimental_spectrum1}}
\end{figure}

\begin{figure}[H]
          \centering
          \includegraphics[height=0.5\textwidth]{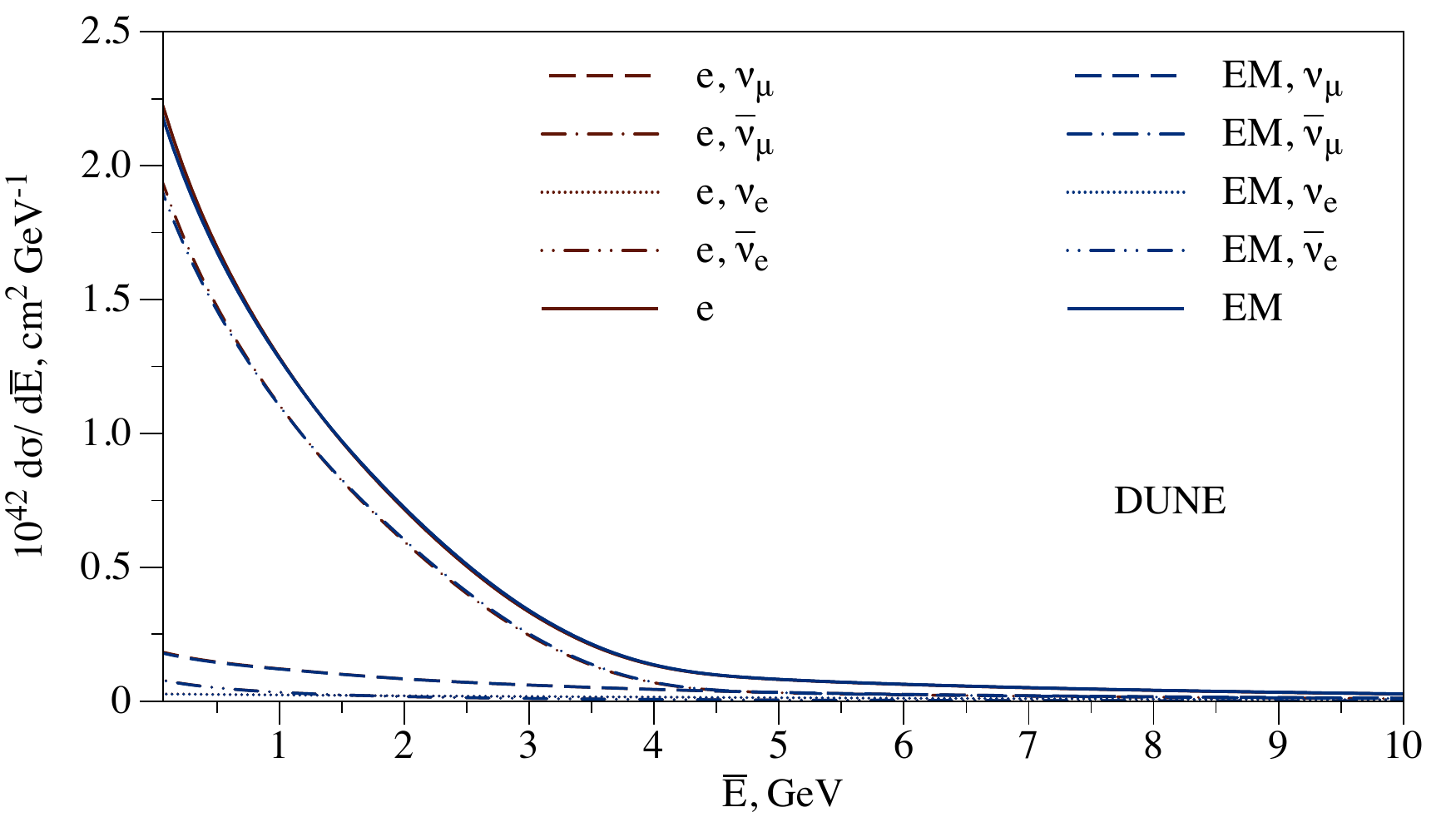}              
          \caption{Same as Fig. \ref{fig:experimental_spectrum1} for the antineutrino beam mode.
    \label{fig:experimental_spectrum2}}
\end{figure}

In the following, we average the energy spectrum with anticipated flux profiles of the DUNE Near Detector~\cite{Alion:2016uaj,dune_page} at Fermilab neglecting detector details. In Figs. \ref{fig:experimental_spectrum1} and \ref{fig:experimental_spectrum2}, we show the resulting electron and electromagnetic energy spectra for neutrino and antineutrino beam modes.

\begin{figure}[H]
          \centering
          \includegraphics[height=0.5\textwidth]{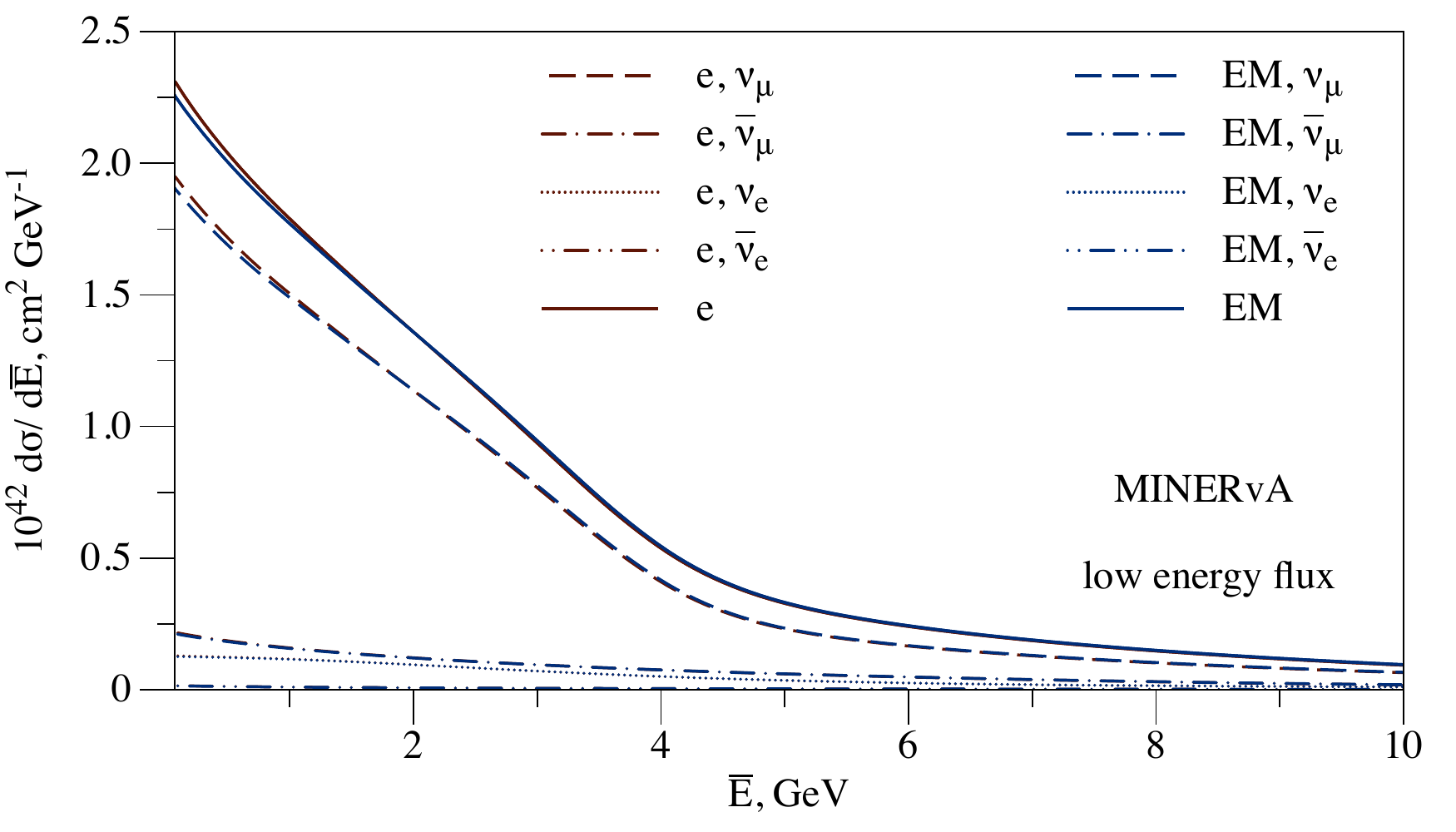}              
          \caption{Same as Fig. \ref{fig:experimental_spectrum1} for the MINERvA experiment.
    \label{fig:experimental_spectrum1_MINERvA}}
\end{figure}

\begin{figure}[H]
          \centering
          \includegraphics[height=0.5\textwidth]{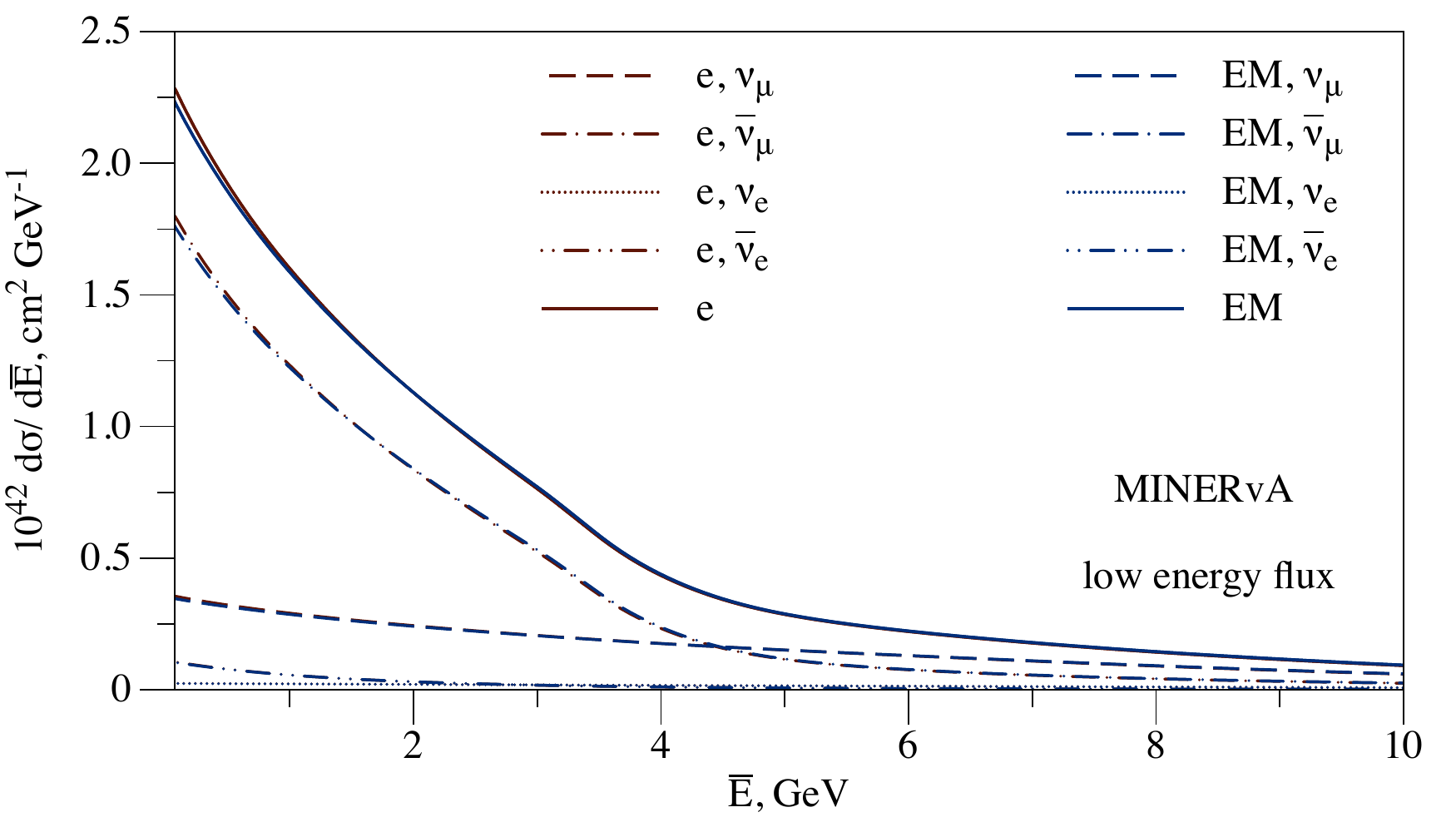}              
          \caption{Same as Fig. \ref{fig:experimental_spectrum2} for the MINERvA experiment.
    \label{fig:experimental_spectrum2_MINERvA}}
\end{figure}

\begin{figure}[H]
          \centering
          \includegraphics[height=0.5\textwidth]{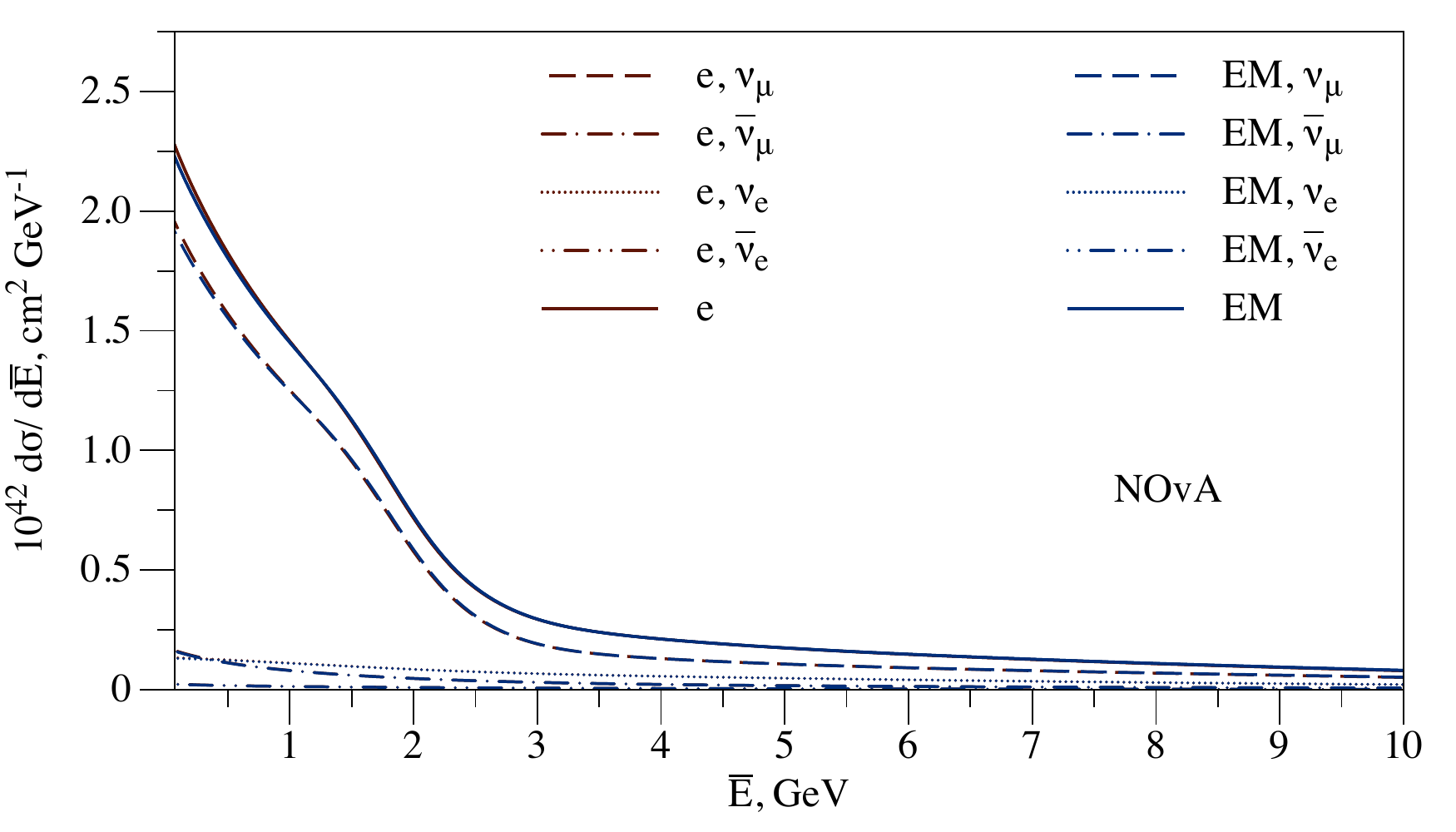}              
          \caption{Same as Fig. \ref{fig:experimental_spectrum1} for the NOvA experiment.
    \label{fig:experimental_spectrum1_NOvA}}
\end{figure}

\begin{figure}[H]
          \centering
          \includegraphics[height=0.5\textwidth]{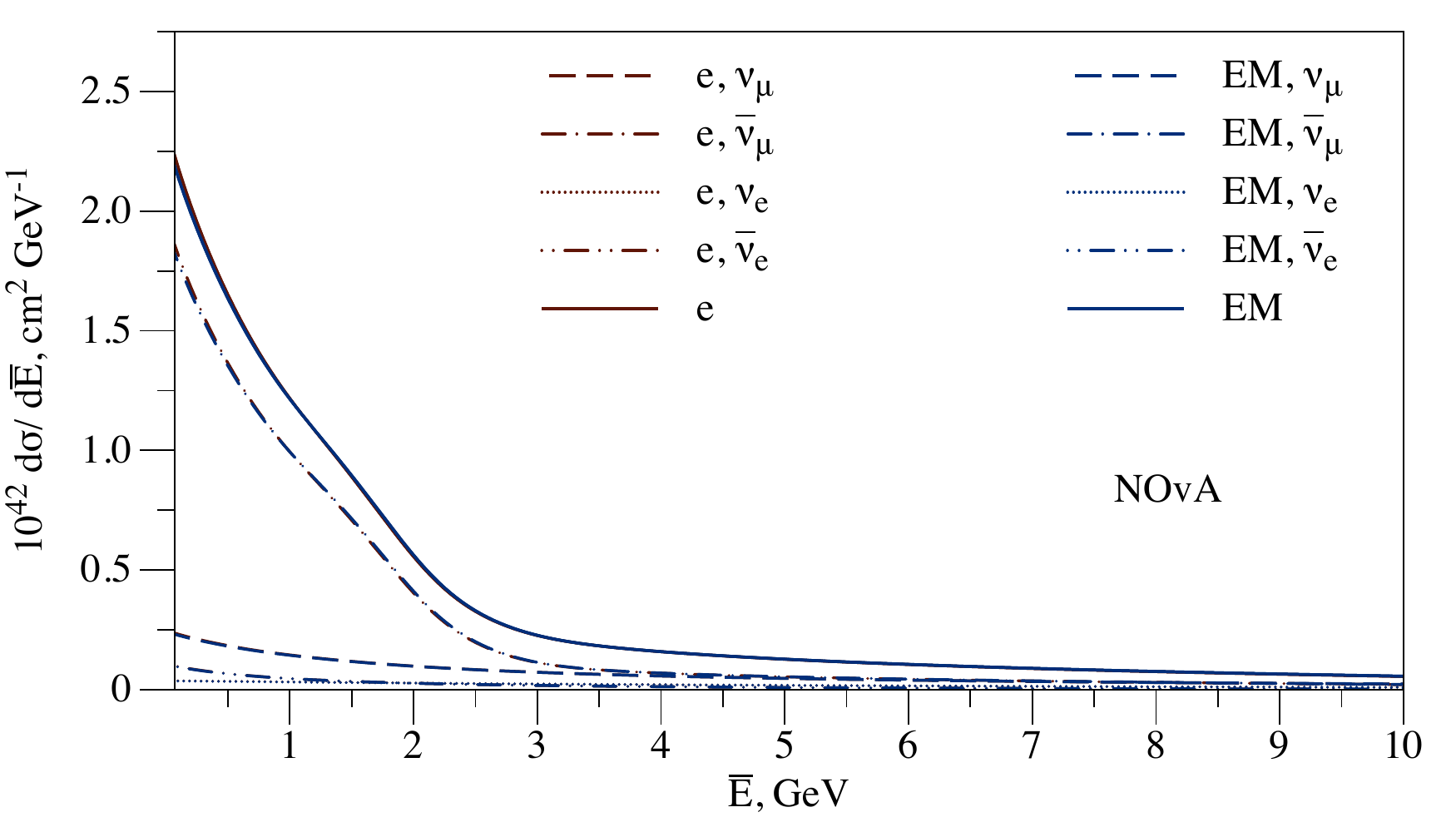}              
          \caption{Same as Fig. \ref{fig:experimental_spectrum2} for the NOvA experiment.
    \label{fig:experimental_spectrum2_NOvA}}
\end{figure}

\begin{figure}[H]
          \centering
          \includegraphics[height=0.5\textwidth]{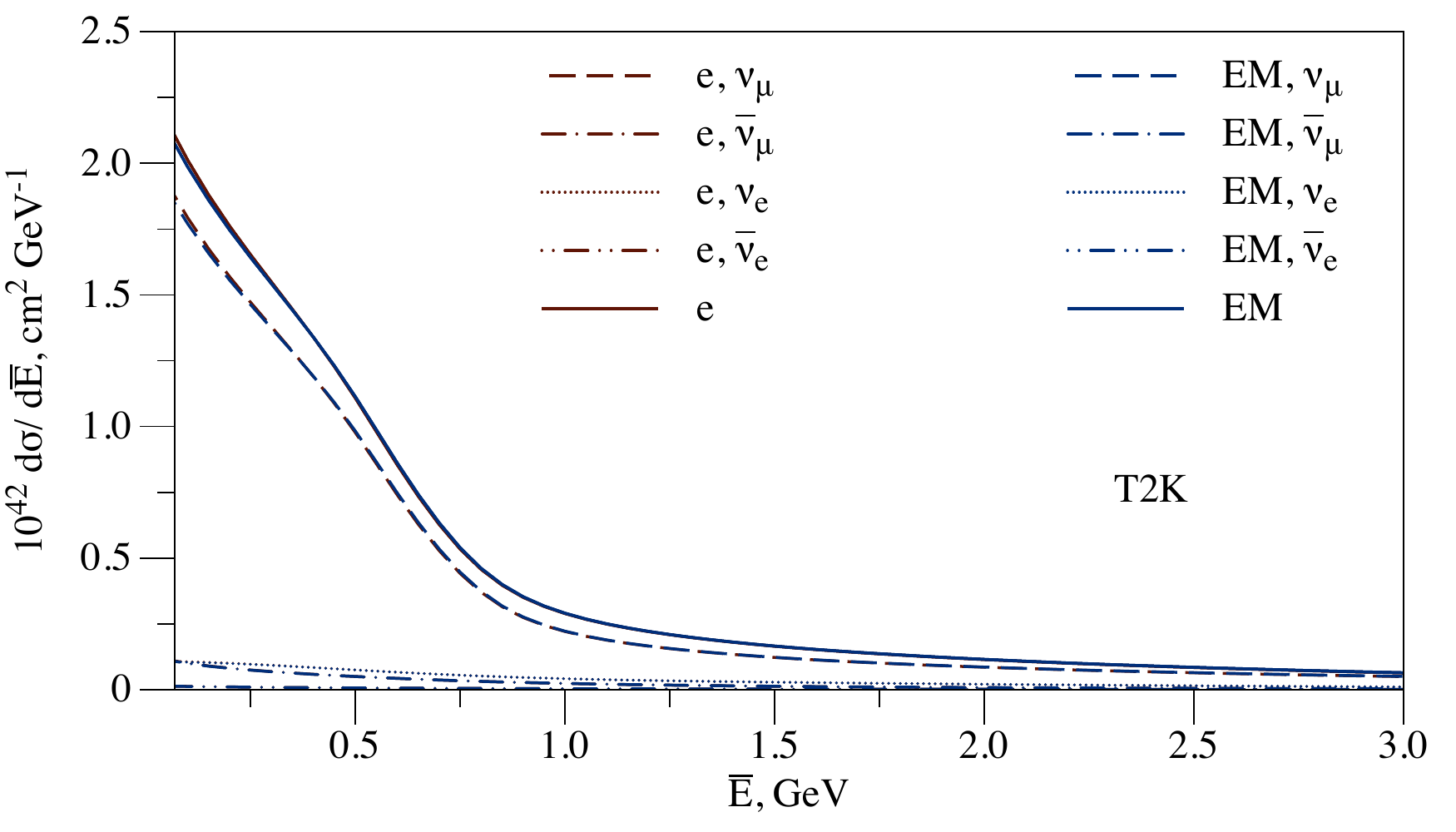}              
          \caption{Same as Fig.\ref{fig:experimental_spectrum1} for the T2K experiment.
    \label{fig:experimental_spectrum1_T2K}}
\end{figure}

\begin{figure}[H]
          \centering
          \includegraphics[height=0.5\textwidth]{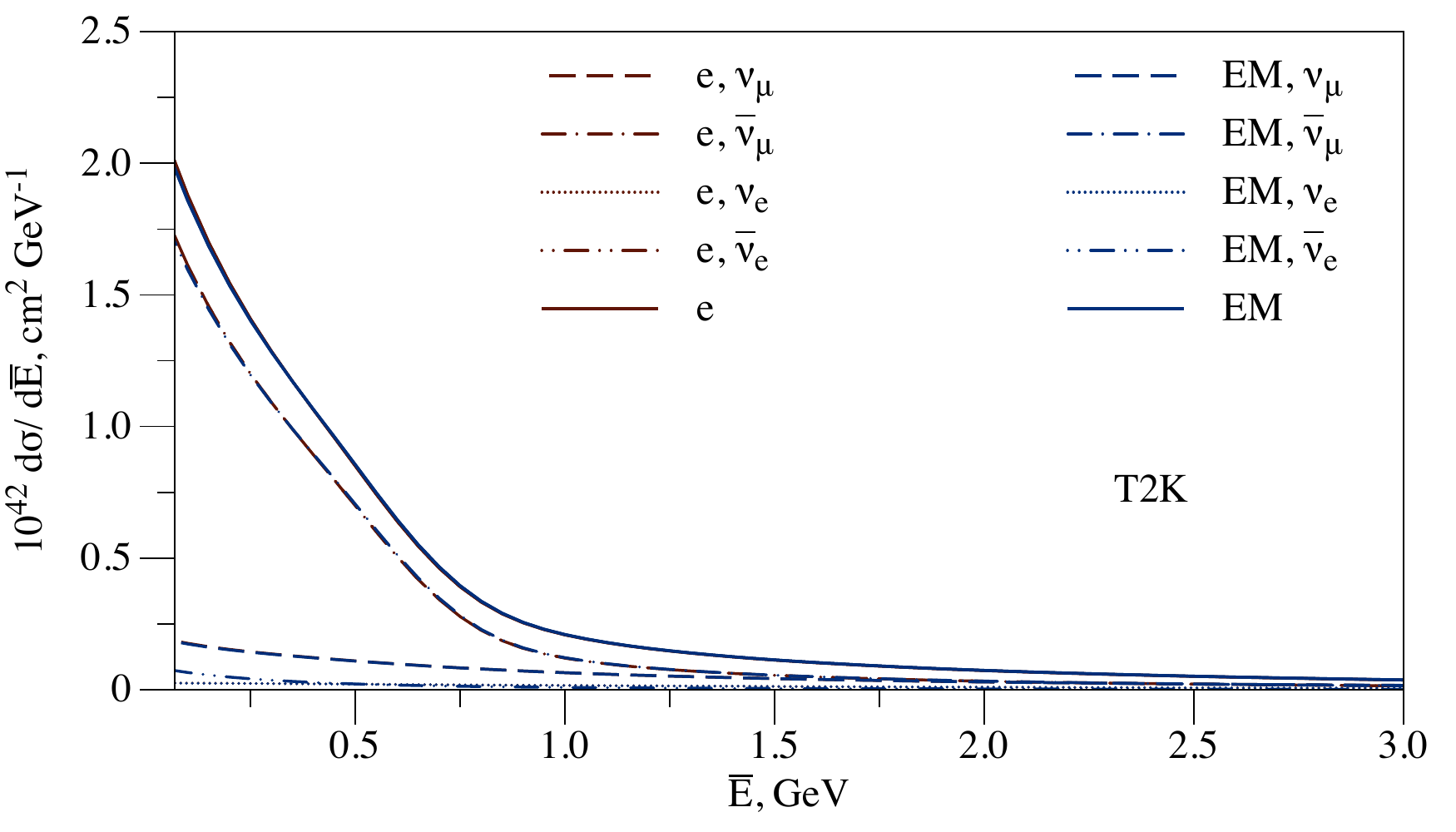}              
          \caption{Same as Fig. \ref{fig:experimental_spectrum2} for the T2K experiment.
    \label{fig:experimental_spectrum2_T2K}}
\end{figure}

The corresponding figures for
MINERvA~\cite{Park:2015eqa,Valencia:2019mkf,DeVan:2016rkm,Aliaga:2016oaz,Soplin:2016PhDthesis},
NOvA~\cite{NOvA_page},
and
T2K~\cite{Abe:2012av,Abe:2015awa} experiments are shown in
Figs.~\ref{fig:experimental_spectrum1_MINERvA}-\ref{fig:experimental_spectrum2_T2K}.
The difference between the electron and electromagnetic energy spectra slightly washes out after averaging over the typical neutrino flux. It is larger at low energies, where it can reach an effect of the relative order 1\%-3\%, and smaller at higher energies reflecting the dependence in Fig.~\ref{fig:nu_e_Etheta2}.


\newpage
\bibliography{NeutrinoElectron}{}

\end{document}